\documentclass[12pt]{article}
\usepackage{hyperref}
\usepackage{graphicx}
\usepackage{rotating}

\usepackage{amssymb}

 \def\a{\alpha}
\def\b{\beta}
\def\g{\gamma}
\def\G{\Gamma}

\def\d{\delta}
\def\e{\varepsilon}

\def\s{\sigma}

\def\u{\upsilon}

\def\beq{\begin{equation}}
\def\eeq{\end{equation}}
\def\beqn{\begin{eqnarray}}
\def\eeqn{\end{eqnarray}}
\def\ba{\begin{eqnarray}}
\def\ea{\end{eqnarray}}

\def\m{{\tt -}}
\def\p{{\tt +}}

\def\l{\langle}
\def\r{\rangle}

\def\xprim2bar{\overline{x}^{\prime\prime}}

\def\beq{\begin{equation}}
\def\eeq{\end{equation}}

\def\ppq{PQ\hspace{-12pt}/ ~\hspace{-2pt}/}

\newcommand{\beqa}{\begin{eqnarray}}
\newcommand{\eeqa}{\end{eqnarray}}

\let\a=\alpha   \let\b=\beta   \let\g=\gamma   \let\d=\delta
\let\e=\epsilon         
      \let\l=\lambda  \let\m=\mu
\let\n=\nu           \let\p=\pi      \let\r=\rho
\let\s=\sigma       \let\u=\upsilon 
         \let\w=\omega
\let\G=\Gamma     \let\L=\Lambda

\def\nn{\nonumber}

\def\sp{\;\;,\;\;}

\let\a=\alpha   \let\b=\beta   \let\g=\gamma   \let\d=\delta
\let\e=\epsilon         
      \let\l=\lambda  \let\m=\mu
\let\n=\nu           \let\p=\pi      \let\r=\rho
\let\s=\sigma       \let\u=\upsilon 
         \let\w=\omega
\let\G=\Gamma     \let\L=\Lambda

\def\nn{\nonumber}

\def\sp{\;\;,\;\;}

\newcommand{\be}{\begin{equation}}
\newcommand{\ee}{\end{equation}}
\newcommand{\bea}{\begin{eqnarray}}
\newcommand{\eea}{\end{eqnarray}}

\newcommand{\bb}[1]{\bar{\textbf{#1}}}
\newcommand{\bbb}[1]{\textbf{#1}}
\renewcommand{\u}[1]{{U(#1)}}

\usepackage{graphics}
\begin{document}
\begin{titlepage}
\setcounter{page}{1}
\rightline{hep-ph/0510332}
\rightline{CPHT-RR 063.1005}
\rightline{LTH 678}
\vspace{1.0cm}
\begin{center}{\Large \bf On the Effective Theory of Low-Scale Orientifold String Vacua}
\vspace{.17in}

\vspace{2cm}

       {\bf\large Claudio Corian\`{o} $^{1,2}$,$\;$  Nikos Irges $^{3}$
and Elias Kiritsis $^{3,4}$}
\vspace{.12in}

\vspace{1cm}

{\it  $^1$Dipartimento di Fisica, Universit\`{a} di Lecce, and \\
 INFN Sezione di Lecce,  Via Arnesano 73100 Lecce, Italy}\\
~\\
{\it $^2$ Department of Mathematical Sciences, University of Liverpool\\
 Liverpool L69 3BX, U.K.}\\
 ~\\
{\it
$^3$Department of Physics, University of Crete, 71003 Heraklion, Greece\\}
~\\
{\it
$^4$CPHT, Ecole Polytechnique, UMR du CNRS 7644, 91128, Palaiseau, FRANCE\\}
\end{center}
\vspace{.5cm}

\begin{abstract}
The effective field theory  of the minimal Low Scale Orientifold Models
is developed. It describes universal features of related orientifold vacua
in string theory.
It contains, beyond the Standard Model fields, an MSSM-like
Higgs sector and three anomalous (massive) U(1) gauge bosons.
All renormalizable couplings are included as well as some
dimension-five couplings that are important for anomaly
cancellation.
The Higgs symmetry breaking induces mixing
between the anomalous U(1) gauge bosons and the photon and $Z^0$.
This mixing as well as the anomaly generated cubic vector boson couplings
is potentially important for discriminating such models from other theories containing
Z's.
Some interesting tree-level processes are also evaluated.

\end{abstract}
\smallskip
\end{titlepage}
\renewcommand{\theequation}{\arabic{section}.\arabic{equation}}
\tableofcontents

%
\section{Introduction}
%

String theory owes its popularity and promise
to the fact that it includes consistently
gravity along with other gauge interactions.
The most explored set of vacua in string theory have a string
scale that is close to the four-dimensional Planck scale.
All successful heterotic vacua, whether perturbative,
or M-theoretic have this property. This is appealing due to indications from
running coupling  unification. On the other hand, no simple
reliable predictions are possible without a detailed vacuum, that
is sufficiently close to the Standard Model (SM) at low energy.
Although there are some heterotic vacua that come close
\cite{hetmod1,hetmod2}, it is fair to say that
none so far has passed all tests in a controllable fashion.
Moreover,
to put it simply, it is hard to see string effects at $E\sim 1$TeV
when $M_s\sim 10^{13}$ TeV.

In the past decade, other perturbative vacua of string theory have
been explored. A particularly interesting class are orientifold
vacua \cite{Sagn,GP,AngSagn}, that in their broad sense are compactifications of the type
I string. Their generic structures involves a compact six
dimensional manifold or orbifold thereof, times a Minkowski space.
The more general option, relevant in the presence of fluxes may
warp Minkowski space.
The internal space is threaded with D$_{p\geq 3}$ branes and
Orientifold planes
that
stretch along Minkowski space and have their potential  extra
dimensions wrapped around cycles of the internal manifold.

Because of this, there is no direct link between the string scale
and the four-dimensional Planck scale. By adjusting (if possible)
the internal volumes any possible number for $M_s<M_P$ can be
obtained \cite{Anton,Lykken}. Of course, especially in the absence of supersymmetry,
volumes along with other moduli acquire potentials, and their
values are determined dynamically. It has been argued that there may
be vacua where the string scale could be as low as 1 TeV although
to the present day, no reliable such vacuum exists.

There has been quite a bit of success though in model building so
far with a high string scale (see \cite{u1,u2,u3} and references therein),
although, as in the heterotic case,
there is no perfect vacuum yet.

Low scale string vacua, have the undeniable charm that there may
be amenable to experimental tests. Even though, as already stated, there may be
no such model at present solving the tadpoles conditions, their
general structure has characteristic features, and experimental
signals that are essentially generic.
The purpose of the present work is to formulate and parameterize  the generic
low energy action of the most interesting class of such vacua, that
we call minimal Low Scale Orientifold Models or mLSOM for
short.
Such an effective action  can help both the string theory search for such vacua, currently under way
\cite{adks}, as well as the
parametrization and computation of experimental observables.

Orientifold vacua have a conceptual simplification build in: there
is a clear separation typically between the open string spectrum,
coming from the D-branes, and the ``bulk" spectrum coming from the
unoriented closed strings. The graviton is part of the bulk
spectrum, whereas the branes give rise to particles at the massless sector
with spin at most one.

The standard model gauge group and other particles is naturally
realized on the D-branes rather than the bulk. There are several
reason for this. A simple and powerful one is that it is not
possible to realize the non-abelian structure of the SM including
its reps in type II string theory \cite{kav}

The gauge group coming from the D-branes is a product of classical
but not exceptional groups, each factor coming from a stack of branes at the same point in
 transverse space. The minimal gauge groups that can accommodate
 the standard model particles are U(3)$\times$U(2)$\times$U(1),  and
 U(3)$\times$U(2)$\times$U(1)$\times$U(1)' \cite{AKT,m1,m2}. There are variants
 where U(2)$\to $Sp(2)$\sim $SU(2), and U(1)$\to$O(2).
 There will be in general a hidden group, that for the whole paper
 we will neglect, although it may be important (depending on the model) for issues of
 supersymmetry breaking.
Of course one may consider more complicated groups. The Pati-Salam
like
group U(4)$\times$U(2)$^2$ is the simplest example.
However unlike field theory, here the minimal groups are more
advantageous, since the larger ones must be eventually broken and
we should be able to describe them directly in their broken phase.

There is an obvious observation: all such embeddings involve U(1)
factors that are more numerous than what we know in the SM, namely
the hypercharge \cite{AKT}. It is also known that many U(1)'s can
be anomalous in orientifolds \cite{IRU}.
Anomalies cancel, although the appropriate charge traces
are non-zero, thanks to variants of the Green-Schwarz mechanism \cite{gs}.
It is known that anomalous U(1)s become eventually massive, and the
associated gauge symmetry is broken. Under certain conditions, the
global symmetry may remain unbroken in perturbation theory (see
\cite{K1} for a review).
It is then hoped that all extra U(1)'s except the hypercharge become hopefully
massive. In fact, this is generically the case\footnote{Aspects of the effective theory of
anomalous U(1)s have been analyzed in \cite{KN1}.}.

The anomalous U(1)
masses, can be calculated unambiguously via a one-loop (annulus)
computation \cite{AKR,A2,A1}. A rich pattern of masses appears. It turns
out that the physical masses are bounded above by the string scale, but
can be arbitrarily low, if some internal dimensions are large.
In the generic case however they turn out to be a few times
smaller than $M_s$, if one includes, $\pi$'s and i's.

In fact,  the anomalous U(1)'s gauge bosons have essentially all
their renormalizable couplings fixed by charges and anomalies.
Apart from their minimal couplings, they mix with appropriate
bulk axions, that couple to other gauge fields via PQ-type
couplings. They also have most of the time, cubic Chern Simons-like interactions
due to anomalies \cite{abdk}.
The effective cubic couplings together with the non-zero triangle diagrams,
 provide an effective cubic vertex to the
anomalous U(1)s gauge bosons.
This effect is absent from usual
non-anomalous Z's. They may have therefore signals that distinguish
them from other Z' candidates,  \cite{CL,L,DND}.
Moreover, due to the fact that the Higgs is always charged under such anomalous U(1)'s
the Z's mix with the $Z^0$ gauge boson. Therefore, the photon and the $Z^0$ acquire a
(suppressed) cubic vertex.

If the string scale is in the TeV range, such anomalous U(1)'s  are prime
candidates for detection. At the same time, they provide many
contributions to known processes, that could exclude ranges of the
parameters (Z-couplings \cite{GIIQ},\cite{G1}, \cite{AS1} and $g-2$ \cite{AK1} being two
examples that have been partially studied so far).

Apart from the SM spectrum and the anomalous U(1) gauge bosons, there are other
 low-energy particles in the orientifold vacua
with low string scale. We will enumerate them below and describe
briefly their characteristics.

\begin{itemize}

\item Additional Higgses. Higgses typically come
in pairs, even if supersymmetry is broken at the
string scale.
The large scale study of \cite{sche} based on
the hypercharge embedding of \cite{BKLO}, \cite{IMR},
shows that there are
different vacua with a  variable number of doublets.
Of course there should be at least one. And for simplicity we
assume that there are no others around.

\item Superpartners of the SM particles. Depending on the way
supersymmetry is broken, they may have masses that are well below,
to around the string scale. In fact, in low scale orientifold
models, the most natural way of breaking supersymmetry is the ``
explicit breaking" which gives $M_s$ as the susy breaking scale.
In such a case the partners have masses at the string scale and
they are typically heavier than the anomalous U(1) gauge bosons,
with the possible exception of the higgsinos.

\item Non-chiral massless states. There are no such states in the SM
therefore they must be somehow lifted in mass. It is possible,
combining intersecting branes with Scherk-Schwarz deformations
to actually remove all such states from the massless spectrum \cite{CN}.

\item  Possible hidden groups, encompassing all other massless-level open string
states that do not directly interact (by construction) with the SM
particles.

\item Open string KK-states. Some of the branes may wrap internal
dimensions. As explained in \cite{AKRT}, when the string scale is
low, the most advantageous configuration has two large dimensions.
All others have size at most twice the string length. Moreover all
SM particles wrap the small dimensions. Therefore, their KK states
have masses at the string scale.
There are two exceptions.
The first is when one of the branes wraps the two large dimensions.
The associated anomalous gauge boson, on the other hand is massive,
 and it should be arranged that the mass
comes from N=2 sectors so that it is of the order of the string scale \cite{AKRT}.
Therefore although its KK states are very tightly spaced, its zero point mass is large.
It is interesting that this type of massive gauge boson, might
have a very particular signal at LHC, because of this very special
property.

The second exception concerns the KK states of the right-handed
neutrinos that come from the above described brane. These mix with the zero
modes and provide an interesting pattern of masses. This was
analyzed in \cite{AKRT}.

\item Stringy states of open strings. All the states above have
stringy excitations (vibration modes) of the associated open
strings with masses at the string scale and above.

\item Massless bulk modes, including the graviton, having
 gravitational strength couplings to the open sector.
 After the breaking of susy, all but the graviton should acquire
 masses.

 \item Bulk KK states. Since there are two compactification
 scales, the one that dominates at low energy is associated to the two larger
 dimensions. The physics of such KK states has been analyzed in
 the  past \cite{low3}

 \item Bulk stringy states, with masses at the string scale or
 more.

\end{itemize}

Typically, apart from the SM particles, the particles that are
lightest from the brane particles, are firstly the anomalous U(1)
gauge bosons and then superpartners. The distribution of masses
depends on the vacuum.
In this paper, we will neglect superpartners, since this is a well
studied sector.
We will focus on the anomalous U(1) gauge bosons, the Higgses
and the SM particles. The bulk axions that are crucial for anomaly
cancellation will also be included.

In successful low scale orientifold vacua, baryon and lepton
number are gauge symmetries. They are in fact some of the
anomalous U(1)s. Their gauge bosons will become massive but the
associated global symmetries
will remain intact in perturbation theory. This is a crucial fact
, since at low $M_s$ baryon and lepton number violating operators
will be hardly suppressed, \cite{Ibanez:1999it}. There will be breaking due to
instantons but this is known to be small.

This general class of models has important open problems that need
to be eventually addressed at the string level,
in order to have concrete successful string vacua
that realize this setup.

\begin{itemize}

\item The setup needs radii much larger than the
string scale. This hierarchy, leading to a low string scale must be
explained/accomodated.

\item It must be arranged that the PQ symmetry is explicitly
broken, in order to avoid a massless axion.

\item The problem of one-loop tadpoles needs to be accommodated
somehow.

\end{itemize}

There are several important effects, in the effective theory
we are describing. A
crucial ingredient is that in all cases, the Higgs gauge bosons
are charged under one linear combination of the anomalous U(1)s.
In fact we can go to a basis (a non-orthogonal one) where the
four generic U(1) symmetries of the low scale orientifold vacua
are hypercharge, Y, baryon number B, lepton number, L, and a
Peccei-Quinn-like symmetry PQ. The Higgs then has B=L=0, and its vacuum
expectation value breaks Y and PQ.

The UV mass matrix of the U(1) gauge bosons is
characterized by three mass eigenvalues of order $M_s$,
(hypercharge is massless) as well as
three mixing angles. The second source of gauge boson masses
is the Higgs symmetry breaking.
Due to the (mild) hierarchy of the Higgs vev and $M_s$ there is
interesting pattern in the gauge boson mass-eigenstates.

The photon is the usual mixture of Y and $W^3$.
However, the $Z^0$, apart from its Y and $W^3$ components,
 it has a small ($\sim {\cal O}(M^2_Z/M^2_s)$)
admixture of the other three anomalous U(1) gauge bosons.
Similarly, the three heavy $Z'$s have a small admixture of Y,$W^3$.
The presence of this mixing affects in an interesting way several
issues:
\begin{itemize}

\item $Z^0$ has non-standard couplings to fermions. This also
affects the $\rho$ parameter.

\item $\gamma$ and $Z^0$ acquire a trilinear vertex, an avatar of
their mixing to the anomalous U(1)s and the triple anomalous U(1)
vertex. This is very interesting for LHC.

\item There are non-standard photon and $Z^0$ couplings to the
Higgs.

\end{itemize}

It is these issues that we will analyse to a certain extent in the present
paper.

We should also briefly mention the parameters of the effective
field theory. We do have, to start with,  all the SM parameters.

The Higgs sector resembles that of the MSSM, in the sense that it has
two Higgses. However, if supersymmetry is broken at the string
scale, the structure of the potential at the level of the
quadratic terms maybe  different.
It depends in fact on the way supersymmetry is broken. However for
orbifold and SS breaking the tree level potential
is of the supersymmetric type.
However, in this paper, for generality we will keep all possible terms.

 We split the terms in the Higgs
potential into those that preserve the PQ symmetry and those who
do not. The PQ-preserving part has four real quartic couplings and
two quadratic ones. The PQ-breaking part has one complex quadratic
coupling and three complex quartic ones. It is essential for
giving a mass to an otherwise massless scalar, the axi-Higgs, a mixture of one if
the Higgs phases and the bulk axions.

The anomalous U(1) sector has a 4$\times$4 UV mass matrix that is
generically not diagonal. One of its eigenvalues is zero
corresponding to the hypercharge. The hypercharge linear
combination is fixed by a set of integers.
For mLSOM, there are two choices. The rest of the matrix can be
parameterized in terms of three mass eigenvalues and three mixing
angles.
The axion-gauge boson mixing, axion-gauge boson CP-odd
couplings as well as the CS-like couplings are then
 determined in terms of the mass matrix and the
charges, that are known.

One of the anomalous U(1) gauge bosons comes from a brane that wraps the
two large dimensions \cite{AKT}. This implies that its UV mass term
as well as the mixing terms with the other U(1)
gauge bosons should be anywhere between $M_s$ and $\sim 10^{-3}$ eV.
We will assume in this paper, for simplicity that its mass comes from an
N=2 sector and therefore its physical mass is
 of the order of the string scale.
 In any case, its mass must be larger than around 50 MeV
 to avoid standard supernova cooling constraints \cite{AKRT}.

In the neutrino sector that is not discussed in this paper, there
are further parameters that enter. If there is a single bulk right
handed neutrino then there are three parameters associated to its
coupling to the three lepton doublets. If there are three bulk
neutrinos, then one has the standard KM-like mixing matrix.
On top of this there is neutrino mixing with the right-handed
neutrino KK modes that are densely spaced.
In the simplest uniform
case of a $T^2$ with the same radius, it is the radius that enters
as an new parameter (constrained at the same time by
fitting the gauge couplings and the Planck scale.)
These issues are discussed in detail in \cite{AKRT}.
In the sequel we choose the innocuous case of three bulk
neutrinos and neglect the mixing with the KK states.

The structure of this paper is as follows: In section \ref{string}
we describe the string theory origin of the effective field theories that we
analyze. They should correspond to string theory vacua with a string scale in the TeV range.

In section \ref{t1} we describe the effective action under study.
We describe in detail the
UV (stringy) gauge boson mass terms, and we also describe convenient
 gauge fixings in the unbroken phase.

In section \ref{t3} we analyze in detail the issue of electroweak
symmetry breaking. This is of importance as
the properties of Z's are affected importantly. We discuss
in particular, the gauge boson masses
the structure of the Higgs sector, the details of the
 Green-Schwarz sector responsible for anomaly cancellation
and finally a convenient gauge-fixing in the broken phase.

Section \ref{t4} contains the computation of various tree level cross
sections that are relevant for constraining the parameter space and
analyzing new physics in this class of models.

Finally \ref{t5} contains our conclusions and further comments.

Appendix \ref{A} contains a comparison of the Higgs sector here with the SM Higgs sector
while appendix \ref{B} contains a rewriting of the Lagrangian in the physical basis.

%
\section{String theory origin of the mLSOM\label{string}}
%

In this section\footnote{Reading this section is a not a prerequisite
in understanding the rest of the paper.
It does however give a motivation for the effective theories,
and also some idea on what parameter choices are easy
to accommodate and which not.} we motivate the type of effective theory we
will be studying in this paper, by linking it to a class of
interesting vacua of string theory.
These are known as orientifold vacua.
Useful reviews introducing this subject and describing
 recent progress can be found in \cite{u1}-\cite{u4}.

The generic structure is as
follows. The ten dimensions of superstring string theory are split
into four flat non-compact dimensions and six compact dimensions,
threaded with other possible background fields (tensor fluxes).
Several groups of D$_{p\geq 3}$-branes are inserted in this vacuum, so that
their 3+1 dimensions are parallel and fill  the four dimensional Minkowski
space. If they have more dimensions, then these wrap appropriately
some cycle of the internal compact manifold. There will be
generically also Orientifold planes, non-dynamical hyperplanes,
with typically negative energy density. Their basic property is to
change the orientation of open and closed strings. They are
typically required for the consistency of the theory, and they
enter crucially both in anomaly cancellation but also in the
conditions for IR stability (or absence of UV divergences).

We shall restrict
ourselves to models in which the closed string sector is
supersymmetric, while supersymmetry is generically broken by the
open strings at the string scale \cite{bsb}.
We intend to have a string scale $M_s$ that is in the TeV range.
$M_s$  is related to the four-dimensional Planck scale as
\be
M_P^2={V_6\over g_s^2}M_s^2,
\ee
where $V_6$ is the volume of the internal six-dimensional manifold
in string units, and $g_s$ the string coupling constant, that is
smaller but not much smaller than one\footnote{This is because it
enters into the gauge couplings constants.
Once there are D$_{3}$ branes , or the volumes higher branes wrap are string-scale sized,
then the gauge couplings at the string scale are essentially determined by $g_s$}.
Therefore a low string scale implies a (very) large volume for the
internal manifold. Its linear dimension is $V_6^{1\over 6}\sim (M_p/M_s)^{1\over 3}$.
For $M_s=1$ TeV, $V_6^{-1/6}\sim 10$ MeV.
We know however that the internal manifold must
not be uniformly large. It must have small cycles, otherwise
some of the standard models fields would have KK states with masses
$\sim 10$ MeV and this is obviously experimentally excluded.

A convenient way to describe this is in orientifold model building
based on orbifolds of $T^6$. There, we can take a number n of radii
to be large and the rest 6-n to be close to the string scale.
There are however cases to be avoided. If only one radius is
large, then it is macroscopic and therefore excluded. Moreover, this
is a highly unstable situation \cite{ab}. If all of
them are large, there is no space to wrap some of the SM branes
since this will produce unacceptable KK descendants of the SM
particles as argued above. In fact we should have as many small
dimensions as possible to allow manoeuvering the SM branes.
This gives the case of two large dimensions (with size in the 1$\m$-1mm range),
as the optimal possibility.

There has been a wide search for D-brane configurations that
provide the standard model in the context of orientifolds, and
have acceptable gauge coupling properties \cite{AKT,AKRT,lust,riz}.
It turns out that
the minimal number of stacks necessary to allow for a low string
scale is 4. One could do with three, but there the string scale
must be close to the Planck scale. This has been analyzed in
\cite{ad}.

The existence of the two large dimensions, provides an immediate
mechanism for light neutrinos \cite{bulknu}: If the right-handed neutrinos
emerge from a U(1) brane wrapping the two large dimensions, then,
as shown in detail in  \cite{AKRT}, the neutrinos will have
masses, with the right order of magnitude.\footnote{There are two options with neutrinos.
The first is that there is a single bulk neutrino which couples to the SM
ones. The KK modes also play a role here. This is a very constrained
situation. In \cite{AKRT} it was shown that this option lies at
the borderline with the current neutrino data. It was pointed out
recently in \cite{msled} that if one weakens the coupling between branes and
bulk, then this option is viable.
The other possibility, involves three bulk neutrinos. This
is much less constrained, but also less predictive.}
We will label this brane as U(1)$_b$ to indicate that it is the
only brane that wraps the two large dimensions.

Therefore within our framework, the minimal ensemble of D-branes needed in
our construction contains the following stacks: a
stack of three coincident branes to generate the color group, a
second stack of two coincident branes to describe the weak
$SU(2)_L$ gauge bosons, and one more brane to generate the
$U(1)_b$ bulk discussed above. The resulting gauge group so far is
$U(3)_c\times U(2)_L\times U(1)_b$, with the three $U(1)$
generators denoted by $Q_c$, $Q_L$ and $Q_b$, respectively.
Since
the string scale will be low, to
ensure proton stability, we require baryon number conservation
with generator $B\equiv Q_c$. The hypercharge $Y$ cannot have a
component along $Q_b$, since this would lead to unrealistically
small gauge coupling, and as explained in \cite{AKT}
the correct assignment of SM quantum numbers requires the presence
of an extra abelian factor, named $U(1)_1$ with generator $Q_1$,
living on an additional brane. In the simplest situation
this brane should lie on top of the
color or the weak stack of branes, as we argue below.
However, one may relax some of the assumptions, and have more
freedom with the U(1)$_1$ coupling constants.

In our framework, supersymmetry is expected to be broken by combinations of (anti)branes
and orientifolds which preserve different subsets of the bulk supersymmetries.
The simplest possibility is that any
pair of D-branes D$p$ and D$p'$ should satisfy $p-p'=0$ mod 4. It follows that a system
with three stacks of mutually orthogonal branes in the six-dimensional internal
(compact) space consists, up to T-dualities, of D9-branes with two different types
of D5-branes, extended in different directions. Specifically, the $U(1)_b$ lives
on the D9-brane, while the $U(3)_c$ and $U(2)_L$ are confined on two stacks of
5-branes, the first along say the 012345 and the other along the 012367 directions
of ten-dimensional space-time. Thus, the (sub-millimeter) bulk is necessarily
two-dimensional (extended along the 89 directions), and the additional $U(1)_1$
brane has to coincide with either $U(3)_c$ or $U(2)_L$. The parameters of the
model are the string scale $M_s$, the string coupling $g_s$ and the volumes
$v_{45}$,
$v_{67}$ and $v_{89}$ of the corresponding subspaces, in string
units. Using T-duality, we choose all internal volumes to be bigger than
unity, $v_{ij}>1$. In terms of those, the four-dimensional Planck mass $M_P$ is
given by
\ba
M_P^2=\frac{8}{g_s^2} v_{45}v_{67}v_{89} M_s^2
\label{mp2}
\ea
and the non-abelian gauge couplings are
\ba
\frac{1}{g_3^2}=\frac{1}{g_s} v_{45}\qquad ;\qquad
\frac{1}{g_2^2}=\frac{1}{g_s} v_{67}
\label{g32}
\ea
It follows that
\ba
M_P^2=\frac{8}{g_3^2 g_2^2}v_{89}M_s^2=
\frac{2}{\alpha_3\alpha_2}{\hat v}_{89}M_s^2 \, ,
\label{mp}
\ea
where $\alpha_i=g_i^2/4\pi$ and ${\hat v}_{89}\equiv v_{89}/(2\pi)^2=R_8R_9$ for a
rectangular torus of radii $R_8, R_9$. The $U(1)_1$ gauge coupling $g_1$ is
equal to $g_3$ ($g_2$), if the $U(1)_1$ brane is on top of the $U(3)_c$
($U(2)_L$).

The gauge coupling $g_b$ of the $U(1)_b$ gauge boson which lives in the bulk is
extremely small since it is suppressed by the volume of the bulk $v_{89}$. For
instance, in the case where the $U(1)_b$ lives on a D9-brane, its coupling is
given by
\ba
\frac{1}{g_b^2}=\frac{1}{g_s}v_{45}v_{67}v_{89}=\frac{g_s}{8}\frac{M_P^2}{M_s^2}\, ,
\label{gb}
\ea
where in the second equality we used eq.~(\ref{mp2}). Using
now the weak coupling condition $g_s<1$ and the inequality
$g_s>g_{3,2}^2$ following from $v_{ij}>1$ in eq.~(\ref{g32}), one finds
\ba
{\sqrt 8}\frac{M_s}{M_P}<g_b<\frac{{\sqrt 8}}{g_3}\frac{M_s}{M_P}\, ,
\label{gbin}
\ea
which implies that $g_b\simeq 10^{-16} - 10^{-14}$ for $M_s\sim 1-10$ TeV.
The corresponding gauge bosons must have a mass larger than $~50$
MeV to avoid supernova constraints that are more stringent than
those for the graviton \cite{AKRT}.

\subsection{The simplest allowed configurations\label{models}}

In this section we will describe the four brane configurations and
hypercharge embeddings, that give models that are compatible with a
low string scale and very basic phenomenological constraints \cite{AKRT}.

In all configurations, the baryon number appears as a
gauged abelian symmetry. This symmetry is broken due to mixed gauge and
gravitational anomalies leaving behind a global symmetry. Baryon number
conservation is essential for low string scale models, since one needs to
eliminate effective operators to very high accuracy in order to avoid fast
proton decay, starting with dimension six operators of the form $Q Q Q L$
which are not sufficiently suppressed
\cite{Ibanez:1999it}.

In addition to baryon number, one should also assure that the
lepton number is a good symmetry of the low energy theory. Lepton number
conservation is also essential for preservation of acceptable neutrino masses, as
it forbids for instance the presence of the dimension 5 operator $L L H H$. Such an
operator would lead to large Majorana neutrino masses, of the order of a few GeV, in
models  where the string scale, typically a few TeV, is too low for the
operation of an effective sea-saw mechanism. Hence, we shall be interested only
in models in which the
lepton number is a good symmetry. Being anomalous, this  symmetry will be broken, but
lepton number will survive as a global symmetry of the effective theory.

In fact, these four models can be derived in a straightforward way by simple
considerations of the quantum numbers. The quark doublet $Q$ is fixed by non abelian
gauge symmetries, while existence of baryon number implies that the anti-quarks
$u^c, d^c$ correspond to strings stretched between the color branes and one each of the
abelian branes $U(1)_1$ and $U(1)_b$. Thus, one has two possibilities leading to
models that we call $A$ ($d^c$ has one end in the bulk) and $B$ ($u^c$ sees the
bulk). Existence of lepton number fixes the lepton doublet as a string stretched
between the weak branes and the $U(1)_b$ brane, while for each of the models
$A$ and $B$ there are two possibilities for the anti-lepton $e^c$ to emerge as a
string stretched between the two abelian branes, or to have both ends on the
weak branes. Thus, we obtain two additional models that we call $A'$ and $B'$.
All these models have tree-level quark and
lepton masses and make use of two Higgs doublets. They also require low energy
string scale for some of the brane coupling conditions.

\medskip
\noindent{\bf Models mLSOM$_A$ and  mLSOM$_A'$}

They are characterized by the common  hypercharge embedding
\ba
Y= -\frac{1}{3}\,Q_c-\frac{1}{2}\,Q_L+Q_1\label{hypa}
\ea
but they differ slightly in their spectra.
The spectrum of model $A$ is

\ba
&~&Q\left(\bbb3,\bbb2,+1,-1,0,0\right)\nn\\
&~&u^c(\bb3,\bbb1,-1,0,-1,0)\nn\\
&~&d^c(\bb3,\bbb1,-1,0,0,-1)\nn\\
&~&L(\bbb1,\bbb2,0,+1,0,-1)\nn\\
&~&e^c(\bbb1,\bbb1,0,0,+1,+1)\nn\\
&~&H_u(\bbb1,\bbb2,0,+1,+1,0)\nn\\
&~&H_d(\bbb1,\bbb2,0,-1,0,-1)\nn
\ea
while in model $A'$ the right-handed electron $e^c$ is replaced by an open string
with both ends on the weak brane stack, and thus $e^c=(\bbb1,\bbb1,0,-2,0,0)$.

\begin{figure}
\center
\includegraphics[width=6cm]{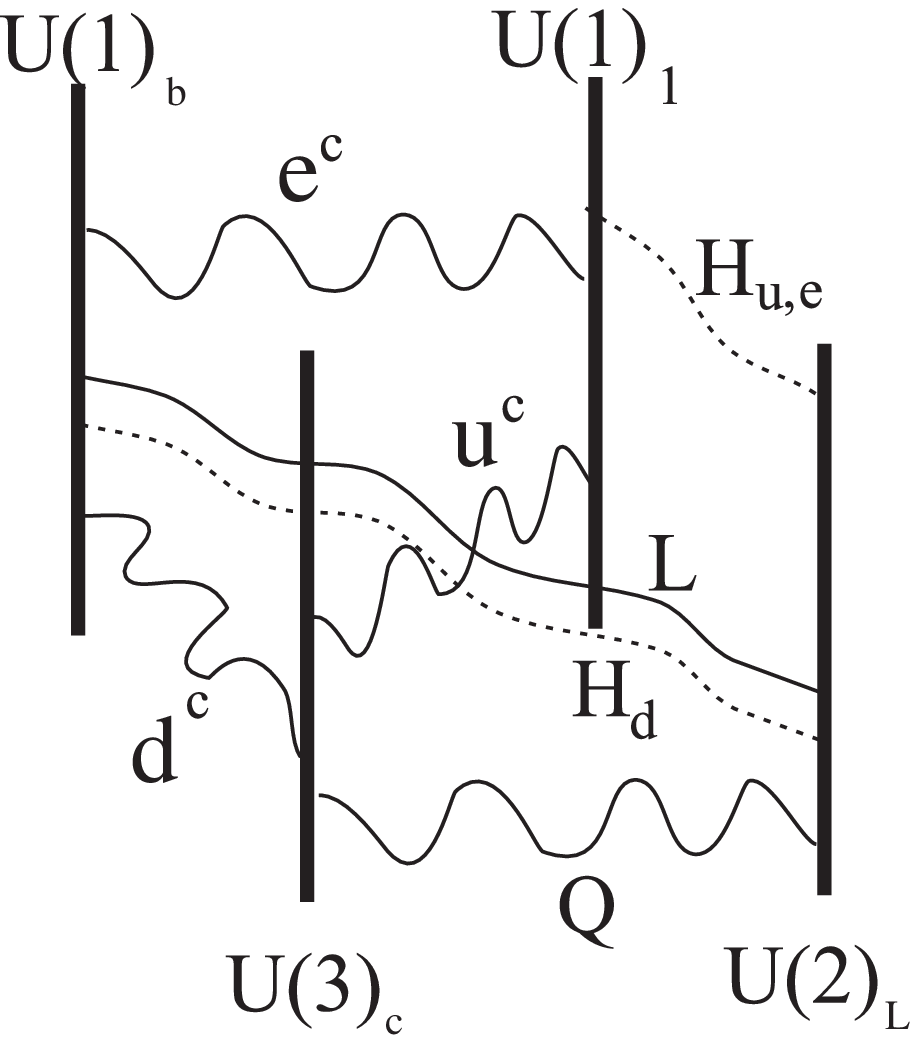}
\includegraphics[width=6cm]{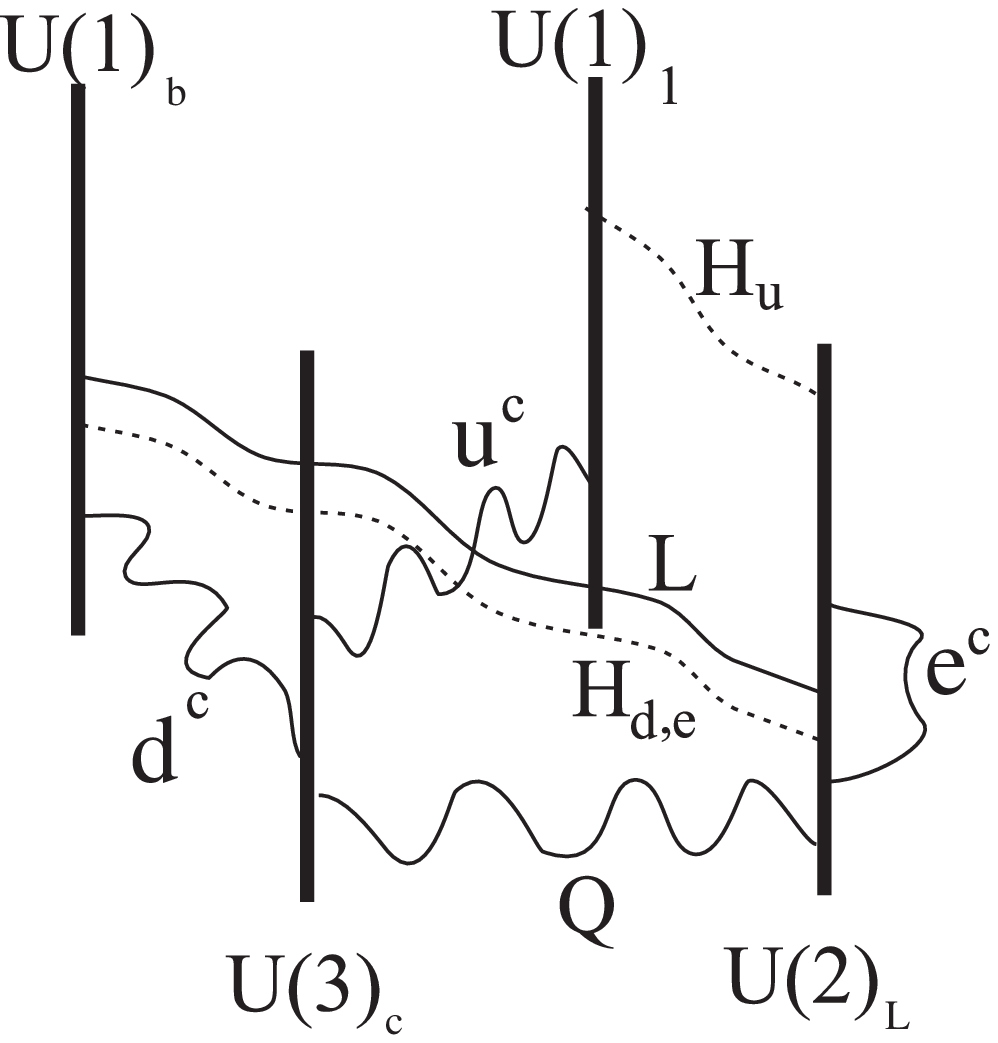}
\caption{\label{figa}{\it Pictorial representation of models} $A, A'$.}
\end{figure}

Apart from the hypercharge combination (\ref{hypa}) all remaining abelian
factors are anomalous. Indeed, for every abelian generator $Q_I, I=(c,L,1,b)$, we can
calculate the mixed gauge anomaly $K_{IJ}\equiv {\rm Tr} Q_I T^2_J$ with
$J=SU(3),SU(2),Y$, and gravitational anomaly $K_{I4}\equiv{\rm Tr} Q_I$
for both models $A$ and $A'$:
\ba
K^{(A)}=
\left(\matrix{
  0 & -1 & -\frac{1}{2} & -\frac{1}{2} \cr
  \frac{3}{2} & -1 & 0 & -\frac{1}{2} \cr
  -\frac{3}{2} & \frac{1}{3} & -\frac{1}{3} & \frac{1}{6} \cr
  0 & -4 & -2 & -4}\right)\ ,\
K^{(A')}=
\left(\matrix{
  0 & -1 & -\frac{1}{2} & -\frac{1}{2} \cr
  \frac{3}{2} & -1 & 0 & -\frac{1}{2} \cr
  -\frac{3}{2} & -\frac{5}{3} & -\frac{4}{3} & -\frac{5}{6} \cr
  0 & -6 & -3 & -5}\right)\label{anoma}
\ea It is easy to check that the matrices $K K^T$ for both models
have only one zero eigenvalue corresponding to the hypercharge
combination (\ref{hypa}) and three non vanishing ones
corresponding to the orthogonal $U(1)$ anomalous combinations. In
the context of type I string theory, these anomalies are canceled
by a generalized Green-Schwarz mechanism which makes use of three
axions that are shifted under the corresponding $U(1)$ anomalous
gauge transformations. As a result, the three extra
gauge bosons become massive, leaving behind the corresponding
global symmetries unbroken in perturbation theory
\cite{Poppitz:1998dj}. The three extra $U(1)$'s can be expressed
in terms of known SM symmetries: \ba
\mbox{Baryon number}\,\ \ \ B&=&\frac{1}{3} Q_c\nn\\
\mbox{Lepton number}\,\,\ \ \ L&=&\frac{1}{2}\left(Q_c+Q_L-Q_1-Q_b\right)\label{glob}\\
\mbox{Peccei--Quinn}\ \ \ Q_{PQ}&=&-\frac{1}{2}\left(Q_c-Q_L-3\,Q_1-3\,Q_b\right)\nn
\ea
Thus, our effective SM inherits baryon and lepton number
as well as Peccei--Quinn (PQ) global symmetries from the anomaly cancellation
mechanism. Note however that $PQ$ is the original Peccei--Quinn symmetry only in
model $A'$, such that all fermions have charges $+1$, while $H_u$ and $H_d$ have
charges $-2$ and $+2$, respectively. In model $A$, the global $PQ$
symmetry defined in (\ref{glob}) is similar but with lepton charge
+3. The reason is that in model $A$ the fermion-Higgs Yukawa couplings are different,
and leptons get masses from $H_u$ and not from $H_d$.

The general one-loop string computation of the masses of anomalous $U(1)$ gauge
bosons, as well as their localization properties in the internal compactified
space, was performed recently for generic orientifold vacua \cite{AKR}.
It was shown that orbifold sectors preserving $N=1$ supersymmetry yield
four-dimensional (4d) contributions, localized in the whole six-dimensional (6d)
internal space, while $N=2$ supersymmetric sectors give 6d contributions
localized only in four internal dimensions. The latter are related to 6d
anomalies. Thus, even $U(1)$s which are apparently anomaly free
may acquire non-zero masses at the one-loop level, as a consequence of 6d
anomalies. These results have the following implications in our case:
\begin{enumerate}
\item
The two $U(1)$ combinations, orthogonal to the hypercharge and localized on the
strong and weak D-brane sets, acquire in general masses of the order of the
string scale from contributions of $N=1$ sectors, in agreement with effective
field theory expectations based on 4d anomalies.
\item
Such contributions are not sufficient though to make heavy the third $U(1)$
propagating in the bulk, since the resulting mass terms are localized and
suppressed by the volume of the bulk. In order to give string scale mass, one
needs instead $N=2$ contributions associated to 6d anomalies along the two large
bulk directions.

\item
Special care is needed to guarantee that the hypercharge remains massless despite
the fact that it is anomaly free.
\end{enumerate}

The presence of massive gauge bosons associated to anomalous abelian gauge symmetries
is generic. Their mass is given by $M_A^2\sim g_s M_s^2$,
up to a numerical model dependent factor and is typically
 smaller by a factor or 2-5 than  the string scale.
When the latter is low, they can affect low energy measurable data, such as $g-2$
for leptons \cite{AK1} and the $\rho$-parameter \cite{GIIQ}, leading to additional bounds
on the string scale.

An extension of the model is the introduction of a right-handed neutrino
in the bulk. A natural candidate state would be an open string ending on the
${U(1)}_b$ brane. Its charge is then fixed to $+2$ by the requirement of
existence of the single possible neutrino mass term $L\,H_d\,\nu_R$. The
suppression of the brane-bulk couplings due to the wave function of $\nu_R$ would
thus provide a natural explanation for the smallness of neutrino masses. Note
that if the zero mode of this bulk neutrino state is chiral, the anomaly
structure of the model changes: $B-L$ becomes anomaly free and as a consequence
the associated gauge boson remains in principle massless. However, as we
discussed above, this is not in general true because of 6d anomalies \cite{AKR}.
 In any case,
this problem is absent if we introduce a vector-like bulk neutrino pair
\ba
&~&{\nu}_R(\bbb1,\bbb1,0,0,0,+2)+{\nu}^c_R(\bbb1,\bbb1,0,0,0,-2)\nn
\ea
that leaves the anomalies (\ref{anoma}) intact. Note that ${\nu}^c_R$ does
not play any role in the subsequent discussion of neutrino masses and oscillations.

\medskip
\noindent{\bf Models mLSOM$_B$ and mLSOM$_B'$}

Another  phenomenologically promising pair of models
consists of two models, named hereafter $B$ and
$B'$, which correspond to the hypercharge embedding
\ba
Y= \frac{2}{3}\,Q_c-\frac{1}{2}\,Q_L+Q_1\, .
\label{hypb}
\ea
The spectrum is
\ba
&~&Q(\bbb3,\bbb2,+1,+1,0,0)\nn\\
&~&u^c(\bb3,\bbb1,-1,0,0,1)\nn\\
&~&d^c(\bb3,\bbb1,-1,0,1,0)\nn\\
&~&L(\bbb1,\bbb2,0,+1,0,-1)\nn\\
&~&e^c(\bbb1,\bbb1,0,0,+1,+1)\nn\\
&~&H_u(\bbb1,\bbb2,0,-1,0,-1)\nn\\
&~&H_d(\bbb1,\bbb2,0,+1,+1,0)\nn
\ea
for model $B$, while in $B'$ $e^c$ is replaced by $e^c(\bbb1,\bbb1,0,-2,0,0)$.

\begin{figure}
\center
\includegraphics[width=6cm]{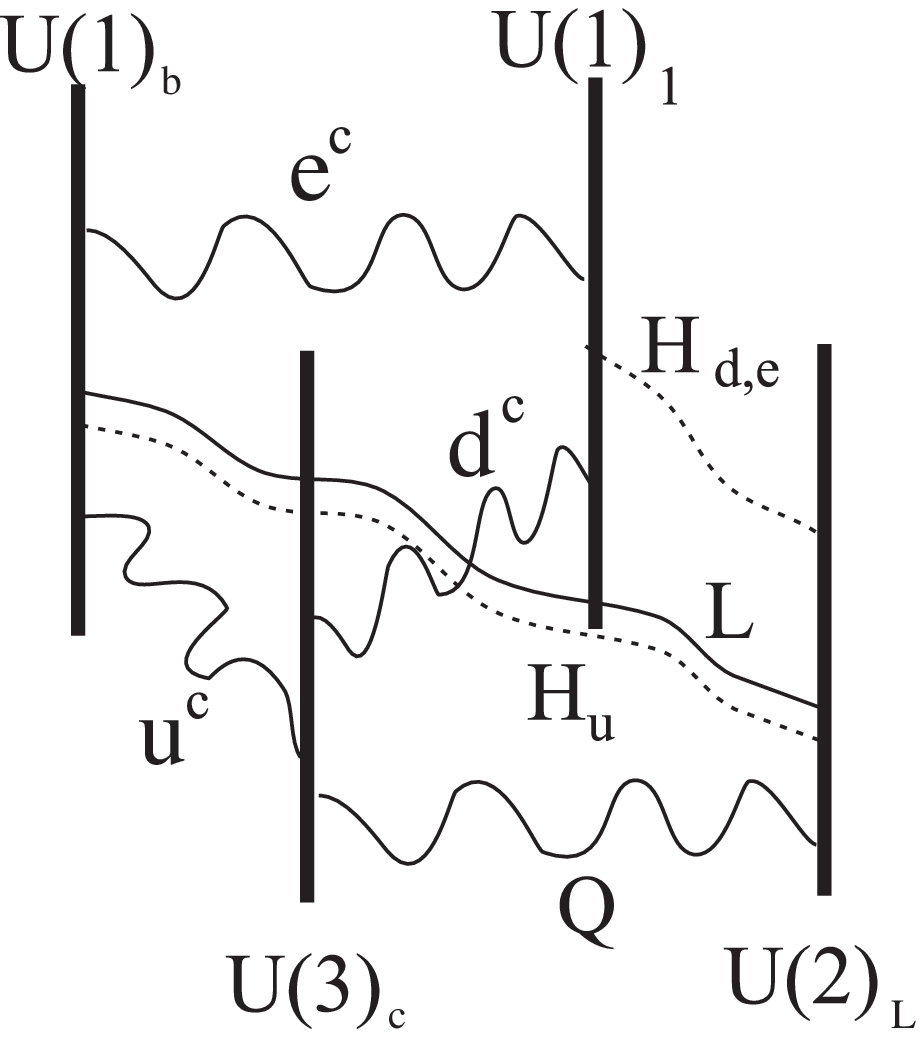}
\includegraphics[width=6cm]{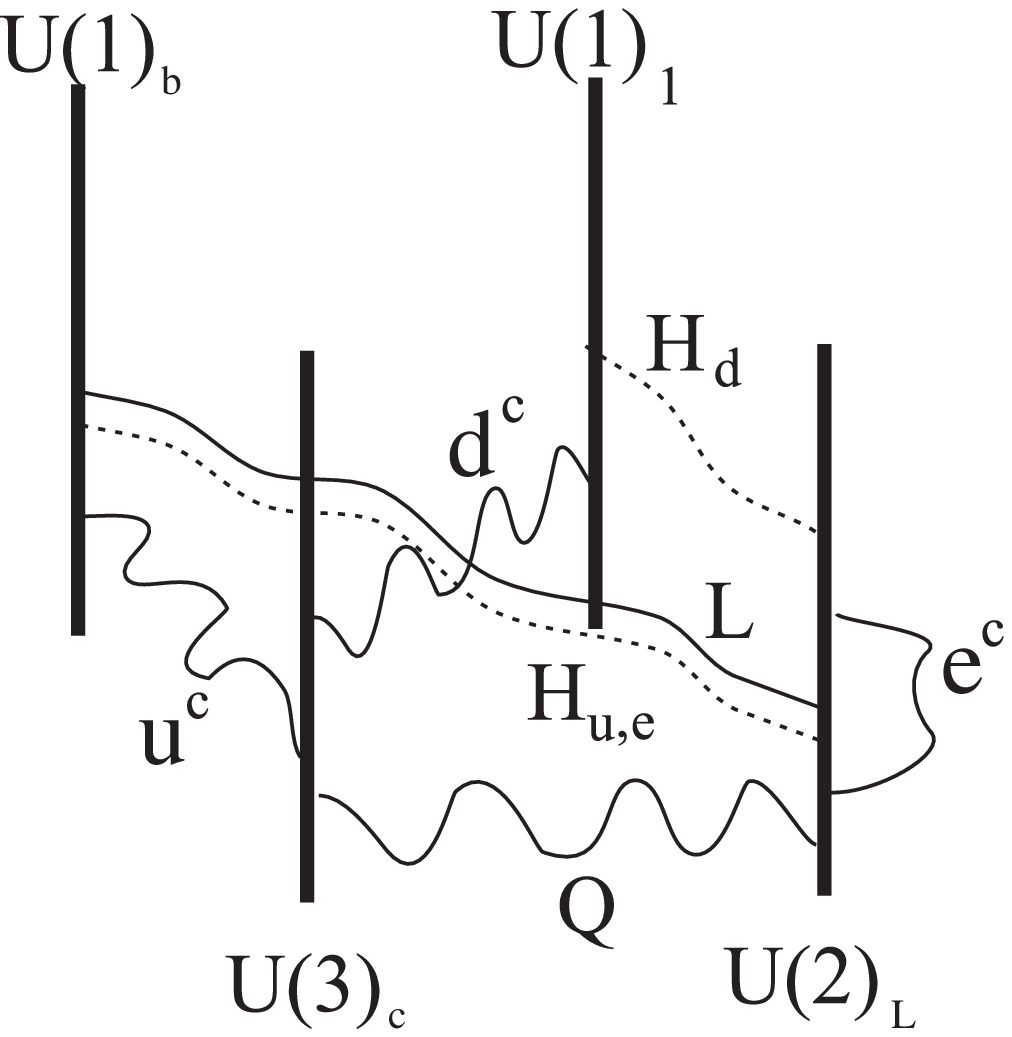}
\caption{\label{figb}\it Pictorial representation of models $B$ and $B'$.}
\end{figure}

The four abelian gauge factors are anomalous. Proceeding as in the analysis
(\ref{anoma}) of models $A$ and $A'$, the mixed gauge and
gravitational anomalies are
\ba
K^{(B)}=
\left(\matrix{
  0 & 1 & \frac{1}{2} & \frac{1}{2} \cr
  \frac{3}{2} & 2 & 0 & -\frac{1}{2} \cr
  -\frac{3}{2} & \frac{2}{3} & \frac{4}{3} & \frac{11}{6} \cr
  0 & 8 & 4 & 2}\right)\ ,\
K^{(B')}=
\left(\matrix{
  0 & 1 & \frac{1}{2} & \frac{1}{2} \cr
  \frac{3}{2} & 2 & 0 & -\frac{1}{2} \cr
  -\frac{3}{2} & -\frac{4}{3} & \frac{1}{3} & \frac{5}{6} \cr
  0 & 6 & 3 & 1}\right)\label{anomb}
\ea
It is easy to see that the only anomaly free combination is the hypercharge (\ref{hypb})
which survives at low energies.
All other abelian gauge factors are anomalous and will be broken by the generalized
Green-Schwarz anomaly cancelation mechanism, leaving behind global symmetries. They can be
expressed in terms of the usual SM global symmetries as the following $\u1$
combinations:
\ba
\mbox{Baryon number}\,\ \ \ B&=&\frac{1}{3}\,Q_c\\
\mbox{Lepton number}\ \ \ \ L&=&-\frac{1}{2}\left(Q_c-Q_L+Q_1+Q_b\right)\\
\mbox{Peccei-Quinn}\ \ \
Q_{PQ}&=&\frac{1}{2}\left(-Q_c+3\,Q_L+Q_1+Q_b\right)
\ea
Similarly to the analysis of models $A$ and $A'$, the $PQ$ charges
defined above are the traditional ones only  for model $B$. In
model $B'$, the lepton charge is $-3$, as a result of the Higgs Yukawa
couplings to the fermions (see below).
The right handed neutrino can also be accommodated as an open string
with both ends on the bulk abelian brane:
\ba
&~&{\nu_R}(\bbb1,\bbb1,0,0,0,+2)+{\nu^c_R}(\bbb1,\bbb1,0,0,0,-2)\nn
\ea

%
\section{The effective action of the mLSOM\label{t1}}
\setcounter{equation}{0}
%

We will consider models which can originate from a non-supersymmetric
string compactification where the Standard Model is localized on
D-branes and/or intersections of D-branes in the presence of orientifold planes.
The low energy limit of such models, assuming that they contain
the Standard Model spectrum, is marked by the presence of
extra $U(1)$ gauge bosons and of a certain
number of scalar fields with axion and St\"{u}ckelberg couplings.
Consistency of these models and more specifically the cancellation of anomalies
requires also certain Chern-Simons type of couplings.

Therefore apart from the Standard model fields, we have three more
U(1)'s, three scalars (axions) that mix with the U(1)'s, and two
Higgs doublets.

The minimal Lagrangian consistent with these features is
\begin{eqnarray}
{\cal L}\; =&-&\frac{1}{2}tr\; G_{\m\n}G^{\m\n}-\frac{1}{2}tr \; W_{\m\n}W^{\m\n}
-\frac{1}{4}F_{\m\n}^{l}F^{\m\n,{l}}\nonumber \\
&-&| (\partial_{\m}+i\frac{g_2}{2}{\tau^a}W_{\m}^a
+iq_{l}^{(H_u)}g_{l} A_{\m}^{l}) H_u|^2
-| (\partial_{\m}+i\frac{g_2}{2}{\tau^a}W_{\m}^a
+iq_{l}^{(H_d)}g_{l} A_{\m}^{l}) H_d|^2\nonumber \\
&+&Q^{\dag}_{Li}\s^{\m}{\cal D}_{\m}Q_ {Li}+
{u}^{\dag}_{Ri}{\overline \s}^{\m}{\cal D}_{\m}{u}_{Ri} +
{d}^{\dag}_{Ri}{\overline \s}^{\m}{\cal D}_{\m}{d}_{Ri} \nonumber \\
&+&L^{\dag}_{Li}\s^{\m}{\cal D}_{\m}L_{Li} +
{e}^{\dag}_{Ri}{\overline \s}^{\m}{\cal D}_{\m}{e}_{Ri} +
{\n}^{\dag}_{Ri}{\overline \s}^{\m}{\cal D}_{\m}{\n}_{Ri}\nonumber \\
&+&  \g^u_{ij} H^T_u \tau^2\; (Q_{Li}^t \s^2 u_{Rj})+
 \g^d_{ij} H^{\dag}_d\; (Q_{Li}^t \s^2 d_{Rj})+ c.c.\nonumber \\
&+&  \g^e_{ij} H^{\dag}_u\; (L_{Li}^t \s^2 e_{Rj})+
 \g^{\n}_{ij} H^T_d \tau^2\; (L_{Li}^t \s^2 {\n}_{Rj})+ c.c.\nonumber \\
&-& \frac{1}{2}\sum_{I} (\partial_{\m}a^{I}+g_{l}{\cal M}^I_{l}A^{l}_{\m})^2+
{E}_{lmn}\; \e^{\m\n\r\s}\; {A}^l_{\m}\, {A}^m_{\n}\, F^n_{\r\s}\nonumber \\
&+& \sum_{I} ({D}_I\, {a}^{I}\; tr\, \{G\wedge {G}\} + {F}_I\, {a}^{I}\; tr\, \{W\wedge {W}\}
+{C}_{Imn}\, {a}^{I}\, F^{m}\wedge {F}^{n})\nonumber \\
&+& V(H_u,H_d,a^{I}).
\label{action}
\end{eqnarray}
where we have introduced two dimensional notations for the fermion interactions, as specified below.
The gauge symmetry under which this Lagrangian is invariant is
\bea
SU(3)_c \times SU(2)_W \times G_1,\hskip 1cm G_1=\prod_{l=1}^{4} U(1)_{l}.
\eea
The $U(1)$ factors are all anomalous in general.
In the above the indices $l,m,n=1,\cdots , 4$ count the $U(1)$s in the D-brane basis.
There is also a sum over $SU(2)$ indices $a=1,2,3$ and a sum over flavor indices $i=1,2,3$.
$G_{\m\n}$ is the field strength for the
gluons and the $W_{\m\n}$ is the field strength of the weak gauge bosons $W_{\m}$.
The fermions in eq. ($\ref{action}$) are either left handed Weyl spinors $f_L$,
or right handed Weyl spinors $f_R$ and they fall in the usual $SU(3)$ and
$SU(2)$ representations of the Standard Model.
The covariant derivatives act on the fermions $f_L,f_R$ as
\begin{eqnarray}
&& {\cal D}_{\m}f_{L} = \left(\partial_{\m} +
i {\bf A}_{\m} + i q_{l}^{(f_L)} g_{l}A_{l,\m} \right)f_{L}\nonumber \\
&& {\cal D}_{\m}f_{R} = \left(\partial_{\m} +
i {\bf A}_{\m} - i q_{l}^{(f_R)} g_{l}A_{l,\m} \right)f_{R}
\end{eqnarray}
where ${\bf A}_{\m}$ is a non abelian Lie algebra element.
The matrices $\s^{\m}=(\s^0,\s^a)$ where $\s^0 = diag(1,1)$, $\s^a$ are the Pauli matrices
and ${\overline \s}^{\m}=(\s^0,-\s^a)$.
We have also introduced two Higgs $SU(2)$ doublets $H_u$ and $H_d$.
The matrices $\tau^j$ are Pauli matrices acting on $SU(2)$ indices.

In the Yukawa sector,
the Pauli matrix $\tau^{2}$ acts on the $SU(2)$ indices
while the Pauli matrix $\s^{2}$ acts on the spinor indices.
The symbol $T$ ($t$) suggests transposition with respect to
$SU(2)$ (spinor) indices.
To lighten the notation we do not show explicitly the $SU(3)$ contraction.
It should be understood however that the quarks are, on the top of
all contractions explicitly shown, contracted as ${\bf q_L}^{\dagger}{\bf q_R}$
in the $SU(3)$ sense.
The $\g^u_{ij}$ etc. are complex three by three matrices.
The standard procedure is to bring them
in a form as close as possible to diagonal.
The result of this is
\begin{eqnarray}
{\cal L}_{\rm Yuk.}&=& \sum_{i,j}H^T_u \tau^2\; (Q_{Li}^t \s^2 {\cal U}^{q}_{ji}\G^u_{jj}u_{Rj})+
\sum_i H^{\dag}_d\; (Q_{Li}^t \s^2 \G^d_{ii}d_{Ri})+c.c.\nonumber \\
&+&  \sum_{i,j}H^{\dag}_u\; (L_{Li}^t \s^2 {\cal U}^{\n}_{ji} \G^e_{jj}e_{Rj})+
\sum_i H^T_d \tau^2\; (L_{Li}^t \s^2 {\G^{\n}_{ii}\n}_{Ri})+c.c.
\end{eqnarray}
where the $\G^{u,d,e,\n}$ are diagonal matrices
and ${\cal U}^{q}$ is the CKM matrix which appears in the Yukawa sector
of this model in a similar way as in the Standard Model.
The ${\cal U}^{\n}$ matrix is the MNS neutrino mixing matrix.
In the electroweak vacuum the Higgs couples universally to the Yukawa sector
and the Yukawa couplings turn into mass terms for the fermions.
The CKM and MNS matrices disappear from the
Yukawa couplings but they appear explicitly in
the gauge boson-fermion-fermion interactions, as we will see later.
Issues of flavor in intersecting D-brane models are discussed in
\cite{CIM1,AMS1,AMS2} and references therein.

The couplings ${\cal M}^I_{m}$, $F_I$, $D_{I}$, $C_{I mn}$ and
$E_{lmn}$ are known once a
specific string vacuum has been chosen. One feature of the action, as we are going to describe below,
is the presence of both dimension-4 and dimension-5 operators, which render it an effective
non-renormalizable extension of the Standard Model. The mechanism of cancellation of the anomalies
which is enforced on the model is different from the Standard Model one and for this reason
all the couplings the $E$, $D$, and $C$ carry an intrinsic power of $h$, the Planck constant, in their
definition.
The index $I=1,\cdots , N_a$ runs over the scalars with axion couplings
whose number is in general different (and usually much larger)
than the number of $U(1)$ fields.
In the mLSOM the number of relevant axions will be taken
to be always one less than the number of $U(1)$s (i.e. the number
of D-brane stacks), in our case $N_a=3$.

Finally, the Higgs potential is one that is consistent with
the symmetries of the theory and breaks the electroweak symmetry
spontaneously down to electromagnetism as in the SM.
In general it can depend on all scalar fields present in the spectrum,
namely both on the Higgs fields and on the axions, provided it is compatible
with the gauge invariances.
We will split the  Higgs potential in two parts. The
one in eq. (\ref{PQ}) which does not depend on the axions,
and the one in eq. (\ref{PQbreak}) which mixes the Higgs doublets with the
pseudo-scalars.

%
\subsection{Changing basis in gauge symmetry space.}
%

A first interesting aspect of such
models is that some of the the gauge bosons can
pick up masses even in the absence of electroweak
(EW) symmetry breaking because of (potential) anomalies.
Indeed, by inspecting eq. (\ref{action}) one can see that there
are couplings that give a tree-level mass to the anomalous $U(1)$
gauge bosons without a Higgs mechanism.
The mass squared matrix of the 4 $U(1)$ gauge bosons is
\be
{\bf M}={\cal M}^T{\cal M}\label{massStuck}
\ee
which in general is a real, symmetric but non-diagonal matrix.
The dimension of ${\bf M}$ is equal to the number of $U(1)$s.
In order to simplify the expressions as much as possible,
we absorb in its elements the corresponding factors of the gauge couplings.
${\bf M}$ is real and symmetric thus it can be
diagonalized by an orthogonal transformation
\be
{\bf M}=O^T {\bf m} \; O\label{transY}
\ee
where $O$ is the appropriate orthogonal matrix. The diagonal matrix ${\bf m}$
contains the eigenvalues of ${\bf M}$, i.e. the masses squared of the gauge bosons.
When $N_a=N_s-1$, where $N_s$ is the number of stacks, ${\bf m}$ contains
at least one zero eigenvalue.

To write the other terms in the action in the new basis,
we start from the $U(1)$ sector in the D-brane basis
and focus for the moment on the gauge kinetic and the gauge-fermion-fermion
interaction terms
\be
\sum_{l}{1\over 4\tilde g_l^2}{F_l}^2+\sum_{l} A_l\, \bar \psi\, q_l\, \psi,
\ee
where the charges are normalized to integers and
\be
\tilde g_l={g_l\over \sqrt{2N}},
\ee
where $g_l$ is the standard normalized coupling of the associated SU(N) group\footnote{
This relation comes from the fact that the full group is U(N), see \cite{AKT}.}.

We will keep three bulk axions, the number that is relevant to
cancel the anomalies of the three anomalous U(1)'s\footnote{In general the number
of bulk axions is larger, but only the linear combinations
that enter into anomaly cancellation is relevant.}.

The (normalized) hypercharge generator can be written as
\be
q_Y=\sum_{l}k_l q_l.
\ee
We will rescale the gauge fields as $A_l\to \tilde g_l A_a$ to obtain
\be
\sum_{l}{1\over 4}F_l^2+\sum_l \tilde g_l~A_l \bar \psi\, q_l\, \psi.
\ee
We will now do an orthogonal transformation to go
to a basis where one of the gauge fields is the hypercharge
while the rest have a diagonal UV mass matrix
\be
A_l=\sum_{l^\prime}O_{l l^\prime}A^{l^\prime}, \sp OO^T=1.
\ee
The index $l^\prime $ in the new basis (referred to as the hypercharge basis from now on)
runs through $l^\prime = Y , I$, with
$I$ running through the last $3$ values.
For the above transformation to be consistent, we must take
\be
\tilde g_l \, O_{l{Y}}\sim k_l ~~~\forall l.
\ee
Normalizing we obtain
\be
O_{l Y}=g_Y{k_l\over \tilde g_l}\sp {1\over g_Y^2}=\sum_l{k_l^2\over \tilde g_l^2}.
\ee
We must now pick the $3$ vectors $\vec n_I= O_{iI}$ so that
they are orthogonal to the hypercharge, they are normalized, and they diagonalize the mass matrix.
They can be parameterized in terms of 3 SO(3) angles but
we will keep it as such for the moment.
The transformation of any of the charges is
\be
q_{l^\prime}=\sum_{l}q_{l}\frac{{\tilde g}_{l}}{g_{l^\prime}}
O_{{l}l^\prime}\equiv \sum_{l} q_{l} U_{{l}^\prime l}.
\ee
We can then use the matrix $U$ defined by the second part
of the above equation to express the couplings
in the new basis in terms of the couplings in the old basis:
\be
\frac{\d_{{l}^\prime{m^\prime}}}{g_{l^\prime}{g_{m^\prime}}}=
\sum_{l}\frac{U_{{l}^\prime l}U_{{m}^\prime l}}{{\tilde g}_{l}^2}.
\ee
Next, we must rotate the Green-Schwarz couplings.
As we will show, in this basis the St\"{u}ckelberg couplings take the simple form
\be
\frac{1}{2}\sum_{I}(\partial a_I^\prime+M_IA^I)^2\label{StuckHyp1}
\ee
with $a_I^\prime$ the hypercharge basis axions and
$M_I$ the (square root of the) non-zero eigenvalues contained in ${\bf m}$.

%
\subsection{Anomalous  couplings}
\label{ancoup}
%

This section is devoted to the discussion of the terms in eq. (\ref{action})
referred to as Green-Schwarz couplings.
It should be clear by now that the low energy
effective action that corresponds to low scale
orientifold vacua has certain distinctive features.
To begin, most of the extensions of the Standard Model
that are widely believed to be experimentally testable, such as the
MSSM or the NMSSM, are essentially usual gauge theories coupled
in a conventional way to
a larger set of matter fields than the one encountered in the SM.
By this we mean that all the couplings that one finds in these extensions are
of the same type as the couplings of the SM.
The reason for qualitatively new types of couplings not being necessary
in these conventional models is the way gauge anomalies cancel.
In the MSSM for example, anomalies cancel in the same way as in the SM:
the anomaly of each gauge factor vanishes separately.
In string theory, however, there is room for an alternative way
to cancel anomalies, via the Green-Schwarz mechanism.

The net effect of the Green-Schwarz mechanism on the four-dimensional effective action
is a number of scalar fields with St\"{u}ckelberg and axion-like
couplings and certain Chern-Simons couplings.
It is also interesting to point out that these unusual couplings are remnants of
the interplay between closed and open strings from the string theory point of view
or the gravitational and
gauge sectors in the language of the low energy effective action.
The pseudoscalar axions originate from (closed string sector)
RR fields coupled to the (open string sector) gauge fields
of the D-brane world volume through the Wess-Zumino effective action.
Besides their theoretical interest, the presence of these terms
may provide us with a unique opportunity to test string theory experimentally.

The D-brane basis St\"{u}ckelberg couplings
in eq. (\ref{action}) can be then written in matrix form as
\bea
{\cal L}^{Stuck} = \frac{1}{2}\sum_I (\partial_\m\, {a}^I + {\cal M}^I_l{A}_\mu^l)
(\partial_\m\, {a}^I + {\cal M}^I_l{A}_\mu^l)
\eea
and as we have seen in detail, ensure that some of the
$U(1)$s pick up masses of the order of the string scale.

The other Green-Schwarz couplings in eq. (\ref{action}) consist of
the axion-like terms
\bea
{\cal L}^{axion} = {D}_I\, {a}^I\; tr\, \{G\wedge {G}\} + {F}_I\, {a}^I\; tr\, \{W\wedge {W}\}
+{C}_{Imn}\, {a}^I\, F^{m}\wedge {F}^{n}\label{GS1}
\eea
where we have introduced the dimensionfull couplings ${D}_I$, ${F}_I$ and ${C}_{Imn}$
and the Chern-Simons terms \cite{abdk},
\bea
{\cal L}^{C-S} = {E}_{lmn}\; \e^{\m\n\r\s}\; {A}^l_{\m} {A}^m_{\n} F^n_{\r\s}\label{GS2}.
\eea
In the above the sum over $l,m,n$ is implied.
Under the $U(1)$ gauge transformation
\bea
{A}_\m^l \longrightarrow {A}_\m^l + \partial_\m {\e}^l
\eea
with ${\bf \e}$ the gauge transformation parameters,
the anomalous variation of the Lagrangian is\footnote{We use a
symmetric regularization scheme.}
\bea
{{\cal L}^{1-loop} = \bf \e}^l
\left[g_l g_3^2 \, \L_3\, {t}^{(3)}_l\, G\wedge {G}
+ g_l g_2^2 \, \L_2\, {t}^{(2)}_l\, W\wedge {W} +
\L_1\, g_l g_m g_n \, {t}^{(1)}_{lmn}\, F^{m}\wedge {F}^{n} \right]\label{anomaly}
\eea
where
\bea
t^{(1)}_{lmn} = tr (q_l q_m q_n), \hskip .5cm t^{(1)}_{llm}=\frac{1}{2!} tr (q_l^2 q_m),
\hskip .5cm t^{(1)}_{lll}=\frac{1}{3!} tr (q_l^3)
\eea
and
\bea
t^{(3)}_{l} = tr (q_l T^AT^A),\hskip 1cm t^{(2)}_{l} = tr (q_l T^jT^j).
\eea
Here the index $A$ ($a$) labels the generators of $SU(3)$ ($SU(2)$).
The nature and meaning of the quantities $\L_1$, $\L_2$ and $\L_3$ is
clear once the anomaly diagrams are explicitly computed in momentum space.
They can be seen to be the shift
necessary to be performed in the momentum integration of the triangle anomaly diagram
so that the Green-Schwarz anomaly cancellation mechanism is reflected by the
Ward identities.

The axions transform under the $U(1)$ transformations as
\bea
{a}^I \longrightarrow {a}^I - {\cal M}^I_l\, {\e^l}.
\eea
The St\"{u}ckelberg and the axion-gauge-gauge couplings are
gauge invariant separately but the Chern-Simons term is not.
The gauge variation of the latter is cancelled by the
anomaly. By comparing the different gauge variations,
we can easily read off the four dimensional version of the
Green-Schwarz anomaly cancellation conditions
\bea
&& {D}_I\; {\cal M}^I_l = \L_3\; g_l g_3^2\, t_l^{(3)} \label{GScond1}\\
&& {F}_I\; {\cal M}^I_l = \L_2\; g_l g_2^2\, t_l^{(2)} \label{GScond3}\\
&& {C}_{Imn}\; {\cal M}^I_l + (E_{lmn} - E_{mln}) = \L_1\, g_lg_mg_n\, t^{(1)}_{lmn}.\label{GScond2}
\eea
The first two of the above, eqs. (\ref{GScond1}) and (\ref{GScond3})
represent the cancellation of the anomalous triangle graph
with a $U(1)_l$ gauge boson and two gluons and
$SU(2)$ gauge bosons for external legs respectively.
The third, eq. (\ref{GScond2}) represents the mixed $U(1)$ anomaly
cancellation.

We can put some restrictions on the couplings $E_{lmn}$.
Define
\bea
S^{lmn} \equiv \int \e^{\m\n\r\s} A_\m^l A_\n^m F_{\r\s}^n
\eea
which transforms as
\bea
\d S^{lmn} = \int (-\e^l F^m \wedge  F^n + \e^m F^l \wedge  F^n ).
\eea
It is easy to see that $S^{lmn}$ satisfy
\bea
S^{mln} = - S^{lmn}\label{antisym}
\eea
and that the transformation property of the Chern-Simons couplings is
\bea
\d (E_{lmn}S^{lmn} ) = \int (E_{mln}-E_{lmn}) \e^l F^m \wedge  F^n,
\eea
which was used to derive eq. (\ref{GScond2}).
An immediate consequence of eq. (\ref{antisym}) is that $E_{lmn}S^{lmn}$ vanishes
identically unless $E_{lmn}$ is antisymmetric in the first two indices.
Now, if $E_{lmn}$ is totally antisymmetric, then $E_{lmn}S^{lmn}$
can be seen to be again identically zero by using the identity
\bea
S^{lmn} + S^{nlm} + S^{mnl} = 0
\eea
which can be derived by integrating by parts.
Therefore the only choice left is the one where
$E_{lmn}$ is antisymmetric in $lm$.
Then, eq. (\ref{GScond2}) reduces to
\bea
{C}_{Imn}\; {\cal M}^I_l + 2 E_{lmn} = \L_1\, g_lg_mg_n\, t^{(1)}_{lmn}\label{GScond4}
\eea
and the gauge transformation to
\bea
\d (E_{lmn}S^{lmn} ) = -2\, \int E_{lmn} \e^l F^m \wedge  F^n.
\eea

The rotation to the hypercharge basis can be done by means of
eq. (\ref{transY}). The transformation of the vectors and axions
consistent with eq. (\ref{transY}) is
\bea
{A}_l \; = \; O_{l^\prime l}\, {A}_{l^\prime}, \hskip 1cm
{a}^I \; = \; \sum_{J} {\cal M}^I_l \, O_{Jl}\, \frac{a^\prime_J}{M_J} \label{traxvec}
\eea
respectively, with $M_J$ the mass of the $J$ th gauge boson in the
hypercharge basis. The sum over $l$ and $l^\prime$ is implicit but we
show the sum over the indices $I$ explicitly when present.
The proper gauge transformation rules become
\bea
&& {A}_{l^\prime} \longrightarrow {A}_{l^\prime} + \partial\;  {\e^\prime}_{l^\prime}\\
&&  {a}_I^\prime \longrightarrow {a}_I^\prime - {M_I}\, {\bf \e}^{\prime}_I\label{axgauge}
\eea
where $\e ^{\prime} = O \e$
and the St\"{u}ckelberg coupling transforms into
\bea
{\cal L}^{Stuck} = \frac{1}{2}\sum_{I} (\partial_\m {a}^\prime_I + M_I {A}^I_\m)^2.\label{StuckHyp}
\eea
Indeed, eq. (\ref{StuckHyp}) is precisely eq. (\ref{StuckHyp1}), as claimed.

The Green-Schwarz couplings eqs. (\ref{GS1})
and (\ref{GS2}) can be written in the hypercharge basis as
\bea
{\cal L}^{GS} &=& \sum_{I} ({D^\prime_I}\, {a^\prime_I} \, tr\, \{G\wedge {G}\} +
{F^\prime_I}\, {a^\prime_I} \, tr\, \{W\wedge {W}\}
+ {C^\prime}_{I m^\prime n^\prime}\, {a^\prime_I}\, F^{m^\prime}\wedge {F}^{n^\prime})\nonumber\\
&+& E_{l^\prime m^\prime n^\prime}\, \e^{\m\n\r\s}\,
{A}^{l^\prime}_{\m} {A}^{m^\prime}_{\n} {F}^{n^\prime}_{\r\s}\nonumber\\ \label{GS3}
\eea
with
\bea
&& {D^\prime_I} = \sum_{J}\frac{1}{M_I} {D}_{J}{\cal M}^J_l O_{Il},\hskip 1cm
{F^\prime_I} = \sum_{J}\frac{1}{M_I} {F}_{J}{\cal M}^J_l O_{Il}, \nonumber\\
&& {C^\prime_I}_{m^\prime n^\prime} =
\sum_{J}\frac{1}{M_I} {C}_{Jmn}{\cal M}^J_l O_{Il}O_{m^\prime n}O_{n^\prime n}
\eea
and
\bea
E_{l^\prime m^\prime n^\prime} = E_{lmn}O_{l^\prime l}O_{m^\prime m}O_{n^\prime n}.
\eea
It is now straightforward to show that the Green-Schwarz anomaly
cancellation conditions in the hypercharge basis are
\bea
&& {D}^\prime_I\; {M_I} = \L_3\, g_I^2 g_3\, {t^\prime}^{(3)}_I \label{HGScond1}\\
&& {F}^\prime_I\; {M_I} = \L_2\, g_I^2 g_2\, {t^\prime}^{(2)}_I \label{HGScond3}\\
&& {C}^\prime_{Im^\prime n^\prime}\; {M_I} + 2 E_{Im^\prime n^\prime}
= \L_1\, g_I g_{m^\prime} g_{n^\prime}\, {t^\prime}^{(1)}_{Im^\prime n^\prime} \label{HGScond2}
\eea
with the anomaly coefficients ${t^\prime}^{(1,2,3)}$ computed in the hypercharge basis.

%
\subsection{Gauge fixing in the unbroken phase}
%

We work in the $(Y,I)$ basis, with $Y$ the hypercharge
and $I$ the index denoting
the anomalous $U(1)$ gauge bosons in the hypercharge basis.

A useful gauge is the one where the axions become longitudinal components
of the massive anomalous gauge fields. Clearly, in this gauge
there should be no direct axion-gauge boson interactions.
It is not hard to come up with a gauge where the unphysical
couplings of the type
\bea
M_I\; (\partial \cdot A_I)\; a_I^\prime
\eea
are absent. The necessary gauge fixing functions
for $SU(3)$, $SU(2)$, $U(1)_Y$ and the $U(1)_I$ are
\bea
&& {\cal G}^{A} = \partial \cdot G^A\\
&& {\cal G}^{a} = \partial \cdot W^a\\
&& {\cal G}^Y   = \partial \cdot A^Y\\
&& {\cal G}^I   = \partial \cdot A^I + \a_I M_I\; a_I^\prime \label{GaugeAn}
\eea
respectively, where we have introduced gauge fixing
functions with real parameters $\alpha_I$ for the anomalous $U(1)$s.
The gauge fixing terms are
\bea
{\cal L}^{gf} = \frac{1}{2\a_3}{\cal G}^{A} {\cal G}^{A} +
\frac{1}{2\a_2}{\cal G}^{a} {\cal G}^{a} + \frac{1}{2\a_Y}{{\cal G}^Y} {{\cal G}^Y} +
\sum_I {\cal L}_{gf}^I, \hskip .5cm
{\cal L}_{gf}^I = \frac{1}{2\a_I}{{\cal G}^I} {{\cal G}^I}
\eea
and the ghost terms are
\bea
{\cal L}^{gh} =
 \eta^{*A}\frac{\d {\cal G}^A}{\d {w}^B} \eta^B
+ \eta^{*a}\frac{\d {\cal G}^a}{\d {w}^b} \eta^b
+ (\partial \eta^{*Y})\cdot  (\partial \eta^Y)
+ \sum_I {\cal L}_{gh}^I,\nonumber\\
\eea
where in the $SU(3)$ part
\bea
\frac{\d {\cal G}^A}{\d {w}^B} = -
(\partial \cdot \partial)\, \d^{AB} - f^{ABC} (\partial \cdot G^C)
\eea
is the change of the gauge function
under a gauge transformation parameterized by $\w^B$ and $\eta^{*A}$ and
$\eta^{A}$ are the anticommuting ghost fields. Analogous is the
notation for the other gauge groups.
We will now derive ${\cal L}_{\rm gh}^I$.

Under the gauge fixing conditions, the St\"{u}ckelberg
Lagrangian describing the dynamics of each anomalous gauge boson
is given by
\beq
{\cal L}_{Stueck}^I =-\frac14 F_{I\, \mu\nu}^2 - \frac{1}{2} \left(
 \partial^\mu a_I^\prime + M_I A^I_\mu\right) ^2 + \frac1{2\alpha_I}\left(
\partial_\mu A_I^\mu + \alpha_I \,M_I \, a_I^\prime \right)^2.
\eeq
The action is not gauge invariant under the full $U(1)_I$
gauge transformation parameterized by $\e_I^\prime$,
it is however invariant under gauge transformations
that satisfy
\bea
(\partial ^2 - \a_I M_I^2 )\; \e_I^\prime  = 0.\label{KG}
\eea
One should now observe that these are just the
equations of motion of the $a_I^\prime$ and therefore that
gauge transformations performed by the axions playing the role of
the gauge functions are a symmetry of the gauge fixed action.
In fact, the model could be extended
to include an anomalous fermion interaction of the form
\beq {\cal L} _f^I =
\bar\psi
\left[\left(g_V\gamma^\mu - g_A \gamma^\mu \gamma_5\right)
\left( i \partial_\mu +g_I\, A^I_\mu \right)\right] \psi.
\eeq
The total Lagrangian (barring ghosts) is then invariant under the transformations
\beqa
&& \d\, {A^I}_\mu  = \partial_\mu a_I^\prime \nonumber\\
&& \d\, a_I^\prime = - M_I \, a_I^\prime \nonumber\\
&& \psi  \longrightarrow {\rm  e}^{i g_I a_I^\prime} \psi \nonumber\\
&&{\bar\psi} \longrightarrow {\rm e}^{-i g_I a_I^\prime}
\bar\psi.\label{BRST1}
\eeqa

Let us now derive the remnant symmetry when we include ghosts.
If we denote by $\eta_I(x)$ and $\eta_I^*(x)$ independent anticommuting  scalar fields,
the Lagrangian is by construction invariant under the transformation
 ${\bf s}$: \ba &&{\bf s}\,A^I_\mu
=\partial_\mu \eta_I \nonumber\\
&&{\bf s}\,a_I^\prime = - M_I\,\eta_I \nonumber\\
&& {\bf s}\,\psi=i\,g_I\, \eta_I \;\psi \nonumber\\
&&{\bf s}\,\bar\psi= -i\,g_I\,\eta_I \; \bar\psi \nonumber\\
&&{\bf s}\,\eta_I =0
\label{38}
\ea
which can be read off eq. (\ref{BRST1}).
The above BRST transformation is nilpotent even off shell (${\bf s}^2 =0$) and
$\eta_I$ are free, i.e. not constrained by the Klein-Gordon like equation eq. (\ref{KG}).
In terms of the new fields the gauge fixing
plus ghost Lagrangian in the quantum action is of the form
\beq
{\cal L}_{\rm gh}^I + {\cal L}_{\rm gf}^I = -
\eta_I^*({\bf s}\,{\cal G}_I ) + \frac1{2\alpha_I} {\cal G}_I^2
\eeq
for a given gauge fixing function ${\cal G}_I$.
We of course choose eq. (\ref{GaugeAn}) for the gauge fixing functions
and for the ghost part we finally obtain the expression
\beq
{\cal L}_{\rm gh}^I = -
\eta_I^*\left(\partial^2 - \alpha_I \, M_I^2 \right) \eta_I.
\eeq

%
\section{Electroweak symmetry breaking\label{t3}}
\setcounter{equation}{0}
%

The electroweak symmetry breaking in mLSOM is a very interesting
effect, because the Higgses are charged under the anomalous U(1)
gauge symmetries.

In order to discuss EW symmetry breaking we have to be
more specific about the Higgs potential.
Before EW breaking the Abelian gauge symmetry in the $D$-brane basis
is ${G}_{1}$ and the
Higgs potential $V_{PQ}$ is the most general $SU(2)_L\times G_1$ invariant constructed from
the two Higgs $SU(2)$ doublets $H_u$ and $H_d$:

\be
V_{PQ}(H_u,H_d)=\sum_{a=u,d}\Bigl(  \m_a^2  H_a^{\dagger} H_a + \l_{aa} (H_a^{\dagger} H_a)^2\Bigr)
-2\l_{ud}(H_u^{\dagger} H_u)(H_d^{\dagger} H_d)+2{\l^\prime_{ud}} |H_u^T\tau_2H_d|^2. \label{PQ}
\ee
We can parameterize the Higgs fields in terms of 8 real degrees of freedom as
\beqa
H_u=\left(\begin{array}{c}
H_u^+\\
H_u^0
\end{array}\right) \qquad H_d=\left(\begin{array}{c}
H_d^+\\
H_d^0 \end{array}\right)
\eeqa
where $H_u^+$, $H_d^+$ and $H_u^0$, $H_d^0$ are complex fields. Specifically
\beq
H_u^+ =  \frac{H_{uR}^+ + i H_{uI}^+}{\sqrt{2}} ,\qquad
H_d^- =  \frac{H_{dR}^- + iH_{dI}^-}{\sqrt{2}} , \qquad
H_u^- = H_u^{+ *}, \qquad
H_d^+ = H_d^{- *}.
\eeq
Expanding around the vacuum we get for the uncharged components
\beq
H_u^0 =  v_u + \frac{H_{uR}^0 + i H_{uI}^0}{\sqrt{2}} , \qquad
H_d^0 =  v_d + \frac{H_{dR}^0 + iH_{dI}^0}{\sqrt{2}}. \label{Higgsneut}
\eeq
The Weinberg angle is defined via
$\cos\theta_W= g_2/g, \sin\theta_W= g_Y/g$, with $g^2= g_Y^2 + g_2^2$. We also define
$\cos \beta=v_d/v$, $ \sin \beta=v_u/v$ and $v^2=v_d^2 + v_u^2$.

As in the MSSM one can set $H_u^+=0$ at the minimum by an $SU(2)$ transformation.
Then a minimum with $\partial V/\partial H_u^+=0$ must also have $H_d^+=0$.
A necessary condition for the potential $V_{PQ}$ to be bounded from below can be
obtained by requiring that the potential is non-negative
definite around the electroweak breaking vacuum:
\bea
\m_u^2 v_u^2+\m_d^2 v_d^2+\l_{uu}v_u^4+\l_{dd}v_d^4-2\l_{ud}v_u^2v_d^2\; \ge\; 0. \label{quart}
\eea
The above constraint should be satisfied simultaneously with the constraint
coming from the requirement that the
vacuum $<H_u^0>\; =0$, $<H_d^0>\; =0$ (which does not trigger electroweak symmetry breaking)
is an unstable minimum of the potential. This is the case when
\bea
\m_u^2\m_d^2\; \le\; 0. \label{ewnotbroken}
\eea
Minimizing the potential with respect to $H_u^0$ and $H_d^0$ one can see that the Higgs vevs
\be
H_u = v_u\pmatrix {0 \cr 1}, \hskip 2 cm
H_d = v_d\pmatrix {0 \cr 1}   \label{Higgsvev}
\ee
do not break electric charge and minimize $V_{PQ}$ (at tree level) if
\be
\pmatrix{\m_u^2\cr \m_d^2} = 4\; \pmatrix{-\l_{uu} & \l_{ud}\cr \l_{ud} & -\l_{dd}}
\pmatrix{v_u^2\cr v_d^2}\label{vaccondi}
\ee
with $\l_{uu}$, $\l_{dd}$ and $\l_{ud}$ all real.
Using the above conditions, the constraint eq. (\ref{quart}) becomes
\bea
\m_u^2 v_u^2+\m_d^2 v_d^2\; \ge\; 0.\label{vaccondi1}
\eea
Furthermore, the couplings $\l_{uu}$, $\l_{dd}$ and $\l_{ud}$ should be such that
eq. (\ref{ewnotbroken}) and eq. (\ref{vaccondi1}) are also consistent.

%
\subsection{The gauge boson masses}
%

The vevs eq. (\ref{Higgsvev}), in addition to
breaking $SU(2)_L\times {G^\prime_1}$ down to $U(1)_{\g}$,
should not be in contradiction with the low energy EW data.
The previous discussion for the gauge boson masses
still applies with appropriate adjustments
that take into account the effects of EW breaking.
Technically speaking, the neutral $U(1)$ mass matrix ${\bf M}^{EW}$
should have precisely one
zero eigenvalue consistent with an unbroken $U(1)_{\g}$.

${\bf M}^{EW}$ is the $5$ by $5$ matrix that can be read off the quadratic form
\beq
|{\cal D}_\mu H_u|^2 + |{\cal D }_\mu H_d|^2 +
\frac{1}{2}\sum_{I}(\partial a_I^\prime + M_IA^I)^2\label{quadform}
\eeq
and whose eigenvalues and eigenvectors we will now compute.
Notice that as before, we have absorbed in $M_I$ a factor of $g_{I}$
in the St\"{u}ckelberg part of the above formula.
One can easily put it back in the following analysis by doing the rescaling
$M_I\longrightarrow g_{I}M_I $.
We will reinstate the couplings explicitly when we discuss NG bosons.
The covariant derivatives are
\beqa
{\cal D}_{\mu} H_u &=& \left( \partial_\mu + \frac{i}{\sqrt{2}}g_2 \left( T^+ W^+ + T^- W^- \right)
+ \frac{i}{2} g_2\tau_3 W_{3 \mu} + \frac{i}{2} g_Y A_{\m}^Y +
{i\over 2}  \sum_{I}q^I_u g_I \, A^I_{\m}\right) H_u \nonumber \\
{\cal D}_{\mu} H_d &=& \left( \partial_\mu + \frac{i}{\sqrt{2}}g_2 \left( T^+ W^+ + T^- W^- \right)
+ i \frac{g_2}{2} \tau_3 W_{3 \mu} + \frac{i}{2} g_Y A_{\m}^Y +
{i\over 2}  \sum_{I}q^I_d g_I \, A^I_{\m}\right) H_d\nonumber
\eeqa
where $q^l_{u,d}$ are the $U(1)$ charges of the two Higgses in the D-brane
basis.
They can be found for the four distinct configurations in section
\ref{models}.
\bea
q_u^I \equiv \sum_{l}q^l_u\frac{\tilde g_l}{g_I}O_{iI}, \hskip 1cm
q_d^I \equiv \sum_{l}q^l_d\frac{\tilde g_l}{g_I}O_{iI}
\eea
are the Higgs charges in the hypercharge basis.
We have normalized the $U(1)$s so that
\bea
q_u^Y = q_d^Y = \frac{1}{2}.
\eea
The $SU(2)$ generators and gauge bosons are defined as
\beq
T^j = \frac{\tau^j}{2}, \qquad T^\pm= T^1 \pm i T^2,\qquad W^\pm = \frac{1}{\sqrt{2}}
\left( W^1 \mp i W^2\right),
\eeq
we obtain explicitly
\be
{\cal D}_\mu H_u=\pmatrix{
\partial_{\mu} + \frac{i}{2}g_2 W_{3 \mu} + \frac{i}{2} {g_Y}A_\mu^Y +
{i\over 2}  \sum_{I}q^I_u g_I \, A^I_{\m}
& \frac{i}{\sqrt{2}} {g_2} {W^+} \cr
\frac{i}{\sqrt{2}} {g_2} {W^-} &
\partial_\mu - \frac{i}{2}g_2 W_{3 \mu} + \frac{i}{2} {g_Y}A_\mu^Y +
{i\over 2}  \sum_{I}q^I_u g_I \, A^I_{\m}}H_u
\ee
and a similar expression for the covariant derivative of $H_d$.
The mass matrix in the mixing of the neutral gauge bosons
can be then computed from
\bea
&& \frac{1}{2}\sum_{I}M_I^2 (A^I_{\m})^2+{1\over 4}(-g_2 W_{3 \mu} + {g_Y}A_\mu^Y +
\sum_{I}q^I_u\, g_I \, A^I_{\m})^2v_u^2\nonumber\\
&& +{1\over 4}(-g_2 W_{3 \mu} + {g_Y}A_\mu^Y +
\sum_{I}q^I_d\, g_I \, A^I_{\m})^2v_d^2,
\eea
and it reads
\be
{\bf M}^2 = {1\over 4}\pmatrix{
{g_2}^2 v^2 & - {g_2} {g_Y} v^2 &  -{g_2} x_I \cr
 - {g_2} {g_Y} v^2 &  {g_Y}^2 v^2 & {g_Y}  x_I \cr
 -{g_2}  x_I &{g_Y}  x_I  & 2M_I^2\delta_{IJ}+N_{IJ}}
\ee
where
\be
N_{IJ}= (q_u^Iq_u^Jv_u^2 + q_d^Iq_d^Jv_d^2)\, g_Ig_J
\ee
\be
x_I= (q_u^I v_u^2 + q_d^I v_d^2)\, g_I.
\ee
The zero eigenvalue corresponds to the photon
\be
A_\g={g_Y\over {g}}W_3+{g_2\over {g}}A^Y.
\ee
We will now assume that the UV masses $M_I$ are much larger than other mass scales
as expected in realistic orientifold vacua. Then we can treat all other
parameters of the mass matrix as of order one.

Parameterize the eigenvectors as
\be
\xi_1 W^3+\xi_2 A^Y + \xi_I A^I
\ee
and the mas eigenvalue as $m^2$ to obtain
\be
 \xi_1=-g_2{x\cdot\xi\over 4m^2-g^2 v^2}\sp
  \xi_2=g_Y{x\cdot\xi\over 4m^2-g^2 v^2}
  \ee
 \be
g^2{x\cdot\xi \over 4m^2-g^2 v^2}x_I+4(\frac{1}{2}M_I^2-m^2)\, \xi_I
+N_{IJ}\xi_J=0.
\ee
There is an extra non-zero eigenvalue which is of order one, corresponding to the Z gauge boson:
\be
Z=\xi_1 W^3+\xi_2 A^Y+\xi_I A^I
\ee
with
\be
\xi_1={g_2\over g}+{\cal O}\left(M_I^{-2}\right)
\sp \xi_2=-{g_Y\over g}+{\cal O}\left(M_I^{-2}\right)\ee
\be
\xi_I= {g \over {2}}\e_I +{\cal O}\left(M_I^{-4}\right)\ee
and a mass
\be
m^2_Z=m^2_{Z^0} - \frac{1}{4} g^2 \sum_I \e_I\, x_I + {\cal O}\left(M_I^{-4}\right)
\ee
with
\bea
m^2_{Z^0}=\frac{1}{4}\, g^2\, v^2
\eea
the SM value of the neutral gauge boson mass and

\bea
\epsilon_I = \frac{x_I}{M_I^2}
\eea
small expansion parameters.
The other eigenvalues are of order $M_I$
\be
({m_{Z^\prime}^I})^2=\frac{1}{2} M_I^2+{1\over 4}N_{II}
\ee
and correspond to the eigenstates
\be
Z^\prime_I=A_I-\e_I {1\over 2}\left(g_2 W^3-g_Y A^Y\right)+
\sum_{J\not= I}{N_{JI}\over 2(M_J^2-M_I^2)}A_J,
\ee
where we have assumed that $|M_J^2-M_I^2| >> v^2$ so that the perturbation theory is non-degenerate.
Here and in the following analysis we will assume that the smallness of the
corrections originating from new physics
is exclusively due to the large value of $M_I$, in other words
we will avoid the ¨accidental¨ $x_I = 0$ and
$N_{IJ} = 0$ regions of the parameter space.
Notice that then $N_{IJ}$ are expected to be of the same order of magnitude as $\e_I$.
We can now read off the rotation matrix:
\bea
\pmatrix{A_\g \cr Z \cr {{Z^\prime_I}}} =
O^A\, \pmatrix{W_3 \cr A^Y \cr {A}_I}\label{OA}
\eea
where
\bea
O^A = \pmatrix{ \frac{g_Y}{g} & \frac{g_2}{g} & {0^I} \cr
      \frac{g_2}{g} & -\frac{g_Y}{g} & \frac{g}{2}\e_I \cr
-\frac{g_2}{2}\e_I  &  \frac{g_Y}{2}\e_I  & X_{IJ}},\label{rotgaugeOA}
\eea
\bea
X_{IJ} = \d_{IJ} + \sum_{I\not= J}{N_{IJ}\over 2(M_I^2-M_J^2)}.
\eea

The decoupling limit can be studied in terms of the parameters $\e_I$.
In order to identify the modifications introduced by the new model on the masses of the
W and Z bosons and to the Standard Model $\rho$ parameter we recall that in any 2-Higgs doublet
extensions of the Standard
Model  the kinetic terms for the $W^\pm$ and $Z$ gauge bosons are given by
\beq
{\cal L}_{kin} = \frac{g_2^2}{4 \cos^2\theta_W}\, v^2 Z^0_\mu Z^{0\, \mu}
+ \frac{g_2^2}{4}\, v^2  W^{+ \mu} {W^-}_\mu + \frac{g_2^2}{4}\, v^2 W^{- \mu} {W_\mu}^+
\eeq
which bring in the identifications
\beqa
m_W^2 &=&\frac{g_2^2}{4}v^2 \nonumber \\
m_{Z^0}^2 &=&\frac{g_2^2}{4 \cos^2\theta_W} v^2.
\label{standardmass}
\eeqa
We can now compute the tree-level
corrections to the $\rho$ parameter, which are given by
\beq
\rho= \frac{m_W^2 g^2}{ m^2_{Z} g_2^2}= 1 + \frac{1}{4}
\sum_I \e_I \frac{x_I}{v^2} + O(M_I^{-4}).
\eeq
Using the experimental fact that the deviation of the $\rho$ parameter
from unity should be $\lesssim 2\times 10^{-4}$, we can obtain constraints
on the UV parameters of the theory
which should be understood as an approximate lower bound on the
$Z'_I$ gauge boson´s mass and consequently on the string scale $M_{str}$ \cite{GIIQ}.

The small $\e_I$ limit can be also studied directly in the mixing matrix
which however yields typically
similar, but weaker constraints than the ones derived from the $\rho$ parameter.

%
\subsection{The Higgs masses}
%

The physical Higgs and axion masses can be found by inserting
eq. (\ref{Higgsneut}) into the scalar potential eq. (\ref{PQ}), collecting the quadratic
terms and then diagonalizing.

We extract the quadratic part $V_q(H)$ of $V_{PQ}$, which is given by

\beqa
V_q(H)&=&\left({H_u}^-, {H_d}^-\right){\cal M}_1\left(\begin{array}{c}
{H_u}^+\\
{H_d}^+ \\
\end{array}\right) + \left(Re{H_u}^0, Re{H_d}^0\right){\cal M}_2\left(\begin{array}{c}
Re{H_u}^0\\
Re{H_d}^0 \\
\end{array}\right) \nonumber \\
&& + \left(Im {H_u}^0, Im{H_d}^0\right){\cal M}_3\left(\begin{array}{c}
Im{H_u}^0\\
Im{H_d}^0 \\
\end{array}\right).
\eeqa
A direct computation shows that ${\cal M}_3\, \equiv \, 0$.
One of the linear combinations of $Im{H_u}^0$ and $Im{H_d}^0$ is a massless
physical Higgs field called $A^0$, and the orthogonal linear combination,
$G^0$, is a NG boson.
In the other sectors we obtain
\beq
{\cal M}_1= 4 {\lambda'}_{u d} v^2
\pmatrix{ \cos^2\beta & -\sin\beta \cos\beta \cr
 - \sin\beta \cos\beta & \sin^2 \beta},
\eeq
and
\beq
{\cal M}_2=8 v^2
\pmatrix{ {\lambda_{uu}} \sin^2 \beta & -{\lambda_{ud}} \sin\beta \cos\beta \cr
 -{\lambda_{ud}} \sin\beta \cos\beta  & {\lambda_{dd}} \cos^2\beta}.
\eeq
The rotation matrix in the charged sector is
\beq
\left(\begin{array}{c}
{H_u}^+\\
{H_d}^+ \\
\end{array}\right)=
\pmatrix{ \sin \beta  & -\cos {\beta} \cr
 \cos {\beta} & \sin {\beta}}
\left(\begin{array}{c}
{G}^+\\
{H}^+ \\
\end{array}\right)\label{chargedrot}
\eeq
and the mass of the physical charged Higgs $H^+$ is given by
\beq
m_{H^+}^2 = 4  {\lambda'}_{ud} v^2.\label{chargedH}
\eeq
The other state in the mass eigenstate basis ${G}^+$ is a NG-boson.
In the CP-even neutral sector, the rotation to the mass-diagonal basis is given by
\beq
\left(\begin{array}{c}
{Re H_u}^0\\
{Re H_d}^0 \\
\end{array}\right)
=\pmatrix{ \sin \a & - \cos\a \cr
 \cos\a & \sin\a}
\left(\begin{array}{c}
{h_0}\\
{H_0} \\
\end{array}\right),
\eeq
where the rotation angle is
\be
\cos\a = \frac{1}{\sqrt{{a}^2+1}},
\ee
and
\be
a=\frac{1}{8 \lambda_{u d} \sin 2 \beta}
\left( \frac{m_{H^0}^2}{v^2} - 16 \lambda_{uu} \sin^2\beta \right)=
-8 \lambda_{u d} \sin 2 \beta \left( \frac{m_{h^0}^2}{v^2} - 16 \lambda_{uu} \sin^2\beta \right)^{-1}.
\ee
We have expressed $a$ in terms of the two mass eigenvalues $m_{h^0} < m_{H^0}$
of the two neutral physical Higgs fields $h_0$ (lighter) and
$H_0$ (heavier)
\beqa
m_{h^0}^2 &=& 8 v^2 \cos^2\beta \left[ (\lambda_{u u} - {\cal K}(\beta)) \tan^2\beta + \lambda_{u u}
\right] \nonumber \\
m_{H^0}^2 &=& 8 v^2 \cos^2\beta \left[ (\lambda_{u u} + {\cal K}(\beta))
\tan^2\beta + \lambda_{u u}\right] \label{neutHmass}
\eeqa
with
\beq
{\cal K}(\beta)=\left[ \lambda_{uu}^2 \tan^4\beta +
( 4 \lambda_{ud} - 2 \lambda_{uu}\lambda_{dd})\tan^2\beta + \lambda_{dd}^2\right]^{1/2}.
\eeq
We do not get any NG-bosons from this sector.

In summary, in the Higgs sector there are four massive
physical Higgs fields, one massless physical Higgs field
and 3 NG-bosons, two from the charged sector and one from the CP-odd sector.
The axion sector leaves an additional $3$ Goldstone modes.
Since a state with the properties of $A^0$ is phenomenologically unacceptable,
we postpone a thorough discussion of the scalar spectrum and NG-bosons until
sections \ref{sect7} and \ref{sect8} where we will modify the model in such a way so that
the presently massless physical Higgs field gets a mass from a more general potential.

%
\subsection{Higgs-axion mixing and NG-bosons}
\label{sect7}
%

{}From the third and fourth lines of eq. (\ref{action})
and more specifically from the parts linear in the
partial derivatives, we can extract the
linear combinations of fields that are physical and
the linear combinations that are NG-bosons.
A new feature with respect to conventional extensions of the SM
is the mixing of the axions with the fields appearing in the Higgs sector.
It is important therefore to describe the unitary gauge of this model in detail.
The results from this analysis will be useful also in the gauge fixing process.

We do a naive counting to see what we can expect:
There are 8 real Higgs scalar degrees of freedom and $3$ axions, for a total of
$11$ degrees of freedom.
There are two different symmetry breaking mechanisms taking place in these models.
One is the UV St\"{u}ckelberg mechanism which breaks $3$ local symmetries
down to their global subgroups.
Recall that in the class of models under consideration where
${\cal M}_l^I =0 $ for $I\ge 4$ the St\"{u}ckelberg mechanism
does not affect hypercharge and therefore it can break only up to
$3$ symmetries.
Thus we expect to find at least an equal number of
NG bosons associated with the broken generators. In fact, in the absence
of a mixing of the axions with the Higgs fields
the identification of these NG bosons would be very simple.
To see why, recall that in order to identify NG bosons
one looks for unphysical couplings in the action that,
after expanding the fields away from their
vacuum values and keeping terms linear in the fluctuations,
trigger a tree level transformation of a gauge field into a scalar field.
Then a generic such term linear in the derivative can be always written as
\be
C\, A^{\m} (\partial_{\m} G),
\ee
where $A^{\m}$ is some gauge boson, $C$ is a constant
with dimension of mass and $G$ is the NG boson.
On the other hand, the cross term from the St\"{u}ckelberg coupling
in the mass diagonal basis has the form
\be
m\,  A^{\m} (\partial_{\m} a), \label{NGax}
\ee
where $m$ is a mass and $a$ is one of the axions. Comparing these two expressions
immediately identifies $a$ as a NG-boson. In some sense this is not unexpected
in view of the fact that, as we have seen, the axions transform under a gauge transformation
by a shift, a signature of NG-bosons.

In the presence of an Higgs-axion mixing the identification is more involved.
This happens when there are axion corrections to the Higgs potential.
The Higgs potential will break as many generators as in the SM,
one linear combination of hypercharge and $W_3$ and $W^\pm$ for a total of
3 generators. In addition,
when the Higgs fields are charged also under some of the other $U(1)$s
they can break them spontaneously too
which means that there will be additional contributions to eq. (\ref{NGax})
from the Higgs kinetic terms. Then there is Higgs-axion mixing and
some linear combinations will be NG-bosons and some others physical fields.

In any case,
the vacuum should break 3+3=6 generators, corresponding to the symmetry breaking pattern
\be
SU(2)_L\times G_1^\prime \;\;\;
\longrightarrow \;\;\; U(1)_{\g}.
\ee
Thus, we expect 5 real physical fields to appear and
corresponding to the 6 broken generators we expect to find in total 6 NG-bosons.
Even though we did our counting for the case of 3 extra $U(1)$s,
it would be straightforward to generalize it for arbitrary
$N_s$ (as for any of the other formulas shown here for $N_s=4$).

We now apply these general arguments to our model.
Defining
\be
C_{1\m}^a = -{H_a^+}^*(\partial_{\m}H_a^+)+{H_a^+}(\partial_{\m}{H_a^+}^*)
+{H_a^0}^*(\partial_{\m}H_a^0)-{H_a^0}(\partial_{\m}{H_a^0}^*),\hskip .3cm
a=u,d
\ee
\be
C_{2\m}^a = -{H_a^+}^*(\partial_{\m}H_a^+)+{H_a^+}(\partial_{\m}{H_a^+}^*)
-{H_a^0}^*(\partial_{\m}H_a^0)+{H_a^0}(\partial_{\m}{H_a^0}^*),\hskip .3cm
a=u,d
\ee
\be
C_{-\m}^a = {H_a^0}(\partial_{\m}{H_a^+}^*)-{H_a^+}^*(\partial_{\m}{H_a^0}),\hskip .3cm
C_{+\m}^a={C_{-\m}^a}^*,\hskip .3cm a=u,d
\ee
the terms contained in the Higgs kinetic terms linear in the derivatives can
be written as
\begin{eqnarray}
&-& \frac{i}{2}\Bigl(
g_2W^+\cdot (C_-^u+C_-^d)-g_2W^-\cdot (C_+^u+C_+^d)
+g_2W^3\cdot (C_1^u+C_1^d)+g_YA_Y\cdot (C_2^u+C_2^d)\nonumber\\
&+&\sum_I \left[ g_IA^I (q_u^I C_2^u + g_d^IC_2^d)
-2ig_{I}M_I A_{I}\cdot (\partial a_I^\prime)\right] \Bigr),\nonumber\\ \label{mix1}
\end{eqnarray}
where in the terms proportional to $M_I$ we
have put back the factor of $g_{I}$.

Let us first look at the charged terms.
They can be written as
\be
-\frac{i}{2} (g_2 W^{+\m} v\, \partial_{\m} G^- - g_2 W^{-\m} v\, \partial_{\m} G^+)
\ee
where
\bea
G^- =  \sin{\b}\, {H_u^+}^* + \cos{\b}\, {H_d^+}^*,
\hskip 1cm  G^+ = (G^-)^* .\label{chargedNGrot}
\eea
The above definition is consistent with the rotation eq. (\ref{chargedrot})
that transforms to the basis where the $G^{\pm}$ are massless.
Two NG bosons gave been accounted for $(G^+ \;{\rm and}\; G^-)$ and therefore we expect
to find the other 4 NG bosons in the neutral sector.

Using eq. (\ref{OA}) we can bring eq. (\ref{mix1}) into the form
\bea
&-&\frac{i}{2} A_{\g}\cdot \left\{ g_2 O^A_{\g W_3}(C^u_1+C^d_1)+ g_Y O^A_{\g Y}(C^u_2+C^d_2)\right\}\nonumber\\
&-&\frac{i}{2} {Z}\cdot \Bigl\{ g_2 O^A_{Z W_3}(C^u_1+C^d_1)+ g_Y O^A_{Z Y}(C^u_2+C^d_2)\nonumber\\
&+&\sum_I \left[g_IO^A_{ZI}(q_u^I C_2^u + g_d^Iv_d)
-2ig_{I}M_I O^A_{ZI} A^{\m}_{I}(\partial_{\m}a_I^\prime)\right]\Bigr\}\nonumber\\
&-&\frac{i}{2} \sum_J {Z^\prime _J}\cdot
\Bigl\{ g_2 O^A_{Z^\prime_J W_3}(C^u_1+C^d_1)+ g_Y O^A_{Z^\prime_I Y}(C^u_2+C^d_2)\nonumber\\
&+&\sum_I \left[g_IO^A_{Z^\prime_J I}(q_u^I C_2^u + g_d^Iv_d)-2ig_{I}M_I O^A_{Z^\prime_J I}
A^{\m}_{I}(\partial_{\m}a_I^\prime)\right]\Bigr\}.\nonumber\\ \label{mix2}
\eea
Notice now that when we expand the Higgs away from its vacuum value
and keep terms linear in the fluctuations we find that
the coefficient of $A_{\g}$ vanishes identically and
\be
C_{2\m}^a = - C_{1\m}^a = 2\, i\, {\rm Im}[v_a\, (\partial_{\m} {H_a^0}^*)]
\equiv 2\,i\, \partial_{\m} C^a,\hskip .5cm
C^a=v_a{\rm Im}{H_a^0}, \hskip 1cm
a=u,d.\label{linfluc}
\ee
We can then rewrite eq. (\ref{mix2}) as
\bea
{Z}^\m\; \partial _\m \left\{ f_u C^u + f_d C^d + \sum_I g_I M_I O^A_{ZI} a_I^\prime \right\}
+ \sum_J {Z^\prime_J}^\m\; \partial _\m \left\{ f_{u,J} C^u + f_{d,J} C^d +
\sum_I g_I M_I O^A_{Z^\prime_J I} a_I^\prime \right\},
\nonumber\\ \label{mix3}
\eea
where
\bea
&& f_u = g_2 O^A_{Z W_3} - g_Y O^A_{Z Y} - \sum_I q_u^I g_IO^A_{Z I},\hskip 1.7cm
f_d = g_2 O^A_{Z W_3} - g_Y O^A_{Z Y} - \sum_I q_d^I g_I O^A_{Z I}\nonumber\\
&& f_{u,J} = g_2 O^A_{Z^\prime_J W_3} - g_Y O^A_{Z^\prime_J Y} -
\sum_I q_u^I g_IO^A_{Z^\prime_J I} ,\hskip 1cm
f_{d,J} = g_2 O^A_{Z^\prime_J W_3} - g_Y O^A_{Z^\prime_J Y}
 - \sum_I q_d^I g_IO^A_{Z^\prime_J I}.\nonumber\\
\eea
By means of the orthogonal rotation
\be
\pmatrix{{\rm Im}H_u^0\cr {\rm Im}H_d^0\cr . \cr a_I^\prime \cr .}=\; O^{\chi} \;
\pmatrix{\chi \cr G_1^0\cr G_2^0 \cr . \cr .}\label{rotunit}
\ee
with $O^{\chi}$ an 5 dimensional orthogonal matrix,
we can transform to the mass eigenstate basis.
We have denoted the physical field by $\chi$ and the 4 NG-bosons
by $G_{1,\cdots ,4}^0$.
Then, eq. (\ref{mix3}) becomes
\bea
&{Z}^\m& \partial _\m \left\{
\chi  \Bigl[ f_u v_u O_{11}^{\chi} + f_d v_d O_{21}^{\chi}
+  \sum _I g_I M_I O^A_{ZI} O^{\chi}_{I+2,1}\Bigr]
+ m_{Z^0}\, G^Z \right\}\nonumber \\
\sum _{J} &{Z^\prime_J}^\m& \partial _\m \left\{
\chi  \Bigl[ f_{u,J} v_u O_{11}^{\chi} + f_{d,J} v_d O_{21}^{\chi}
+  \sum _I g_I M_I O^A_{Z^\prime_J I} O^{\chi}_{I+2,1}\Bigr]
+ m_{{Z^\prime}_J}\, G^{Z^\prime}_J\right\},
\label{mix4}
\eea
where
\bea
G^{Z} =
& G^0_1 & \Bigl[ f_u \frac{v_u}{m_{Z^0}} O_{12}^{\chi} + f_d \frac{v_d}{m_{Z^0}} O_{22}^{\chi}
+  \sum _I g_I \frac{M_I}{m_{Z^0}} O^A_{ZI} O^{\chi}_{I+2,2}\Bigr] + \nonumber \\
&.& \nonumber \\ &.& \nonumber \\
& G^0_{4} & \Bigl[ f_u \frac{v_u}{m_{Z^0}} O_{1,5}^{\chi} + f_d \frac{v_d}{m_{Z^0}} O_{2,5}^{\chi}
+  \sum _I g_I \frac{M_I}{m_{Z^0}} O^A_{ZI} O^{\chi}_{I+2,N_s+1}\Bigr] \label{mix41} \\
G^{Z^\prime}_J =
& G^0_1 & \Bigl[ f_{u,J}  \frac{v_u}{m_{{Z^\prime}_J}} O_{12}^{\chi} + f_{d,J}
\frac{v_d}{m_{{Z^\prime}_J}} O_{22}^{\chi}
+  \sum _I g_I \frac{M_I}{m_{{Z^\prime}_J}} O^A_{Z^\prime_J I} O^{\chi}_{I+2,2}\Bigr] + \nonumber \\
&.& \nonumber \\ &.& \nonumber \\
& G^0_{4} & \Bigl[ f_{u,J} \frac{v_u}{m_{{Z^\prime}_J}} O_{1,N_s+1}^{\chi} +
f_{d,J} \frac{v_d}{m_{{Z^\prime}_J}} O_{2,5}^{\chi}
+  \sum _I g_I \frac{M_I}{m_{{Z^\prime}_J}} O^A_{Z^\prime_J I} O^{\chi}_{I+2,5}\Bigr] .
\nonumber \\ \label{mix42}
\eea
Let us try to elucidate a bit these apparently complicated expressions.
The simplest example is the case of the potential $V_{PQ}$ of eq. (\ref{PQ}) where
the axions do not couple to the Higgs fields,
which translates into applying eq. (\ref{rotunit}) with all but
the upper left two by two sub-matrix of $O^{\chi}$ set to zero:
\bea
\pmatrix{{\rm Im}H_u^0\cr {\rm Im}H_d^0}=\; O_2^{\chi} \;
\pmatrix{A^0 \cr G^0}.\label{rotunit2}
\eea
In the above we have added a subscript to the rotation matrix in order to emphasize its dimension
and called the physical mass eigenstate $A^0$ (as in the MSSM) instead of $\chi$,
a term reserved for fields with axion-like couplings.
{}From eq. (\ref{rotunit2}) it is clear that since in $V_{PQ}$ the
Higgs fields do not mix with axions
the physical state in the CP-odd sector does not acquire an axion coupling.
Furthermore, since the mass matrix in the CP-odd sector ${\cal M}_3$
is identically zero not only the NG-boson $G^0$ but also $A^0$ remains massless.
On the other hand, according to our general discussion, from the axion sector all
$a_I^\prime=G_{I}^0$ are (3) massless Goldstone modes.
The total number of fields is then 5 physical Higgs fields
(four massive and one massless) and $6$ NG-bosons
taking into account also the 2 NG modes from the charged sector.

For the potential $V_{PQ}+V_{\ppq}$ in eq. (\ref{PQbreak}) of the next section
the situation is quite different.
The fields $G_{1}^0,\cdots , G_{4}^0$ will turn out to be massless
accounting for that many NG bosons and the field
$\chi$ will turn out to be a massive physical field with an axion coupling
because of the mixing of the D-brane basis axions with the CP-odd
Higgs sector. Again, the counting is 5 physical fields, four Higgs and one axion
(all five massive) and $6$ NG-bosons.

The expression eq. (\ref{mix4}) contains unphysical couplings.
Requiring that the gauge fields mix only with NG-bosons,
introduces the constraints
\begin{eqnarray}
& & f_u v_u O_{11}^{\chi} + f_d v_d O_{21}^{\chi}
+  \sum _I g_I M_I O^A_{ZI} O^{\chi}_{I+2,1}=0 \nonumber\\
& & f_{u,J} v_u O_{11}^{\chi} + f_{d,J} v_d O_{21}^{\chi}
+  \sum _I g_I M_I O^A_{Z^\prime_J I} O^{\chi}_{I+2,1}=0.
\end{eqnarray}
These have the simple solution
\bea
&& O_{11}^{\chi} = - N \cos\b, \hskip 1cm O_{21}^{\chi} = N \sin\b \label{solax}\\
&& O_{I+2,1}^{\chi} = - \frac{q_u^I - q_d^I}{2} \frac{v}{M_I} N \sin{2\b}
\equiv {\bf \Theta }_I \label{axcoupl1}
\eea
normalized as
\bea
N=\frac{1}{\sqrt{1+\frac{v^2\sin^2{2\b}}{4}{\sum_I
\left( \frac{q_u^I-q_d^I}{M_I} \right)^2}}}.\label{normcoeff}
\eea
Eqs. (\ref{solax}) and (\ref{axcoupl1})
represent the first column of $O^{\chi}$ which is an $SO(5)$ matrix
with $10$ independent rotation angles.
Having fixed its first column (which is essentially a consequence of the fact
that there is one physical linear combination)
leaves a freedom of $SO(4)$ rotations on the vacuum manifold.
The number of NG bosons is therefore
\bea
dim \frac{SO(5)}{SO(4)} = 4,
\eea
corresponding to the $4$ rotation angles parameterizing the
vacuum manifold $S^{4}$, as expected.
We will present the rotation matrix in its full form when we encounter it again
while we are discussing the Higgs and axion masses in the next section
where we will extract the rotation matrix $O^{\chi}$ from the full Higgs potential.
Of course the two methods give the same result which means in particular that
if we had computed the rotation matrix first and then the mixings
between $\chi$ and the $Z$-bosons, we would have found them to be all identically zero.

Evidently, the charged and CP-odd parts of the original Higgs kinetic terms
together with the gauge boson mass terms contained in eq. (\ref{quadform})
can be written in this new basis as
\bea
&& (\partial G^+ -i m_W W^+)(\partial G^- + i m_W W^-) + \nonumber\\
&& (\partial \chi)^2 + (\partial G^Z + m_{Z^0} Z)^2 +
\sum_I (\partial G^{Z^\prime}_I + m_{{Z^\prime}_I} {Z^\prime}_I)^2,\label{NGbasis}
\eea
a form that clearly suggests that indeed the $G^\pm$, $G^Z$ and $G^{Z^\prime}_I$ are the
$6$ NG bosons we are after.

%
\subsection{Higgs-axion mixing in the potential}
\label{sect8}
%

The potential $V_{PQ}$ does not give a mass to one of the scalars.
In order to avoid this, one must take into account
new types of contributions to the scalar potential, where not only the
Higgs fields enter but also the the axion fields $a_I$ which
transform under $U(1)$ transformations as (see eq. (\ref{axgauge}))
\be
a_I^\prime\longrightarrow a_I^\prime - M_I\, \e^\prime_I.
\ee
The gauge invariant Higgs potential is then
\be
V_{PQ} = \sum_{a=u,d}\Bigl(  \m_a^2  H_a^{\dagger} H_a + \l_{aa} (H_a^{\dagger} H_a)^2\Bigr)
-2\l_{ud}(H_u^{\dagger} H_u)(H_d^{\dagger} H_d)+2{\l^\prime_{ud}} |H_u^T\tau_2H_d|^2
\ee
as before, plus the new terms
\begin{eqnarray}
V_{\ppq} = &&b\, (H_u^{\dagger}H_d e^{-i\sum_I(q_u^I-q_d^I)\frac{a_I^\prime}{M_I}})
+\l_1 (H_u^{\dagger}H_de^{-i\sum_I(q_u^I-q_d^I)\frac{a_I^\prime}{M_I}})^2  \nonumber\\
&+&\l_2 (H_u^{\dagger}H_u)(H_u^{\dagger}H_de^{-i\sum_I(q_u^I-q_d^I)\frac{a_I^\prime}{M_I}})
+ \l_3 (H_d^{\dagger}H_d)(H_u^{\dagger}H_de^{-i\sum_I(q_u^I-q_d^I)\frac{a_I^\prime}{M_I}}) + c.c.
\nonumber\\ \label{PQbreak}
\end{eqnarray}
In the above, $b$ has dimension of mass squared and $\l_{1,2,3}$ are dimensionless.
As before, we can set $H_u^+=0$ at the minimum by an $SU(2)$ rotation and then
consistency requires that also $H_d^+=0$ at the minimum.
To avoid a stable vacum with unbroken electroweak symmetry, the (MSSM-like)
condition
\bea
\mu_u^2\mu_d^2\; \le\; {b}^2 \label{EWcon}
\eea
must hold. Contribution of terms proportional to $\l_{1,2,3}$ do not appear in this
condition since they all correspond to terms that
mix neutral and charged components of the Higgs fields.
The potential, around the correct vacum, is non-negative definite when
\bea
\m_u^2 v_u^2 + \m_d^2 v_d^2 + \l_{uu}v_u^4+\l_{dd}v_d^4-2\l_{ud}v_u^2v_d^2 +
2b v_uv_d+2 v_u^2v_d^2 (\l_1 + \l_2 \tan{\b} + \l_3 \cot{\b}) \ge 0.\nonumber\\\label{c1}
\eea
This is a necessary condition so that
the potential is bounded from below.
The vevs eq. (\ref{Higgsvev}) still minimize $V_{PQ}+V_{\ppq}$ if
\begin{eqnarray}
&& \mu_u^2 = -b\frac{v_d}{v_u}-2\l_{uu}v_u^2+2\l_{ud}v_d^2
-2\l_1v_d^2-3\l_2v_uv_d-\l_3\frac{v_d^3}{v_u}  \nonumber\\
&& \mu_d^2 = -b\frac{v_u}{v_d}-2\l_{dd}v_d^2+2\l_{ud}v_u^2
-2\l_1v_u^2-\l_2\frac{v_u^3}{v_d}-3\l_3v_uv_d \label{c2}
\end{eqnarray}
and consistency of the minimum of the potential
requires that the couplings $b\m, \l_{1,2,3}$ are all real.
Furthermore, eq. (\ref{c1}) and eq. (\ref{c2}) are compatible when
\bea
\m_u^2 v_u^2+\m_d^2 v_d^2\; \ge\; -2b\; v_uv_d.\label{c3}
\eea
Finally, the parameter range of the couplings should be such that
eq. (\ref{c3}) is consistent also with eq. (\ref{EWcon}).

One should not forget that these statements about the minimum of the potential
are tree level statements. At 1-loop the potential will change and
in general one has to do the minimization from the beginning and make sure
that the chosen Higgs vacuum expectation values still correspond to
a stable minimum that can break the electroweak symmetry in the desired way.
It is possible that there exists an energy regime where the 1-loop
correction to the effective potential is negligible (as it is the
case in the MSSM) and then the tree level results can be still trusted.
In this paper however we restrict ourselves to the tree level analysis.

In order to find the masses of the physical Higgs fields we have to
expand $V_{PQ}+V_{\ppq}$ away from eq. (\ref{Higgsvev}) and collect
the terms quadratic in the fields. Our discussion here is similar to that of section 4 with the
obvious modifications.

The quadratic sector is given by
\beqa
V_q(H) + V'_q(H,a_I^\prime)&=&\left({H_u}^-, {H_d}^-\right){\cal N}_1\left(\begin{array}{c}
{H_u}^+\\
{H_d}^+ \\
\end{array}\right) + \left(Re{H_u}^0, Re{H_d}^0\right){\cal N}_2\left(\begin{array}{c}
Re{H_u}^0\\
Re{H_d}^0 \\
\end{array}\right) \nonumber \\
&& + \left(Im {H_u}^0, Im{H_d}^0, a_I^\prime \right){\cal N}_3\left(\begin{array}{c}
Im{H_u}^0\\
Im{H_d}^0 \\
a_I^\prime\\
\end{array}\right).
\eeqa
In the charged sector, the mass matrix elements are
\beqa
{\cal N}_1(1,1)&=&
-2 \cot \beta \left({\lambda_3} \cos ^2\beta+ ( {\lambda_1} -{\lambda'}_{ud})
\sin 2 \beta +{\lambda_2} \sin ^2\beta\right) v^2 \nonumber \\
&& -2 b  \cot \beta\nonumber \\
{\cal N}_1(1,2)&=&
2 \left({\lambda_3} \cos ^2\beta+ ( {\lambda_1} -{\lambda'}_{ud}) \sin 2 \beta    +{\lambda_2} \sin
   ^2\beta\right) v^2+2 b \nonumber \\
{\cal N}_1(2,2)&=&
-2 \left({\lambda_3} \cos ^2\beta+ ( {\lambda_1} -{\lambda'}_{ud}) \sin 2 \beta     +{\lambda_2} \sin
   ^2\beta\right) v^2 \tan \beta  \nonumber \\
&& -2 b \tan \beta
\eeqa
and we find a zero eigenvalue,
corresponding to the goldstone mode $G^+$ and the nonzero eigenvalue
\beqa
m^2_{H^+} &=&
4{\lambda'}_{ud}v^2 -2 \left(
\frac{2b }{v^2\sin{2\b}} + 2\l_1 + \tan{\b} \l_2 + \cot{\b} \l_3 \right)v^2
\eeqa
corresponding to the charged Higgs mass.
The rotation matrix into the physical eigenstates is
\beq
\left(\begin{array}{c}
{H_u}^+\\
{H_d}^+ \\
\end{array}\right)=
\pmatrix{ \sin \beta  & -\cos {\beta} \cr
 \cos {\beta} & \sin {\beta}}
\left(\begin{array}{c}
{G}^+\\
{H}^+ \\
\end{array}\right),\label{chargedrot5}
\eeq

consistent with eq. (\ref{chargedNGrot}).

In the neutral sector both a CP-even and a CP-odd sector are present.
The CP-even sector is described by ${\cal N}_2$.
The mass matrix in the CP-even sector is given by
\beqa
{\cal N}_2(1,1)&=&
-2 (-4 {v^2\lambda_{uu}} \sin ^2\beta +v^2{\lambda_3} \cos ^2
\beta \cot \beta -\frac{3}{2} v^2{\lambda_2} \sin 2 \beta +b \cot
   \beta ) \nonumber \\
{\cal N}_2(1,2)&=&
2 \left(3 v^2{\lambda_3} \cos ^2\beta +3 v^2{\lambda_2} \sin ^2\beta\ +
2 v^2{\lambda_1} \sin 2 \beta -2 {v^2\lambda_{ud}} \sin 2 \beta +b \right)\nonumber \\
{\cal N}_2(2,2)&=&
-2 \sec \beta \left(-4 {\lambda_{dd}} v^2 \cos ^3\beta-3 {\lambda_3}
v^2 \sin \beta \cos ^2\beta+{\lambda_2} v^2 \sin ^3\beta+b \sin \beta\right) \nonumber \\
\eeqa
and can be diagonalized by an appropriate rotation matrix
in terms of CP-even mass eigenstates $(h^0,H^0)$ as
\beq
\left(\begin{array}{c}
{Re H_u}^0\\
{Re H_d}^0 \\
\end{array}\right)=
\pmatrix{ \sin \a  & -\cos {\a} \cr
 \cos {\a} & \sin {\a}}
\left(\begin{array}{c}
{h}^0\\
{H}^0 \\
\end{array}\right),\label{chargedrotb1}
\eeq
with
\beq
\tan\a= \frac{{\cal N}_2(1,1) - {\cal N}_2(2,2) - \sqrt{\Delta}}{2 {\cal N}_2(1,2)}\label{alpha}
\eeq

and
\beq
\Delta=\left({\cal N}_2(1,1)\right)^2 - 2 {\cal N}_2(2,2){\cal N}_2(1,1)
+ 4 \left({\cal N}_2(1,2)\right)^2 +
\left({\cal N}_2(2,2)\right)^2.
\eeq
The eigenvalues corresponding to the physical neutral Higgs fields are given by
\beqa
m_{h^0}^2 &=& \frac{1}{2}\left( {\cal N}_2(1,1)+{\cal N}_2(2,2) - \sqrt{\Delta}\right)\label{SMHiggs}
\nonumber \\
m_{H^0}^2 &=& \frac{1}{2}\left( {\cal N}_2(1,1)+{\cal N}_2(2,2) + \sqrt{\Delta}\right).
\eeqa
The lighter of the two, $h^0$, is the state which is expected to be
the one that corresponds to the Standard Model Higgs field.

Finally, the symmetric matrix describing the mixing of the CP-odd
Higgs sector with the axion fields $a_I^\prime$ reads
\bea
{\cal N}_3= - \frac{1}{2} v_uv_d \, c_{\chi^\prime}
\pmatrix{\cot{\b} & -1 & v_d\frac{q_u^1-q_d^1}{M_1} & v_d\frac{q_u^2-q_d^2}{M_2}
& v_d\frac{q_u^3-q_d^3}{M_3}  & \cr
         -1 & \tan{\b} & -v_u\frac{q_u^1-q_d^1}{M_1} & -v_u\frac{q_u^2-q_d^2}{M_2}
& -v_u\frac{q_u^3-q_d^3}{M_3} & \cr
v_d\frac{q_u^1-q_d^1}{M_1} & -v_u\frac{q_u^1-q_d^1}{M_1} \cr
v_d\frac{q_u^2-q_d^2}{M_2} & -v_u\frac{q_u^2-q_d^2}{M_2} & &
v_uv_d \frac{(q_u^I-q_d^I)(q_u^J-q_d^J)}{M_IM_J}
\cr v_d\frac{q_u^3-q_d^3}{M_3}  & -v_u\frac{q_u^3-q_d^3}{M_3}}
\eea
with
\bea
c_{\chi^\prime} = \frac{ 4b}{v^2\sin{2\b}} +4\l_1 + \l_2 \tan{\b} + \l_3 \cot{\b}.
\eea
The rotation from the interaction to the mass eigenstates
in the CP-odd sector is given by eq. (\ref{rotunit}).
To construct the rotation matrix we start from the matrix whose
columns are the normalized eigenvectors of ${\cal M}_3$:
\beqa
E^{\chi} = \pmatrix{
-N\cos{\b} & \sin{\b} & N_1 Q_1 & N_2 Q_2 & N_3 Q_3 \cr
 N\sin{\b} & \cos{\b}  & 0 & 0 & 0 \cr
N Q_1 \cos{\b} & 0 & N_1 & 0 & 0 \cr
N Q_2 \cos{\b} & 0 & 0 & N_2 & 0 \cr
N Q_3 \cos{\b} & 0 & 0 & 0 & N_3 },\nonumber\\
\eeqa
where we have defined
\bea
Q_I = -(q_u^I-q_d^I) \frac{v}{M_I}\sin{\b},
\eea
$N$ is given by eq. (\ref{normcoeff}) and
\bea
N_I = \frac{1}{\sqrt{1+Q_I^2}}.
\eea
This is not an orthogonal matrix yet
since it corresponds to $3$ degenerate eigenvalues.
One can construct the orthogonal
matrix $O^{\chi}$ by starting from the first two columns
which are already orthonormal and apply the
Gram-Schmidt orthogonalization method to the rest. Doing this
one obtains the matrix
\beqa
O^{\chi} = \pmatrix{
-N\cos{\b} & \sin{\b} & {\overline N}_1 {\overline Q}_1 \cos{\b} &
{\overline N}_1{\overline N}_2\, {\overline Q}_2 \cos{\b} &
{\overline N}_1{\overline N}_2{\overline N}_3\, {\overline Q}_3 \cos{\b}\cr
 N\sin{\b} & \cos{\b} & -{\overline N}_1 {\overline Q}_1 \sin{\b} &
-{\overline N}_1{\overline N}_2\, {\overline Q}_2 \sin{\b} &
-{\overline N}_1{\overline N}_2{\overline N}_3\, {\overline Q}_3 \sin{\b}\cr
 NQ_1 \cos{\b}& 0 & {\overline N}_1 &
- {\overline N}_1{\overline N}_2\, {\overline Q}_1{\overline Q}_2&
-{\overline N}_1{\overline N}_2{\overline N}_3\, {\overline Q}_1{\overline Q}_3\cr
 NQ_2 \cos{\b}& 0 & 0 & {\overline N}_2 &
-{\overline N}_2{\overline N}_3\, {\overline Q}_2{\overline Q}_3\cr
 NQ_3 \cos{\b} & 0 & 0 & 0& {\overline N}_3},\nonumber\\
\eeqa
where we defined
\bea
{\overline Q}_1 = Q_1 \cos{\b}, \hskip 1cm {\overline Q}_2 = Q_2 {\overline N}_1\cos{\b},
\hskip 1cm {\overline Q}_3 = Q_3 {\overline N}_1{\overline N}_2 \cos{\b}
\eea
and
\bea
{\overline N}_1 = \frac{1}{\sqrt{1+{\overline Q}_1^2}},\hskip .5cm
{\overline N}_2 = \frac{1}{\sqrt{1+{\overline Q}_2^2}},\hskip .5cm
{\overline N}_3 = \frac{1}{\sqrt{1+{\overline Q}_3^2}}.
\eea

The fact that this is indeed the same as the matrix $O^{\chi}$ of
eq. (\ref{rotunit}) of the previous section is
an important self consistency check of the model.
The mass matrix ${\cal M}_3$ has 4 zero eigenvalues representing the NG-bosons
that parameterize the corresponding 4-dimensional branch of the vacuum manifold
and one non-zero eigenvalue that corresponds to a physical
axion field $- \chi -$ with mass
\bea
m_{\chi}^2 =
-\frac{1}{2}\left[1+\sum_I \left( \frac{q_u^I-q_d^I}{2}
\frac{v}{M_I}\sin{2\b}\right)^2 \right] c_{\chi^\prime}\; v^2.\label{axionmass}
\eea
The mass of this state is positive if $c_{\chi^\prime} < 0$.

%
\subsection{The Green-Schwarz sector in the broken phase}
%

The anomalous couplings computed in sect. \ref{ancoup}
imply in the broken phase a number of interesting processes.
After electroweak symmetry breaking some of the $U(1)$s get rotated to the
basis where electromagnetism is the good quantum number. In particular,
the $W_3, Y$ and $A_I$
gauge bosons become linear combinations of the physical states
$A_\g, Z, Z^\prime_I$, as we have seen in detail.
The rotation to the physical mass eigenstate basis is done by the
5 by 5 orthogonal matrix $O^{A}$ of eq. (\ref{rotgaugeOA}):
\bea
{A}\, =\, O^A\, {\cal B},\label{rotAA}
\eea
where in components
\bea
{A}^{\overline p} = \{A_\g,Z,{Z_I^\prime} \},\hskip 1cm
{\cal B}^p = \{W_3, {A}^{l^\prime} \}.
\eea
The rotation of the $l^\prime$th and $W_3$ components then reads
\bea
{A}^{l^\prime}=O^{A}_{{\overline p}l^\prime} {A}^{\overline p},\hskip 1cm
W_3 = O^{A}_{{\overline p}W_3} {A}^{\overline p}.\label{rot11}
\eea
with the sum over $\overline p$ implied.
In order to analyze the theory in the broken phase it is also convenient to
separate the quadratic from the cubic and quartic terms in the product of the
field strengths of the gauge fields. We define
\beqa
W^\pm_{\mu\nu}&\equiv&\frac{1}{\sqrt{2}}\left( W^1_{\mu\nu} \mp i W^2_{\mu\nu}\right)
\nonumber \\
&=&\overline{W}^{\pm}_{\mu\nu} \pm \overline{Q}^{\pm}_{\mu\nu},\nonumber \\
W^3_{\mu\nu}&=& \overline{W}^{3}_{\mu\nu} + \overline{Q}^{3}_{\mu\nu},\nonumber \\
\eeqa
where
\beqa
\overline{W}^{\pm}_{\mu\nu}&\equiv&\partial_\mu W^\pm_\nu -\partial_\nu W^\pm_\mu \nonumber \\
\overline{Q}^{\pm}_{\mu\nu}&\equiv& i g_2 \left( W^3_\mu W^\pm_\nu - W^3_\nu W^\pm_\mu  \right)
\nonumber \\
\overline{W}^{3}_{\mu\nu}&=&\partial_\mu W^3_\nu - \partial_\nu W^3_\mu \nonumber \\
\overline{Q}^{3}_{\mu\nu} &=& i g_2 \left( W^+_\mu W^-_\nu - W^-_\mu W^+_\nu\right).
\eeqa
Also, as we have showed, the hypercharge basis axions must be rotated as
\bea
a^\prime_I = {\bf \Theta}_I\, \chi + \sum _{i=1}^{4} c_i^{(I)} G_i^0 \label{rot12}
\eea
where ${\bf \Theta}_I$ and $c_i^{(I)}$ are dimensionless,
computable but model dependent coefficients.
Putting eqs. (\ref{rot11}) and (\ref{rot12}) into eq. (\ref{GS3}),
we obtain the Green-Schwarz terms in the photon eigenstate basis
\bea
{\cal L}^{GS} &=&
g^{\chi gg} \, \chi\, tr\, \{G\wedge {G}\} +
\sum_{I} {D_I^\prime} \sum _{i=1}^{4} c_i^{(I)} G_i^0\, tr\, \{G\wedge {G}\} \nonumber\\
&+& g^{\chi +-} \, \chi\, tr\, \{W^+\wedge {W^-}\} +
\sum_{I} {F_I^\prime} \sum _{i=1}^{4} c_i^{(I)} G_i^0\, tr\, \{W^+\wedge {W^-}\}\nonumber\\
&+& g^{\chi}_{{\overline p}{\overline q}} \, \chi\, F^{\overline p}\wedge F^{\overline q}+
\sum_{I} (F_I^\prime O^A_{{\overline p}W_3} O^A_{{\overline q}W_3} +
C^\prime_{Im^\prime n^\prime} O^A_{{\overline p}m^\prime}
O^A_{{\overline q}n^\prime} ) \sum _{i=1}^{4} c_i^{(I)} G_i^0\,F^{\overline p}\wedge F^{\overline q}\nonumber\\
&+& g_{{\overline p}{\overline q}{\overline r}}\, \e^{\m\n\r\s}\,
{A}^{\overline p}_{\m} {A}^{\overline q}_{\n} {F}^{\overline r}_{\r\s}\nonumber\\
\eea
where we have separated the physical couplings
\bea
&& g^{\chi gg}\,= \, \sum_{I} {D_I^\prime}\, {\bf \Theta}_I \nonumber\\
&& g^{\chi +-}\, = \, \sum_{I} {F_I^\prime}\, {\bf \Theta}_I\nonumber\\
&& g^{\chi}_{{\overline p}{\overline q}}\, = \,
\sum_{I} (F_I^\prime O^A_{{\overline p}W_3} O^A_{{\overline q}W_3} +
C^\prime_{Im^\prime n^\prime} O^A_{{\overline p}m^\prime}
O^A_{{\overline q}n^\prime} )\, {\bf \Theta}_I \nonumber\\
&& g_{{\overline p}{\overline q}{\overline r}}\, = \,
E_{l^\prime m^\prime n^\prime}O^A_{{\overline p}l^\prime}
O^A_{{\overline q}m^\prime} O^A_{{\overline r}n^\prime} \label{GSEW}
\eea
from the interactions of the NG-bosons with the gauge bosons.

%
\subsection{Gauge fixing in the broken phase}
%
The gauge fixing functions can be straightforwardly obtained, as in the SM.
The $SU(3)$ part of the gauge fixing terms is without any
modification since the symmetry is not broken.
For the rest, we now have in the charged sector the gauge fixing functions
\bea
&& {\cal G}^+ = \partial \cdot W^+ + \frac{i}{2} g_2 v \a^+ G^+ \nonumber\\
&& {\cal G}^- = \partial \cdot W^- - \frac{i}{2} g_2 v \a^- G^-
\eea
where $G^\pm$ are as in (\ref{chargedNGrot}) and $\a^- = (\a^+)^*$.
In the neutral sector we have the gauge fixing functions
\bea
&& {\cal G}^\g = \partial \cdot A^\g  \nonumber\\
&& {\cal G}^Z = \partial \cdot Z + \a^Z G^Z  \nonumber\\
&& {\cal G}^{Z^\prime}_I = \partial \cdot Z^\prime_I + \a^{Z^\prime}_I G^{Z^\prime}_I
\eea
with $G^Z$ and $G^{Z^\prime}_I$ given in (\ref{mix41}) and (\ref{mix42}) respectively.
The gauge fixing terms are then
\bea
{\cal L}^{gf} = \frac{1}{2 \a^\g} {\cal G}^\g {\cal G}^\g
+ \frac{1}{\sqrt{\a^+\a^-}} {\cal G}^+ {\cal G}^-
+ \frac{1}{2 \a^Z} {\cal G}^Z {\cal G}^Z
+ \sum_{I} \frac{1}{2 \a^{Z^\prime}_I} {\cal G}^{Z^\prime}_I {\cal G}^{Z^\prime}_I.
\eea
The $SU(3)$ ghosts remain the same as before and
for the broken part we have
\bea
{\cal L}^{gh} &=& \Bigl\{ (\partial \eta_\g^*)\cdot (\partial \eta_\g) +
(\partial \eta_+^*)\cdot (\partial \eta_-)  + (\partial \eta_-^*)\cdot (\partial \eta_+)
+ (\partial \eta_Z^*)\cdot (\partial \eta_Z) +
\sum_{I}(\partial \eta_{Z^\prime_I}^*)\cdot (\partial \eta_{Z^\prime_I}) \nonumber\\
&+& m_W^2 (\a^+ \eta_+^*\eta_- + \a^- \eta_-^*\eta_+) +
\a^Z m_Z^2 \eta_Z^* \eta_Z +
\sum_{I} \a^{Z^\prime}_I (m_{Z^\prime_I}^I)^2 \eta_{Z^\prime_I}^* \eta_{Z^\prime_I}\Bigr\}.
\eea
In the limit $\a^+, \a^-, \a^Z, \a^{Z^\prime}_I\longrightarrow \infty$ the gauge fixing conditions
reduce to the unitary gauge conditions
\bea
G^+, G^-, G^Z, G^{Z^\prime}_I\longrightarrow 0.
\eea
Indeed, one should be able to exploit the
original $SU(2)_W\times U(1)^{4}$ gauge symmetry to transform into
a gauge where the NG-bosons vanish.
Denoting by $\xi^{+-3}$, $\xi^Z$ and $\xi^{Z^\prime_I}$
the corresponding gauge transformation parameters, we can choose
them in such a way so that they act on the gauge fields as
\bea
&& \d W^- = \partial \xi^- + \xi^+ W^3 - \xi^3 W^- = \frac{i}{m_W} \partial G^+\nonumber\\
&& \d W^+ = \partial \xi^+ + \xi^- W^3 - \xi^3 W^+ = -\frac{i}{m_W} \partial G^-\nonumber\\
&& \d W^3 = \partial \xi^3 +2(\xi^+ W^- + \xi^-W^+) = 0 \nonumber\\
&& \d Z = \partial \xi^Z = \frac{1}{m_{Z}}\partial G^Z\nonumber\\
&& \d {Z^\prime_I} = \partial \xi^{Z^\prime_I} =
\frac{1}{m_{{Z^\prime_I}}}\partial G^{Z^\prime_I}.\nonumber\\
\eea
These transformations act on the
NG-bosons in such a way so that in the new basis they vanish.

It is expected that in this gauge the ghosts decouple, as in the SM.
For certain computations this is a useful gauge but in general
computations are done in more practical gauges.
In this work the processes we will be interested in are simple enough
so that we can perform all calculations in the unitary gauge without any problems.

%
\section{Tree level decay rates and cross sections\label{t4} }
\setcounter{equation}{0}
%
In this section we are going to present selected
tree-level decays rates that are useful on two counts.

$\bullet$ They are important in order top constraint
the parameters of mLSOM using current data.

$\bullet$ They are crucial for uncovering new physics
 in forthcoming colliders.

%
\subsection{Minimal gauge  interactions}
%

The interaction of two fermions with a gauge boson can be found in the
fermion kinetic terms and more precisely in the part linear
in the gauge fields.

Let us first look at the interaction terms contained in
the interaction Lagrangian ${\cal L}_{int}$.
We will use the hypercharge values
%
\hskip 2cm
\begin{center}
\begin{tabular}{|c|c|c|c|c|c|c|}
\hline
$ f $ & $Q$ &  $ u_{R} $ &  $ d_{R} $ & $ L $ & $e_R$ & $\n_R$ \\
\hline \hline
$q_Y$  &  $1/6$  & $-2/3$  &  $1/3$ & $-1/2$ & $1$ & $0$\\ \hline
\end{tabular}
\end{center}
%
Writing the lepton doublet as
\bea
L = \pmatrix {{\n}_{Li}\cr e_{Li}}
\eea
we have the terms
\beqa
{\cal L}_{int}
&=& i L^{\dag}_{i}\s^{\m}{\cal D}_{\m}L_{i} +
i{e}_{Ri}^{\dag}{\overline \s}^{\m}{\cal D}_{\m}{e}_{Ri}
+ i{\n}_{Ri}^{\dag}{\overline \s}^{\m}{\cal D}_{\m}{\n}_{Ri}=\nonumber
\nonumber
\eeqa
\beqa
&&-\pmatrix{{\n}_{Li}^{\dag} & e_{Li}^{\dag}} \s^{\m}
\left[g_2 \tau^aW_{\m}^a +q^{(L)}_Yg_YA^Y_{\m}+
\sum_{I} q^{(L)}_Ig_IA^I_{\m}\right]\pmatrix{\n_{Li} \cr e_{Li}} \nonumber\\
&& +  \; e^{\dag}_{Ri} \; {\overline \s}^{\m}\left[
q^{(e_{R})}_Yg_YA^Y_{\m}+\sum_{I} q^{(e_{R})}_Ig_IA^I_{\m}\right]e_{Ri}\nonumber\\
&& +  \; {\n}^{\dag}_{Ri} \; {\overline \s}^{\m}\left[
q^{({\n}_{R})}_Yg_YA^Y_{\m}+\sum_{I} q^{(\n_{R})}_Ig_IA^I_{\m}\right]{\n}_{Ri}.
\end{eqnarray}
The interaction terms in terms of the currents are given by
\bea
{\cal L}_l&=& - \frac{g_Yg_2}{g}A^{\g}_{\m}J^{{\m}(SM)}_{\g}
-\frac{1}{\sqrt{2}}g_2 W_{\m}^+J^{{\m}(SM)}_{-}
-\frac{1}{\sqrt{2}}g_2 W_{\m}^-J^{{\m}(SM)}_{+} \nonumber\\
&-& g\, {Z}_{\m} J^\m_{Z}
 - \sum_{I} {Z'_I}_{\m} J^{{\m}}_{Z'_I},\label{interlepton}
\eea
where the electromagnetic and charged currents are
\begin{eqnarray}
&& J^{{\m}(SM)}_{\g}=-(e_{Li}^{\dag}\s^{\m}e_{Li} +
e_{Ri}^{\dag} {\overline \s}^{\m} e_{Ri})\\
&& J^{{\m}(SM)}_{+}=e^{\dag}_{Lj} \ \, {\cal U}^{\n}_{ji}\, \s^{\m} {\n}_{Li}\\
&& J^{{\m}(SM)}_{-}={\n}_{Lj}^{\dag} \, {\cal U}^{\n\dag}_{ji}\, \s^{\m}e_{Li}.
\end{eqnarray}
The latter are the same as in the Standard Model
modulo the presence of the MNS matrix due to the presence of
the right handed neutrinos.
In particular, eq. (\ref{interlepton}) implies that the electric charge can be defined as
\bea
e\; =\; \frac{g_Yg_2}{g}.
\eea
The neutral currents
have, in addition to the usual Standard Model values, corrections due to
the additional $U(1)$ structure. They can be expressed as
\beqa
J^{{\m}}_{Z} &=& C^{Z}_{\n_{Li}}\, {\n}_{Li}^{\dag}\s^{\m}{\n}_{Li}
+ C^{Z}_{\e_{Li}} \, e_{Li}^{\dag}\s^{\m}e_{Li}  +
C^{Z}_{\n_{Ri}}\, {\n}_{Ri}^{\dag}{\overline \s}^{\m}{\n}_{Ri}
+ C^{Z}_{\e_{Ri}} \, e_{Ri}^{\dag}{\overline \s}^{\m}e_{Ri},
\eeqa
where
\beqa
C^{Z}_{\n_{Li}}&=& =\frac{1}{2} +\frac{1}{2}\sum_{I} \e_I\, q_I^{(L)} g_I +\cdots  \nonumber \\
C^{Z}_{\e_{Li}}&=& =\frac{g_Y^2 - g_2^2}{2 g^2}+\frac{1}{2}
\sum_{I} \e_I\, q_I^{(L)} g_I +\cdots \nonumber \\
C^{Z}_{\e_{Ri}}&=& = -\frac{g_Y^2}{g^2}+\frac{1}{2}
\sum_{I} \e_I\, q_I^{(e_R)} g_I +\cdots\nonumber \\
C^{Z}_{\n_{Ri}}&=& +\frac{1}{2}
\sum_{I} \e_I\, q_I^{(\n_R)} g_I +\cdots\nonumber \\
\eeqa
It is convenient to organize to lowest order the currents as follows
\beq
J^{{\m}}_{Z}=J^{{\m}(SM)}_{Z} + \frac{1}{2} \sum _{I} \e_I\, g_I J^{{\m}(D)}_{Z,I} ,
\eeq
where we have introduced a standard model contribution (SM) and a D-brane correction $(D)$.
The SM contribution is obtained in the $M_I \to \infty$ limit:
\bea
&&J^{{\m}(SM)}_{Z}= J^{{\m}}_{Z^0}= \frac{1}{2}{\n}_{Li}^{\dag}\s^{\m}{\n}_{Li}
+\frac{1}{2}\frac{g_Y^2-g_2^2}{g^2}e_{Li}^{\dag}\s^{\m}e_{Li} -
\frac{g_Y^2}{g^2} e_{Ri}^{\dag}{\overline \s}^{\m}e_{Ri}
\eea
and the corrections induced by the extra gauge bosons are given by
\be
J^{{\m}(D)}_{Z,I} = q_I^{(L)} {\n}_{Li}^{\dag}\s^{\m}{\n}_{Li} +
q_I^{(L)} e_{Li}^{\dag}\s^{\m}e_{Li}
 + q_I^{(e_R)} e_{Ri}^{\dag}{\overline \s}^{\m}e_{Ri}
+ q_I^{(\n_R)} {\n}_{Ri}^{\dag}{\overline \s}^{\m}{\n}_{Ri}.\nonumber \\
\ee
The currents $J^{{\m}(D)}_{Z,I}$ are new interactions not predicted in the Standard Model
and therefore should put constraints on the model.

A similar computation for the $J_{Z'_I}$ current gives
\beqa
J^{{\m}}_{Z'_I} &=& C^{Z'_I}_{\n_{Li}}\, {\n}_{Li}^{\dag}\s^{\m}{\n}_{Li}
+ C^{Z'_I}_{\e_{Li}} \, e_{Li}^{\dag}\s^{\m}e_{Li}  +
C^{Z'_I}_{\n_{Ri}}\, {\n}_{Ri}^{\dag}{\overline \s}^{\m}{\n}_{Ri}
+ C^{Z'_I}_{\e_{Ri}} \, e_{Ri}^{\dag}{\overline\s}^{\m}e_{Ri}
\eeqa
where
\beqa
C^{Z'_I}_{\n_{Li}}&=& -\frac{1}{4} g^2 \e_I +
\sum_{J} q_J^{(L)}g_J X_{JI}+\cdots \nonumber \\
C^{Z'_I}_{\e_{Li}}&=& -\frac{1}{4} (g_Y^2-g_2^2)\e_I +
\sum_{J} q_J^{(L)}g_J X_{JI} +\cdots \nonumber \\
C^{Z'_I}_{\e_{Ri}}&=& \frac{g_Y^2}{2}\e_I +
\sum_{J} q_J^{(e_R)}g_J X_{JI} +\cdots \nonumber \\
C^{Z'_I}_{\n_{Ri}}&=& \sum_{J} q_J^{(\n_R)}g_J X_{JI} +\cdots ,
\eeqa
which we can express as
\beq
J^{{\m}}_{Z'_I} = g_I J^{{\m}(0)}_{Z'_I} + J^{{\m}(1)}_{Z'_I} + \cdots
\eeq
where
\beqa
J^{{\m}(0)}_{Z'_I} &=& q_I^{(L)} {\n}_{Li}^{\dag}\s^{\m}{\n}_{Li}
+  q_I^{(L)} e_{Li}^{\dag}\s^{\m}e_{Li}
+  q_I^{(e_R)} e_{Ri}^{\dag}{\overline\s}^{\m}e_{Ri}
+  q_I^{(\n_R)} {\n}_{Ri}^{\dag}{\overline \s}^{\m}{\n}_{Ri}
\eeqa
and
\beqa
J^{{\m}(1)}_{Z'_I} &=&
\left(-\frac{1}{4} g^2 \e_I +
\sum_{J\ne I} q_J^{(L)}g_J \frac{N_{JI}}{4(M_J^2-M_I^2)}\right){\n}_{Li}^{\dag}\s^{\m}{\n}_{Li}\nonumber\\
&+& \left(-\frac{1}{4} (g_Y^2-g_2^2)\e_I +
\sum_{J\ne I} q_J^{(e_R)}g_J \frac{N_{JI}}{4(M_J^2-M_I^2)} \right)e_{Li}^{\dag}\s^{\m}e_{Li}\nonumber\\
&+& \left(\frac{g_Y^2}{2}\e_I +
\sum_{J\ne I} q_J^{(e_R)}g_J \frac{N_{JI}}{4(M_J^2-M_I^2)} \right)e_{Ri}^{\dag}{\overline \s}^{\m}e_{Ri}\nonumber\\
&+& \left(
\sum_{J\ne I} q_J^{(\n_R)}g_J \frac{N_{JI}}{4(M_J^2-M_I^2)} \right)\n_{Ri}^{\dag}{\overline \s}^{\m}\n_{Ri}.
\eea
As we can see from these results, in the limit $\e_I \to 0$,
the $Z'_I$ gauge boson
interacts with a strength which is of order of the coupling $g_{I}$
and can be identified with the original $A_{I}$ gauge boson, modulo corrections
of order $v/M_I$. The $Z$ couplings to the leptons
tend to the usual Standard Model coupling of the $Z^0$.

We now turn to the Higgs-quark-quark interactions contained in
the quark kinetic terms
$i Q^{\dag}_{Li}\s^{\m}{\cal D}_{\m}Q_{Li} +
i{u}_{Ri}^{\dag}{\overline \s}^{\m}{\cal D}_{\m}{u}_{Ri} +
i{d}_{Ri}^{\dag}{\overline \s}^{\m}{\cal D}_{\m}{d}_{Rj} $.
Writing the quark doublet as
\bea
Q_{Li} = \pmatrix {u_{Li}\cr d_{Li}},
\eea
after some algebra we find the interactions
\begin{eqnarray}
{\cal L}_q&= &- \frac{g_Yg_2}{g}A^{\g}_{\m}I^{{\m}(SM)}_{\g}
-\frac{1}{\sqrt{2}}g_2 W_{\m}^+I^{{\m}(SM)}_{-} -
\frac{1}{\sqrt{2}}g_2 W_{\m}^-I^{{\m}(SM)}_{+} \\
&&- g\, {Z}_{\m} I^{\mu}_Z
- \sum_{I} g_{I} {Z'_I}_{\m} I^{{\m}}_{Z'_I},
\end{eqnarray}
where the SM hadronic electromagnetic and charged currents given by
\be
I^{{\m}(SM)}_{\g}=\frac{2}{3}(u_{Li}^{\dag}\s^{\m}u_{Li}+u_{Ri}^{\dag}{\overline \s}^{\m}u_{Ri})
-\frac{1}{3}(d_{Li}^{\dag}\s^{\m}d_{Li}+d_{Ri}^{\dag}{\overline \s}^{\m}d_{Ri})
\ee
\be
I^{{\m}(SM)}_{-}=u^{\dag}_{Lj} \, {\cal U}^{q}_{ji}\, \s^{\m} {d}_{Li}\label{CKM1}
\ee
\be
I^{{\m}(SM)}_{+}=d^{\dag}_{Lj} \, {\cal U}^{q\dag}_{ji}\, \s^{\m} {u}_{Li}.\label{CKM2}
\ee
In eqs. (\ref{CKM1}) and (\ref{CKM2}) we introduced in the quark mass eigenstate basis
the CKM matrix which again enters
in an analogous way as in the Standard Model.

The quarks couple to the Z-boson by the current
\beqa
I^{{\m}}_{Z} &=&
C^{Z}_{u{Li}}\, {u}_{Li}^{\dag}\s^{\m}{u}_{Li} +
C^{Z}_{d_{Li}} \, d_{Li}^{\dag}\s^{\m}d_{Li}  +
C^{Z}_{u_{Ri}}\, {u}_{Ri}^{\dag}{\overline \s}^{\m}{u}_{Ri} +
C^{Z}_{d_{Ri}} \, d_{Ri}^{\dag}{\overline \s}^{\m}d_{Ri}
\eeqa
where
\beqa
C^{Z}_{u_{Li}}&=& = \frac{g_2^2}{2 g^2} - \frac{g_Y^2}{6 g^2}
+\frac{1}{2}\sum_{I}\e_I\, q_I^{(Q)} g_I +\cdots\nonumber \\
C^{Z}_{d_{Li}}&=& - \frac{g_2^2}{2 g^2} - \frac{g_Y^2}{6 g^2}
+\frac{1}{2}\sum_{I}\e_I\, q_I^{(Q)} g_I +\cdots\nonumber \\
C^{Z}_{u_{Ri}}&=& - \frac{2}{3}\frac{g_Y^2}{g^2}
+\frac{1}{2}\sum_{I}\e_I\, q_I^{(u_R)} g_I +\cdots\nonumber \\
C^{Z}_{d_{Ri}}&=&   \frac{g_Y^2}{3 g^2}
+\frac{1}{2}\sum_{I}\e_I\, q_I^{(d_R)} g_I +\cdots  .
\eeqa
We decompose the $Z$ current also in this case in terms of an ordinary Standard Model contribution
and a second term coming from the extra contributions
\beq
I^{\m}_Z=I^{{\m}(SM)}_{Z}+\frac{1}{2} \sum _{I} \e_I\, g_I I^{{\m}(D)}_{Z,I}.
\eeq
The SM contribution is obtained in the $\e_I\to 0$ limit
\bea
I^{{\m}(SM)}_{Z} = I^{{\m}}_{Z^0} &=&
\left(\frac{g_2^2}{2g^2}-\frac{g_Y^2}{6g^2}\right)u_{Li}^{\dag}\s^{\m}u_{Li}
- \left(\frac{g_2^2}{2g^2}+\frac{g_Y^2}{6g^2}\right)d_{Li}^{\dag}\s^{\m}d_{Li}\\
&-& \frac{2}{3}\frac{g_Y^2}{g^2}\; u_{Ri}^{\dag}{\overline \s}^{\m}u_{Ri}
+\frac{1}{3}\frac{g_Y^2}{g^2}\; d_{Ri}^{\dag}{\overline \s}^{\m}d_{Ri}
\eea
while the corrections induced by the extra gauge bosons at lowest order are given by
\bea
I^{{\m}(D)}_{Z,I} &=& q_I^{(Q)}{u}_{Li}^{\dag}\s^{\m}{u}_{Li} +
q_I^{(Q)}d_{Li}^{\dag}\s^{\m}d_{Li}
+ q_I^{(u_R)}{u}_{Ri}^{\dag}{\overline \s}^{\m}{u}_{Ri}
+q_I^{(d_R)}d_{Ri}^{\dag}{\overline \s}^{\m}d_{Ri}.
\eea
Again, as for the lepton currents, the corrections to the SM  neutral currents
$I^{{\m}(D)}_{Z}$ are suppressed in the limit of large $M_I$.

Finally, the coupling of the $Z'_I$ gauge boson to the quarks is given by
\beqa
I^{{\m}}_{Z'_I} &=& C^{Z'_I}_{u{Li}}\, {u}_{Li}^{\dag}\s^{\m}{u}_{Li}
+ C^{Z'_I}_{d_{Li}} \, d_{Li}^{\dag}\s^{\m}d_{Li}  +
C^{Z'_I}_{u_{Ri}}\, {u}_{Ri}^{\dag}{\overline \s}^{\m}{u}_{Ri}
+ C^{Z'_I}_{d_{Ri}} \, d_{Ri}^{\dag}{\overline \s}^{\m}d_{Ri}
\eeqa
where
\beqa
C^{Z'_I}_{u_{Li}}&=& -\frac{1}{4} (g_2^2-\frac{1}{3}g_Y^2)\e_I +
\sum_{J} q_J^{(Q)}g_J X_{JI}+\cdots \nonumber \\
C^{Z'_I}_{d_{Li}}&=& \frac{1}{4} (g_2^2+\frac{1}{3}g_Y^2)\e_I +
\sum_{J} q_J^{(Q)}g_J X_{JI}+\cdots \nonumber \\
C^{Z'_I}_{u_{Ri}}&=& -\frac{1}{3} g_Y^2\e_I +
\sum_{J} q_J^{(u_R)}g_J X_{JI}+\cdots \nonumber \\
C^{Z'_I}_{d_{Ri}}&=& \frac{1}{6} g_Y^2\e_I +
\sum_{J} q_J^{(d_R)}g_J X_{JI}+\cdots  .
\eeqa
As in the leptonic coupling of the $Z'_I$ also in this case we split
the quark current writing
\beq
I^{\m}_{Z'_I}= g_I I^{{\m (0)}}_{Z'_I} + I^{{\m}(1)}_{Z'_I}
\eeq
with
\beqa
I^{{\m}(0)}_{Z'_I} &=&  q_I^{(Q)}{u}_{Li}^{\dag}\s^{\m}{u}_{Li} + q_I^{(Q)}d_{Li}^{\dag}\s^{\m}d_{Li}  +
q_I^{(u_R)}{u}_{Ri}^{\dag}{\overline \s}^{\m}{u}_{Ri} +q_I^{(d_R)} d_{Ri}^{\dag}{\overline \s}^{\m}d_{Ri}
\eeqa
and
\beqa
I^{{\m}(1)}_{Z'_I} &=&
\left(-\frac{1}{4} (g_2^2-\frac{1}{3}g_Y^2)\e_I +
\sum_{J\ne I} q_J^{(Q)}g_J \frac{N_{JI}}{4(M_J^2-M_I^2)}\right)
{u}_{Li}^{\dag}\s^{\m}{u}_{Li} \nonumber \\
&+& \left( \frac{1}{4} (g_2^2+\frac{1}{3}g_Y^2)\e_I
+\sum_{J\ne I} q_J^{(Q)}g_J \frac{N_{JI}}{4(M_J^2-M_I^2)}\right)
d_{Li}^{\dag}\s^{\m}d_{Li} \nonumber \\
&+& \left( -\frac{1}{3} g_Y^2\e_I +
\sum_{J\ne I} q_J^{(u_R)}g_J \frac{N_{JI}}{4(M_J^2-M_I^2)}\right)
{u}_{Ri}^{\dag}{\overline \s}^{\m}{u}_{Ri} \nonumber \\
&+& \left( \frac{1}{6} g_Y^2\e_I +
\sum_{J\ne I} q_J^{(d_R)}g_J \frac{N_{JI}}{4(M_J^2-M_I^2)}\right)
d_{Ri}^{\dag}{\overline \s}^{\m}d_{Ri}.
\eeqa

Having the explicit form for the currents and the couplings, one can derive easily
certain tree level decay rates and cross sections.
Since the charge currents are the same as in the
Standard Model, the tree level decay rates of the $W$'s do not get any new contributions.
On the other hand the neutral currents associated with
the remaining gauge bosons receive (small) corrections with respect to their Standard Model values,
so we expect to have comparable corrections to the tree level
Standard Model decay rates of the $Z^0$.
Here we will provide a rather general analysis
of the main decays of the $Z$ and $Z'_I$ gauge bosons into leptons and quarks and then we will
present predictions for the Drell-Yan cross sections. We will also compare at a quantitative
level the predictions coming from the D-brane model with other models containing, for instance,
sequential $U(1)$'s, with charge assignments different from the ones we discuss here.

We present a few examples of some possible phenomenological interest.
For future convenience and for a direct comparison with the literature on the Standard Model,
we rewrite here the interaction Lagrangian of quarks and leptons with the neutral gauge bosons
$Z$ and $Z'_I$ in 4-component form. We obtain for the leptons
\beqa
{\cal L}_l &=& e\overline{e}_i\gamma^\m e_i
-\frac{g_2}{2 \sqrt{2}} W^+ \overline{e}_j\gamma^\m(1 - \gamma_5) \nu_i\, {\cal U}^\nu_{ji}
-\frac{g_2}{2 \sqrt{2}} W^- \overline{\nu}_j\gamma^\m(1 - \gamma_5) e_i\,
{\cal U}^{\nu\dag}_{ji} \nonumber \\
&& - \frac{g_2}{2\cos\theta_W} Z^\mu \overline{\nu}_i
\left( g^{\nu-Z}_V\gamma^\mu - g^{\nu-Z}_A\gamma^\mu \gamma^5\right)\nu_i
- \frac{g_2}{2\cos\theta_W} Z^\mu \overline{e}_i
\left( g^{e-Z}_V\gamma^\mu - g^{e-Z}_A\gamma^\mu \gamma^5\right)e_i
\nonumber \\
&& - g_{I} {Z'_I}^\mu \overline{\nu}_i\left( g^{\nu-Z'_I}_V\gamma^\mu
- g^{\nu-Z'_I}_A
\gamma^\mu \gamma^5\right)\nu_i
 - g_{I} {Z'_I}^\mu \overline{e}_i\left( g^{e-Z'_I}_V\gamma^\mu
- g^{e-Z'_I}_A\gamma^\mu \gamma^5\right)
e_i \nonumber \\
\eeqa
and
\beqa
{\cal L}_q &=& -e \left( \frac{2}{3} \overline{u}_i\gamma^\m u_i -
\frac{1}{3} \overline{d}_i\gamma^\m d_i \right)A_\mu \nonumber \\
&& - \frac{g_2}{2 \sqrt{2}} W^+ \overline{u}_j\gamma^\m(1 - \gamma_5) d_i\, {\cal U}^q_{ji}
-\frac{g_2}{2 \sqrt{2}} W^- \overline{d}_j\gamma^\m
(1 - \gamma_5) u_i\, {\cal U}^{q\dag}_{ji} \nonumber \\
&& - \frac{g_2}{2\cos\theta_W} Z^\mu \overline{u}_i
\left( g^{u-Z}_V\gamma^\mu - g^{u-Z}_A\gamma^\mu \gamma^5\right)u_i
-  \frac{g_2}{2\cos\theta_W} Z^\mu \overline{d}_i
\left( g^{d-Z}_V\gamma^\mu - g^{d-Z}_A\gamma^\mu \gamma^5\right)d_i
\nonumber \\
&& - g_{I} {Z'_I}^\mu \overline{u}_i\left( g^{u-Z'_I}_V\gamma^\mu - g^{u-Z'_I}_A
\gamma^\mu \gamma^5\right)u_i
 - g_{I} {Z'_I}^\mu \overline{d}_i\left( g^{d-Z'_I}_V\gamma^\mu -
g^{d-Z'_i}_A\gamma^\mu \gamma^5\right)
d_i \nonumber \\
\eeqa
for the quark interaction.
We recall that the Standard Model result for the neutral currents
\beq
{\cal L}_{Z^0}= -\frac{g_2}{2 \cos\theta_W}{Z^0}_\mu
\overline\psi_f \left(g_V^{f-Z^0}\gamma^\mu - {g_A}^{f-Z^0}\gamma_\mu \gamma_5\right)\psi_f
\eeq
where
\beq
g_V^{f-Z^0}=T_{w3}^{(f)}- 2 \sin^2\theta_W Q_{EM}^{(f)}
\qquad g_A^{f-Z^0}=T_{w3}^{(f)}
\eeq
and $g_2/\cos \theta_W=g$, is now generalized to include corrections of first order in
$v/M_I$.
We obtain for the coupling of the $Z$ gauge boson to the leptons
\beqa
g_V^{\nu-Z} &=&\frac{1}{2}\left( C^{Z}_{\nu_{Li}} + C^{Z}_{\nu_{Ri}}\right)
\nonumber \\
g_A^{\nu-Z} &=& \frac{1}{2}\left( C^{Z}_{\nu_{Li}} - C^{Z}_{\nu_{Ri}}\right)
\nonumber \\
g_V^{e-Z} &=& \frac{1}{2}\left( C^{Z}_{e_{Li}} + C^{Z}_{e_{Ri}}\right)
\nonumber \\
g_A^{e-Z} &=&  \frac{1}{2}\left( C^{Z}_{e_{Li}} - C^{Z}_{e_{Ri}}\right)
\nonumber \\
\eeqa
while the $Z'_I$ couple as
\beqa
g_V^{\nu-Z'_I} &=&  \frac{1}{2}\left( C^{Z'_I}_{\nu_{Li}} +
C^{Z'_I}_{\nu_{Ri}}\right)\nonumber \\
g_A^{\nu-Z'_I} &=& \frac{1}{2}\left( C^{Z'_I}_{\nu_{Li}} -
C^{Z'_I}_{\nu_{Ri}}\right)
\nonumber \\
 g_V^{e-Z'_I} &=&  \frac{1}{2}\left( C^{Z'_I}_{e_{Li}} +
C^{Z'_I}_{e_{Ri}}\right)
\nonumber \\
g_A^{e-Z'_I} &=&
 \frac{1}{2}\left( C^{Z'_I}_{e_{Li}} -
C^{Z'_I}_{e_{Ri}}\right).
\nonumber \\
\eeqa
For the coupling of the quarks to the $Z$ boson we obtain
\beqa
g_V^{u-Z} &=&  \frac{1}{2}\left( C^{Z}_{u_{Li}} +
C^{Z}_{u_{Ri}}\right)
\nonumber \\
g_A^{u-Z} &=&  \frac{1}{2}\left( C^{Z}_{u_{Li}} -
C^{Z}_{u_{Ri}}\right)
\nonumber \\
g_V^{d-Z} &=&  \frac{1}{2}\left( C^{Z}_{d_{Li}} +
C^{Z}_{d_{Ri}}\right)
\nonumber \\
g_A^{d-Z} &=& \frac{1}{2}\left( C^{Z}_{d_{Li}} -
C^{Z}_{d_{Ri}}\right)
\nonumber \\
\eeqa
and the corresponding coupling of the $Z'_I$ are
\beqa
g_V^{u-Z'_I} &=&  \frac{1}{2}\left( C^{Z}_{u_{Li}} +
C^{Z}_{u_{Ri}}\right)
\nonumber \\
 g_A^{u-Z'_I} &=&  \frac{1}{2}\left( C^{Z'_I}_{u_{Li}} +
C^{Z'_I}_{u_{Ri}}\right)
\nonumber \\
g_V^{d-Z'_I} &=&  \frac{1}{2}\left( C^{Z'_I}_{d_{Li}} +
C^{Z'_I}_{d_{Ri}}\right)
\nonumber \\
g_A^{d-Z'_I} &=& \frac{1}{2}\left( C^{Z'_I}_{d_{Li}} +
C^{Z'_I}_{d_{Ri}}\right).
\nonumber \\
\eeqa
%

%
\subsection{$Z$ and $Z^\prime _I$ decays into fermions}
%

In the case of the decay of the neutral gauge boson $Z$
into leptons we obtain in the massless limit
\beq
\Gamma\left(Z\to l \overline{l}\right) =
\frac{g_2^2 m_Z}{48 \pi \cos^2\theta_W} \left( (g_A^{l-Z})^2 + (g_V^{l-Z})^2 \right)
\label{gammazl}
\eeq
and for its decay into massive $(m_q)$ quarks
\beq
\Gamma\left(Z\to q \overline{q}\right) =\frac{g_2^2 m_Z}{48 \pi \cos^2\theta_W}
 \left( (g_A^{q-Z})^2 + (g_V^{q-Z})^2 +2 \frac{m_q^2}{m_Z^2}
\left((g_A^{q-Z'_I})^2 -2 (g_V^{q-Z'_I})^2\right)\right)
\left( 1 - 4 \frac{m_q^2}{m_Z^2}\right)^{1/2}.
\label{gammazq}
\eeq
Similar results hold for the $Z'_I$ gauge boson
\beq
\Gamma\left(Z'_I\to l \overline{l}\right) =\frac{g_{PQ}^2 m_{Z'_I}}{12 \pi}
\left( (g_A^{l-Z'_I})^2 + (g_V^{l-Z'_I})^2 \right)
\label{gammazprimel}
\eeq
and for its decay into massive $(m_q)$ quarks
\beq
\Gamma\left(Z'_I\to q \overline{q}\right) =
\frac{g_{PQ}^2 m_{Z'_I}}{12 \pi} \left( (g_A^{q-Z'_I})^2 + (g_V^{q-Z'_I})^2 +2 \frac{m_q^2}
{m_{Z'_I}^2} \left((g_A^{q-Z'_I})^2 -2 (g_V^{q-Z'_I})^2\right)\right)
\left( 1 - 4 \frac{m_q^2}{m_{Z'_I}^2}\right)^{1/2}.
\label{gammazprimeq}
\eeq
%

%
\subsection{The Drell-Yan cross section}
%

In $e^+ e^-$ annihilations and in $p \,p$ collisions there are some standard signatures
for the new gauge interactions which can be tested against the Standard Model
background, as we are going to discuss below. For simplicity
we consider the case where only one of the $Z'_I$ gauge bosons is
relevant in the spectrum, being the heavier contributions suppressed.

The contribution of the 3 diagrams in the computation of the Drell-Yan cross section
near the resonances is summarized here.
There are three diagrams
containing an s-channel gauge boson, respectively the photon $\gamma$, the gauge boson $Z$
and the gauge boson $Z'$ - the corresponding squared amplitudes
are denoted by $A_\gamma$, $A_Z$ and $A_{ Z'}$ respectively - plus
there are their interferences $(A_{\gamma-Z},A_{\gamma-Z'},A_{Z-Z'})$.
At amplitude level we have
\beq
M=\frac{J^q_\gamma J^f_\gamma}{s} + \frac{J^q_Z J^f_Z}{s + m_Z^2 - i {Im} \Pi^{1 loop}_{ZZ}(s)}
+ \frac{J^q_{Z'} J^f_{Z'}}{s + m_{Z'}^2 - i {Im} \Pi^{1 loop}_{Z'Z'}(s)}
\eeq
where $f$ here denotes a generation of leptons
and we approximate the width with the imaginary part of the 1 loop self-energies $\Pi(s)$
defined by
\beq
m_{Z} \Gamma_Z=Im \Pi^{1 loop}_{ZZ}(s=m_Z^2)=\sum_q \Gamma_Z(Z\to q \bar{q})\times
3\left( 1 +\frac{\alpha_s(m_Z^2)}{\pi}\right) + \sum_l \Gamma_Z(Z\to l^+ l^-)
\label{self}
\eeq
and with a similar expression for the $Z'$. The decay rates appearing in (\ref{self}) are those
computed in (\ref{gammazq})and (\ref{gammazl}) or
(\ref{gammazprimeq})and (\ref{gammazprimel}) for the $Z'$ case. We have included a correction
factor $3(1 + \alpha_s/\pi)$ in the contribution of the quarks.

The position of the two massive resonances is sensitive to the
ratio $v/M$ and to the other parameters of the
theory, most notably $g_{PQ}$. To simplify the notation
here we denote by $e,$ $g_Z$ and $g_Z'$ the three coupling
constants of the photon, the $Z$ and the $Z'$ gauge bosons.
We let $p_1$ and $p_2$ be the momenta of the
$q \bar{q}$ pair, while $k_1$ and $k_2$ are those of the
final state fermions (leptons).  $P_1$ and $P_2$ are the momenta
of the two incoming protons which in the collinear limit are
expressed in terms of the two Bj\"orken variables $x_1$ and $x_2$:
 $p_1=x_1 P_1$ and $p_2=x_2 P_2$.
We also define $s=x_1 x_2 S$ to be the total energy of the initial
partons. At parton level we define $t_1=(p_1 - k_1)^2$ and
$u_1= (p_2 - k_1)^2$, with $s + t_1 + u_1=0$. The partonic
contributions to the cross section are given by
\beqa
\hat{\sigma}&=&\frac{1}{8 N_c s \pi^2} \int d^4 k \delta_+(k^2 )
\delta_+( s + t_1 + u_1 ) |M|^2 \nonumber \\
&=& \frac{1}{32 \pi s N_c}\int d\cos\theta |M|^2 \nonumber \\
&=& \frac{1}{16 \pi s^2 N_c}\int d\, t_1 |M|^2
\eeqa
A factor $1/N_c$ has been introduced for color average and with
$|M|^2$ being the partonic matrix element given by
\beq
|M|^2= A_\gamma + A_Z + A_{Z'}a + A_{\gamma-Z} +A_{\gamma-Z'} + A_{Z-Z'}.
\eeq
The diagram with the photon exchange gives
\beq
|A_\gamma|^2 = 8 \frac{e^4}{s}\left( t_1^2 + u_1^2\right)
\eeq
while the $Z$ gives
\beqa
|A_Z|^2 &=& 8 g_Z^4  \left({D_0(s)}^2+{D_1(s)}^2\right) \left(\left(({g_A^{q-Z}})^2
+({g_V^{q-Z}})^2\right) \left(s^2+2 {t_1} s+2 {t_1}^2\right) {(g_A^{f-Z}})^2 \right. \nonumber \\
&& \left. 4 {g_A^{q-Z}} {g_V^{f-Z}} {g_V^{q-Z}} s (s+2 {t_1}) {g_A^{f-Z}}+{(g_V^{f-Z}})^2
   \left(({g_A^{q-Z}})^2+({g_V^{q-Z}})^2\right) \left(s^2+2 {t_1} s+2 {t_1}^2\right)\right)\nonumber \\
\eeqa
where we have introduced the Breit-Wigner propagator
$(D_0(s) + i D_1(s) )D_Z^{\mu \nu}$
with
\beqa
D_0(s) &=&\frac{s - m_Z^2}{(s - m_Z^2)^2 + m_Z^2 \Gamma_Z^2} \nonumber \\
D_1(s) &=&\frac{m_Z \Gamma_Z}{(s - m_Z^2)^2 + m_Z^2 \Gamma_Z^2} \nonumber \\
D_Z^{\mu \nu}&=& - g^{\mu\nu} + \frac{q^\mu q^\nu}{m_Z^2}.
\label{breit}
\eeqa
The expansion of the propagator for the $Z'$ gauge boson is similar.
We use the notation $D_0'$ and $D_1'$ with
$m_Z\to m_Z'$ and $\Gamma_Z\to \Gamma_{Z'}$ in (\ref{breit}).
We obtain the interferences
\beq
A_{\gamma-Z}=16 e^2 g_Z^2 \frac{D_0}{s}\left[ g_A^{f-Z} g_A^{q-Z} s
(s + 2 t_1) + g_V^{f-Z} g_V^{q-Z} (s^2 + 2 t_1 s
+ 2 t_1^2)\right]
\eeq
\beq
A_{\gamma-Z'}=16 e^2 g_Z^2 \frac{D_0'}{s}\left[ g_A^{f-Z'} g_A^{q-Z'} s
(s + 2 t_1) + g_V^{f-Z'} g_V^{q-Z'} (s^2 + 2 t_1 s
+ 2 t_1^2)\right],
\eeq
while the $Z-Z'$ interference is
\beq
A_{Z-Z'}= 16 g_Z^2 g_{Z'}^2(D_0 D_0' + D_1 D_1')( v_1 s^2 + v_2 t_1^2 + v_3 s t_1),
\eeq
where
\beqa
v_1 &=&g_A^{q-Z} g_A^{q-Z'} g_V^{f-Z} g_V^{f-Z'} + g_V^{f-Z} g_V^{q-Z} g_V^{q-Z'} g_V^{f-Z'} +
  g_A^{f-Z} (g_A^{q-Z'} g_V^{q-Z} + g_A^{q-Z} g_V^{q-Z'}) g_V^{f-Z'} \nonumber \\
&& + g_A^{f-Z'} g_A^{q-Z'} g_V^{f-Z} g_V^{q-Z} + g_A^{f-Z'} g_A^{q-Z} \
g_V^{f-Z} g_V^{q-Z'} + g_A^{f-Z} g_A^{f-Z'} (g_A^{q-Z} g_A^{q-Z'} + g_V^{q-Z} g_V^{q-Z'})
\nonumber \\
v_2 &=& 2 (g_A^{f-Z} g_A^{f-Z'} + g_V^{f-Z} g_V^{f-Z'})
(g_A^{q-Z} g_A^{q-Z'} + g_V^{q-Z} g_V^{q-Z'}) \nonumber \\
v_3 &=&  2 \left[g_A^{f-Z} (g_A^{f-Z'} g_A^{q-Z} g_A^{q-Z'} +
g_V^{f-Z'} g_V^{q-Z} g_A^{q-Z'} + g_A^{q-Z}g_V^{f-Z'} g_V^{q-Z'}
+ g_A^{f-Z'} +g_V^{q-Z} g_V^{q-Z'})\right. \nonumber \\
&& \left. g_V^{f-Z} (
            g_A^{q-Z} g_A^{q-Z'} g_V^{f-Z'} + g_V^{q-Z} g_V^{q-Z'} g_V^{f-Z'} +
            g_A^{f-Z'} g_A^{q-Z'} g_V^{q-Z} + g_A^{f-Z'} g_A^{q-Z} g_V^{q-Z'})\right]. \nonumber \\
\eeqa
Given the generality of the charges of the model and the presence of additional parameters
such as the masses of the heavy resonances it is not possible, at this stage, to provide a
specific estimate of the leading order process.

%
\subsection{Properties of the axi-Higgs}
%

Let us now discuss the properties of the physical field
that appears in the CP-odd scalar sector.
We  call this field $\chi$ the axi-Higgs since
it is a gauge invariant mixture of the bulk axions and the Higgs phase.
In a unitary gauge it is proportional to
$a_I$ which have axionic  couplings.
As its D-brane basis cousins $a^I$, it appears in the
Lagrangian through a dimension five operator.
We have already computed its coupling
in the broken phase to the gauge bosons.
We can also compute the coupling of $\chi$ to the fermions.

The Yukawa couplings provide
mass terms for the fermions as well as cubic
Higgs-fermion-fermion interactions and axion-fermion-fermion
interactions. All these can be extracted from
\bea
{\cal L}_{\rm Yuk}^{unit.} =
&-&  H_u^0 ({\bf u}_{Li}^t)^{\bf \dagger} \s^2 \G^u_{ii}{\bf u}_{Ri} +
H_u^+ ({\bf d}_{Li}^t)^{\bf \dagger} \s^2 \G^u_{ii}{\bf u}_{Ri}
+ {H_d^+}^* ({\bf u}_{Li}^t)^{\bf \dagger} \s^2 \G^d_{ii}{\bf d}_{Ri} +
{H_d^0}^* ({\bf d}_{Li}^t)^{\bf \dagger} \s^2 \G^d_{ii}{\bf d}_{Ri}\nonumber\\
&+& {H_u^+}^* {\n}_{Li}^t \s^2 \G^e_{ii}e_{Ri} +
{H_u^0}^* {e}_{Li}^t \s^2 \G^e_{ii}e_{Ri}
- {H_d^0} {\n}_{Li}^t \s^2 \G^{\n}_{ii}{\n}_{Ri} +
{H_d^+} {e}_{Li}^t \s^2 \G^{\n}_{ii}{\n}_{Ri}+ c.c.
\nonumber\\\label{Yukun}
\eea
with
\bea
H_u^0 &=& v_u + \frac{1}{\sqrt{2}}(h^0\; \sin{\a}  - H^0\; \cos{\a})  +
i \, O^{\chi}_{11} \;\chi\nonumber\\
H_u^+ &=& -\frac{1}{\sqrt{2}}H^+ \cos{\b} \nonumber\\
H_d^0 &=& v_d +  \frac{1}{\sqrt{2}}(h^0\; \cos{\a}  + H^0\; \sin{\a})  +
i \, O^{\chi}_{21}\;\chi\nonumber\\
H_d^+ &=& \frac{1}{\sqrt{2}}H^+ \sin{\b} \label{Higgsdec}
\eea
the unitary gauge expression for the Higgs fields.
In eq. (\ref{Yukun}) the bold notation and the dagger on the quarks
reflects their non-trivial $SU(3)$ transformation property.
The Higgs field expanded around its vacuum expectation value is
to lowest order
\bea
&& H_u^0 = (v_u + \cdots )\; e^{i \frac{N\cos{\b}}{v_u + \cdots }\chi}
\;\simeq\; v_u + i N \cos{\b} \chi\nonumber\\
&& H_d^0 = (v_d + \cdots )\; e^{i \frac{N\sin{\b}}{v_d + \cdots }\chi}
\;\simeq\; v_d + i N \sin{\b} \chi
\eea
where the dots stand for the contribution of the (small) fluctuations
of $h^0$ and $H^0$ and which are negligible for this discussion.
Defining
\bea
&& m_{ui} = - v_u \G_{ii}^{u},\hskip 1cm m_{ei} = - v_u \G_{ii}^{e}\nonumber\\
&& m_{di} = - v_d \G_{ii}^{d},\hskip 1cm m_{\n i} = - v_d \G_{ii}^{\n}
\eea
we can write the parts of the effective action that the axion appears
(suppressing the spinor contraction) as
\bea
&& m_{ui}\; {\bf u}_{Li}^{\bf \dagger} {\bf u}_{Ri}\; e^{i \frac{N\cos{\b}}{v_u }\chi} +
m_{di}\; {\bf d}_{Li}^{\bf \dagger} {\bf d}_{Ri}\; e^{i \frac{N\sin{\b}}{v_d }\chi} +
m_{ei}\; {e}_{Li} {e}_{Ri}\; e^{i \frac{N\cos{\b}}{v_u }\chi} +
m_{\n i}\; {\n}_{Li} {\n}_{Ri}\; e^{i \frac{N\sin{\b}}{v_d }\chi}+c.c. \nonumber\\
&& \partial_\m \chi\; \partial^\m \chi + g^{\chi gg} \, \chi\, tr\, \{G\wedge {G}\} +
g^{\chi +-} \, \chi\, tr\, \{W^+\wedge {W^-}\}+
g^{\chi}_{{\overline p}{\overline q}} \, \chi\, F^{\overline p}\wedge F^{\overline q}.
\eea
{}From the above equations one can see that the couplings of
the Higgs fields to the fermions in the Yukawa
sector will induce an axion-fermion-fermion
coupling
\be
-i\G_{ii}^{u,e} O^{\chi}_{11} =  i\G_{ii}^{u,e} N {\cos{\b}}\label{aFFc1}
\ee
to the up quark and ¨electron¨ sector respectively and
\be
i\G_{ii}^{d,\n} O^{\chi}_{21} = i\G_{ii}^{d,\n} N {\sin{\b}}\label{aFFc2}
\ee
to the down quark and right handed neutrino sector respectively.
As expected, by doing a chiral rotation on the quarks one can make the
$\theta$--parameter of QCD vanish.

The decay rate of $\chi$ into two gauge bosons $A_1$ and $A_2$ of mass
$m_1$ and $m_2$ is given by
\bea
\Gamma(\chi \longrightarrow A_1A_2) =
\frac{1}{16\p m_{\chi}} \Phi^{1/2} \left[1,
\frac{m_1^2}{m_{\chi}^2},\frac{m_2^2}{m_{\chi}^2}\right]
< \mid {\cal A}\mid ^2 >,
\eea
where
\bea
\Phi(x,y,z) = x^2+y^2+z^2-2xy-2xz-2yz
\eea
and the part involving the amplitude can be computed to be
\bea
\mid {\cal A}\mid ^2 = - (g^{\chi A_1A_2})^2
\left[m_1^2m_2^2-\frac{1}{2}\left(m_{\chi}^2 - m_1^2 - m_2^2 \right)^2 \right].
\eea
In our case, the gauge bosons $A_{1,2}$ can be two gluons, a $W^\pm$ pair or
any of a photon, a $Z$ and a $Z^\prime _I$.
Clearly, in the electroweak channel, the decay that dominates is the one into two photons.
Including a factor of 1/2 when averaging over the final state,
we obtain
\bea
\Gamma(\chi \longrightarrow \g\g) = \frac{(g^{\chi \g\g})^2 m_{\chi}^3}{64 \pi}
\eea
with $g^{\chi \g\g}$ given in eq. (\ref{GSEW}).
This decay rate is to be compared with the rate of the
1 loop decay $h^0\longrightarrow \g\g$
(the only channel for a scalar decaying into two photons available in the SM) which is
\bea
\Gamma(h^0 \longrightarrow \g\g) \sim \frac{e^4 \sin^2{\a}\,
m_{h^0}^3}{1024\, \pi^5\, m_{W}^2},
\eea
with $m_{h^0}$ given in eq. (\ref{SMHiggs}) and $\sin{\a}$ in eq. (\ref{alpha}).
In low scale models these two rates could be comparable in magnitude
or even the axi-Higgs decay could be dominating.
When $\chi$ is off shell, the axi-Higgs-photon-photon vertex gives a tree level
contribution to the $p {\overline p}\longrightarrow \g\g$ cross section.

The state $\chi$ is peculiar since it is neither a typical
PQ-type of axion \cite{PQ} nor a typical Higgs field.
It is something in between. It inherits properties from both
precisely because it is a linear combination of the original Higgs
and axion fields (for a similar situation see \cite{Rubakov}).
We can summarize then by saying that the mLSOM axion
is massive with mass $m_{\chi}$ generated
by the $V_{\ppq}$ part of the scalar potential
(given in eq. (\ref{axionmass}))
\bea
m_{\chi}^2 =
-\frac{1}{2}\left[1+\sum_I \left( \frac{q_u^I-q_d^I}{2}
\frac{v}{M_I}\sin{2\b}\right)^2 \right]
\left[\frac{ 4b}{v^2\sin{2\b}} +4\l_1 + \l_2 \tan{\b} + \l_3 \cot{\b}\right]\; v^2.
\label{axionmass2}\nonumber\\
\eea
Strictly speaking to this one should add the usual mass that is generated
non-perturbatively. This is a small contribution to the mass, proportional to
the coupling of the axion to the gauge bosons.
For the PQ axion
this is the only mass generating source which implies that
if its coupling to the gauge bosons is suppressed then its mass is
automatically tiny. This (strong) correlation between the (small) mass and the
coupling of the PQ axion results in computable cosmological and
astrophysical effects that put severe bounds on models with such axions
\cite{ASik}.
Here, as can be seen from eq. (\ref{axionmass2}) the mLSOM axion
acquires from spontaneous symmetry breaking a rather large mass
which is generically expected to be of the order of
${\cal O}(100\; GeV)$ since it is proportional to $v$.
This property is one inherited by its Higgs nature.
On the other hand, its coupling to the gauge bosons is given by
eqs. (\ref{GSEW}) which show that it is suppressed by
the explicit factor of $1/M_{str}$ contained in $D$, $F$ and $C$
and further suppressed by the factor $v/M_I$ contained in ${\bf \Theta}_I$
defined in eq. (\ref{axcoupl1}).
This property is a remnant of its axion nature.
Evidently the mass is essentially not related
to the gauge boson couplings, i.e. a suppressed coupling
does not imply a tiny mass as in typical axion extensions of the SM.

In the fermion sector the situation is slightly different.
The PQ axion has a coupling to the fermions proportional to
its coupling to the gauge bosons and therefore it is equally
suppressed. The mLSOM axion on the other hand
from eqs. (\ref{aFFc1}) and (\ref{aFFc2})
is seen to have an ${\cal O}(1)$ coupling to the fermions.
Some relative suppression in the latter
due to $\b$ (by one or two orders of magnitude)
is perhaps still allowed.

%
\section{Conclusions\label{t5}}
\setcounter{equation}{0}
%
We have presented the mLSOM effective field theory describing universal features of
orientifold string vacua with a low string scale, in the TeV range.
The basic features of such vacua have been described in \cite{AKT,AKRT}.

Although, the associated string theory has numerous different types of particles,
we have kept here, for simplicity, the ones that are generically the lightest, namely the
two Higgs and three extra anomalous U(1) gauge bosons.

The theory has a gauge group $U(3)\times U(2)\times U(1)\times U(1)'$
generated by appropriate stacks of D-branes in the string theory.
The U(1)' in particular comes from a brane wrapping the two large dimensions, and
it has therefore a tiny gauge coupling of almost gravitational strength.

The hypercharge is a specific linear combination of the U(1)$_3$, U(1)$_2$ and U(1) factors.
It is massless and anomaly free.
The other three U(1)'s can be identified with a gauge version of well known symmetries: baryon number,
lepton number and a Peccei-Quinn-like symmetry.
The extra U(1)'s  have triangle anomalies, that as usual
 in string theory, they are cancelled by generalizations of the Green-Schwarz mechanism. As a result
 the three U(1) gauge symmetries are broken, and the associated gauge bosons have (UV) masses
 that can be computed in string theory.

The theory has a Higgs sector that MSSM-like. The Higgses are charged under the anomalous U(1).
When the Higgses acquire vevs and break the electroweak symmetry:

(a) There are additional sources of mass for the anomalous U(1) gauge bosons

(b) They mix with the $Z^0$ with strength of order $M_Z^2/M_s^2$.

Having seen the ingredients of mLSOM we may reappraise the parametric
freedom of the effective field theory.
We do not include the SM parameters in our counting.

In the U(1) sector, we have a priori four coupling constants, one for each U(1).
However, in the case of the U(3) and U(2) groups, it is related to the associated non-abelian
SM coupling $\alpha_{2,3}$ as \cite{AKT}
\be
\alpha_{N}\equiv {g_{N}^2\over 4\pi}={g^2\over 8\pi N}
\ee
where $g$ is the associated U(1) coupling, normalized so that all U(1) brane charges are integers.
The three-coupling constants corresponding to the U(3), U(2) and U(1) branes, are
therefore fixed from the measured SM coupling constants. The coupling constant of the U(1)' is a free
parameter. Since the U(1)' brane wraps the two large dimensions,
$g'\sim 10^{-7}-10^{-8}$.

Further, the UV mass matrix of the U(1) gauge bosons, can be parameterized
by three mass eigenvalues and three mixing angles.

We have kept the Higgs sector of the mLSOM general.
There are  4 independent quadratic couplings
\bea
\m_u,\;\; \m_d, \;\; b, \;\;  \l_1
\eea
and the 6 independent quartic couplings
\bea
\l_{uu},\;\; \l_{dd},\;\; \l_{ud},\;\; \l_{ud}^\prime,\;\; \l_2,\;\; \l_3.
\eea
Compared to the MSSM Higgs sector
whose quadratic part is parameterized by
the 4 parameters $\m$, $m_{H_u}$, $m_{H_d}$ and $B$
and the quartic part that is parameterized
by the 2 gauge couplings $g_Y$ and $g_2$ we
have  6 additional parameters.
However , in a class of vacua that are non-supersymmetric orbifolds
the tree-level potential is that of MSSM.

There are many interesting issues that we have not addressed
here and that are left for future work.

(i) There are important constraints of parameters coming
from couplings of the $Z^0$ to fermions, limits on the Higgs sectors
as well as other astrophysical and cosmological bounds.

(ii) There are important production cross sections that may be relevant in LHC.
We should mention, $pp\to \gamma,Z^0 \to Z^0\gamma$, $pp\to Z^0 \to \gamma\gamma$,
$pp\to Z' \to Z^0Z^0, \gamma\gamma, \gamma Z^0$ etc.

Another issue is that there are particles that eventually should be included in the effective field theory
as they may  relevant for physics.

$\bullet$  Superpartners. They can be straightforwardly included. The theory will be essentially the MSSM with
three extra U(1) anomalous gauge multiplets. It is interesting that unlike the MSSM
no R parity needs to be imposed as
baryon and lepton number will remain as global symmetries.

$\bullet$
 Right-handed neutrinos in mLSOM originate on the U(1)' brane that wraps the two large dimensions.
This is the reason that the associated masses are small and have the right magnitude \cite{AKRT}.
One should also include their KK states. They are important in the case where there is a single flavor of
right-handed neutrino, since their mixing generated the requisite structure of the neutrino sector,
albeit marginally \cite{AKRT}.
Even in the case where there are three flavors of bulk neutrinos, the mixing with
KK states may have interesting effects.

$\bullet$   The KK states of the U(1)' gauge boson as well as
the graviton are both densely populated, as they are sensitive to the large
dimensions. They may be responsible for ADD-like signals.

Such issues and their experimental implications need to be addressed in the future.

\bigskip

\centerline{\bf Acknowledgements}
N. I and C. C. would like to thank A. Denner, D. Zeppenfeld and T. Tomaras for discussions.
E. K. would like to thank P. Anastasopoulos, I. Antoniadis, M. Bianchi, E. Dudas
and B. Schellekens for discussions.
N.I. would like to thank the Ecole Polytechnique, INFN of Lecce and the
University of Lecce for hospitality.
This work was partially supported by the INTAS grant, 03-51-6346,
RTN contracts MRTN-CT-2004-005104 and MRTN-CT-2004-503369,
CNRS PICS 2530 and 3059 and by a European Excellence Grant,
MEXT-CT-2003-509661.
The work of C.C. is partially supported by INFN of Italy
(grant BA21). He thanks the Crete Theory Group
for hospitality.

\newpage

\appendix

\vskip 10mm
 \renewcommand{\theequation}{\thesection.\arabic{equation}}
\centerline{\Large\bf Appendices}
\addcontentsline{toc}{section}{Appendices}
%
\section{Comparison with the MSSM Higgs sector\label{A}}
\setcounter{equation}{0}
%

In the MSSM there are two Higgs doublets $H_u$ and $H_d$
\be
H_u=\pmatrix{H_u^+ \cr H_u^0},
\hskip 1cm
H_d=\pmatrix{H_d^0 \cr H_d^-}\label{MSSMHiggs}
\ee
with the same hypercharge, accounting for 8 degrees of freedom.
There are 3 broken generators as in the Standard Model and therefore we expect to
find 3 NG-bosons and 5 physical Higgs states.
The potential of the MSSM reads
\bea
&& V^{MSSM}(H_u,H_d) = iB (H_u^T\tau^2H_d) + c.c. +
\m_1^2 H_u^\dagger H_u + \mu_2^2 H_d^\dagger H_d\nonumber\\
&& +\frac{1}{8} g_1^2(H_u^\dagger H_u-H_d^\dagger H_d)^2+
\frac{1}{8} g_2^2(H_u^\dagger  \tau^a H_u+H_d^\dagger  \tau^a H_d)^2\label{MSSMpot}
\eea
where the quadratic terms contain F-term as well as soft supersymmetry breaking term
contributions
\bea
\m_1^2 = |\m|^2 + m_{H_u}^2,\hskip .5cm \m_2^2 = |\m|^2 + m_{H_d}^2,
\eea
while the quartic terms represent D-term contributions.
The dimensionfull parameter $B$ is real.
The vacuum that does not break electromagnetism and minimizes the potential is
\bea
H_u=\pmatrix{0 \cr v_u},
\hskip 1cm
H_d=\pmatrix{v_d \cr 0}.
\eea
Expanding around the vacuum we find indeed 3 massless NG-bosons,
a neutral CP odd mass eigenstate $A^0$ with mass
\bea
m_{A^0}^2 = \frac{2B}{\sin{2\b}},
\eea
a pair of charged Higgs eigenstates with mass
\bea
m_{H^{\pm}}^2 = m_{A^0}^2 + m_W^2
\eea
and two neutral CP-even mass eigenstates $h^0$ and $H^0$ with masses
\bea
m_{h^0}^2 = \frac{1}{2}\left[(m_{A^0}^2+
m_{Z^0}^2)-\sqrt{(m_{A^0}^2+m_{Z^0}^2)^2-(2m_{Z^0}m_{A^0}\cos{2\b})^2}\right]\label{lightH}
\eea
and
\bea
m_{H^0}^2 = \frac{1}{2}\left[(m_{A^0}^2+m_{Z^0}^2)+\sqrt{(m_{A^0}^2+
m_{Z^0}^2)^2-(2m_{Z^0}m_{A^0}\cos{2\b})^2}\right].
\eea
Expanding eq. (\ref{lightH}) in inverse powers of $m_{A^0}$ one
concludes that the $h^0$ Higgs boson mass is smaller than the mass
of $Z^0$ and therefore radiative effects have to be taken into account
to avoid conflict with experiment. This is indeed possible due to
the large Yukawa coupling of the top quark.
It is instructive to make a comparison of our potential eq. (\ref{PQ}) with the
potential eq. (\ref{MSSMpot}) since the structures are quite similar.
For the comparison, the following identities are proven to be useful:
\bea
&& (H_a^\dagger \tau^j H_a)^2 = (H_a^\dagger H_a)^2, \hskip .5cm a=u,d  \nonumber\\
&& \left|H_u^\dagger H_d \right|^2 =
(H_u^\dagger H_u)(H_d^\dagger H_d)
-\left|H_u^T \tau^2 H_d\right|^2 \nonumber\\
&& (H_u^\dagger \tau^j H_u)(H_d^\dagger \tau^j H_d) =
(H_u^\dagger H_u)(H_d^\dagger H_d)
-2\left|H_u^T \tau^2 H_d\right|^2 .\label{SU2id}
\eea
In order to translate to our convention where
the Higgs doublets have the same hypercharge, one has to make the
transformations
\bea
H_u \longrightarrow H_u \hskip 1cm H_d \longrightarrow -i\tau^2H_d^*.
\eea
Using the identities eq. (\ref{SU2id}) and
the above transformation, the MSSM potential can be brought in the form
\bea
&& V^{MSSM}_D(H_u,H_d) = B H_u^\dagger H_d + c.c.
+ \m_1^2H_u^\dagger H_u + \m_2 ^2 H_d^\dagger H_d\nonumber\\
&& +\frac{1}{8}(g_1^2+g_2^2)(H_u^\dagger H_u)^2+\frac{1}{8}(g_1^2+g_2^2)(H_d^\dagger H_d)^2
-\frac{1}{4}(g_1^2+g_2^2) (H_u^\dagger H_u)(H_d^\dagger H_d) \nonumber\\
&& +\frac{g_2^2}{2}\left|H_u^T\tau^2H_d\right|^2.\label{MSSMpotD}
\eea
The identifications can be then read off the above potential and eqs. (\ref{PQ})
and (\ref{PQbreak}):
\bea
\m_1^2 \rightarrow \m_u^2, \hskip .5cm  \m_2^2 \rightarrow \m_d^2, \hskip .5cm
\frac{1}{8}(g_1^2+g_2^2) \rightarrow \l_{uu},\; \l_{dd},\; \l_{ud},   \hskip .5cm
\frac{1}{4} g_2^2 \rightarrow \l_{ud}^{\prime},
\label{MSSMtr}
\eea
\bea
B \longrightarrow b.
\eea
The complex term in eq. (\ref{MSSMpotD}) breaks PQ and therefore it does not appear
in eq. (\ref{PQ}).
It appears though in the PQ breaking potential eq. (\ref{PQbreak}) as we have seen.
The rest of the terms in eq. (\ref{PQbreak}) evidently are not present in the MSSM.

The first thing that one can immediately see is why the potential eq. (\ref{PQ})
does not give mass to the axion. The CP-odd Higgs eigenstate $A^0$ of
the MSSM has been traded for the CP-odd axion $\chi$ in our model.
The mass of $A^0$ originates exclusively from the term proportional to $B$
in eq. (\ref{MSSMpot}) and since such a term is not part of eq. (\ref{PQ})
the axion does not get a mass. It is also now clear why the mass of the charged
Higgs states is given by eq. (\ref{chargedH}).
Inspecting eq. (\ref{MSSMtr}) one can see that it corresponds to
the part in the MSSM charged Higgs mass, proportional to the mass of the $W$-boson.
The only difference is that instead of the $SU(2)$ D-term origin of
the coupling $g_2$ in the MSSM charged Higgs mass, in our model
we have an independent coupling $\l^\prime_{ud}$
and therefore the mass is not directly related to the $W$-boson mass.
When the potential eq. (\ref{PQbreak}) is added, then the charged Higgs mass
acquires a part proportional to $b$ which corresponds to the part
proportional to $B$ in the MSSM.
The axion on the other hand, in the presence of eq. (\ref{PQbreak}) acquires
a mass proportional to $b$ (just as  $A^0$ in the MSSM acquires a mass from the term
proportional to $B$) and additional contributions proportional
to $\l_{1,2,3}$. These latter contributions are new, not present in the MSSM.
Finally, from eqs. (\ref{neutHmass}) we see that
for the potential eq. (\ref{PQ}), the masses of
the neutral Higgs states are not related to the $Z$-boson mass as it is the case
for the $B=0$ limit of the MSSM because the couplings $\l_{uu}$, $\l_{dd}$ and
$\l_{ud}$ are not related in our model to gauge couplings.
In particular, the light neutral Higgs mass vanishes
and the heavy neutral Higgs mass is proportional to that of the Z-boson
in the $B=0$ limit in the MSSM, contrary to our case.
For the case of the potential eq. (\ref{PQbreak}) the neutral Higgs bosons
masses acquire their MSSM-like contributions plus additional terms proportional to
$\l_{1,2,3}$.

\section{The Lagrangian in the physical basis\label{B}}
\setcounter{equation}{0}

In this appendix we provide the Lagrangian expressed in terms of the
physical base.
The kinetic  Lagrangian of the gauge fields is given by

\beqa
&&{\cal L}_{\it kin\,\, gauge}=\nonumber \\
&& -\frac{1}{4}\left( \partial_\mu A_{\gamma\nu} - \partial_\nu A_{\gamma\mu}\right)^2
-\frac{1}{4}\left( \partial_\mu Z_\nu - \partial_\nu Z_{\mu}\right)^2
-\frac{1}{4}\sum_I\left( \partial_\mu Z_{I\nu}' - \partial_\nu Z_{I\mu}'\right)^2 \nonumber \\
&&-\frac{1}{2}\left( \partial^\mu W^{+\nu} - \partial^\nu W^{+\mu}\right)
\left(\partial_\mu W^-_{\nu} - \partial_\nu W^-_{\mu}\right) \nonumber \\
&& + i\frac{ e}{\sin\theta_W}\left( \sin\theta_W A_{\gamma}^\mu + \cos\theta_W Z^\mu -
\frac{g_2}{2} \epsilon_I {Z_I'}^\mu \right) \left[ W^\nu(\partial_\nu
W^-_\mu  - \partial_\mu W^-_\nu)\right. \nonumber \\
&&\left. - W^-_\nu(\partial_\nu W^+_\mu - \partial_\mu W^+_\nu)\right]
 +\frac{1}{2}\frac{e^2}{\sin^2\theta_W}
\left[ (W^-_\mu W^{-\mu})(W^-_\nu W^{-\nu})- (W^-_\mu W^{-\mu})^2\right]\nonumber \\
&& -\frac{e^2}{\sin^2\theta_W}\left[ \sin\theta_W A_{\gamma\mu}+ \cos\theta_W Z_\mu
-\frac{g_2}{2}\epsilon_I {Z_I'}_\mu\right] \left[ \sin\theta_W A_{\gamma}^{\mu}
+ \cos\theta_W Z_\mu -\frac{g_2}{2}\epsilon_I {Z_I'}_\mu\right] \nonumber \\
&&\times W^{+\rho}W_{-\rho} -
\left[ \sin\theta_W A_{\gamma\mu}+ \cos\theta_W Z_\mu
-\frac{g_2}{2}\epsilon_I {Z_I'}_\mu\right] \nonumber \\
&& \times\left[ \sin\theta_W A_{\gamma}^{\mu}
+ \cos\theta_W Z_\mu -\frac{g_2}{2}\epsilon_I {Z_I'}_\mu \right]W^\mu W^{-\rho};
\eeqa

\beqa
x_I &=& (q_u^I v_u^2 + q_d^I v_d^2)\, g_I \nonumber \\
\epsilon_I &=& \frac{x_I}{M_I^2} \nonumber \\
\eeqa
and
\beqa
Z^\mu &=& \cos\theta_W W^3 -\sin\theta_W A_Y + \xi_I A_I\nonumber \\
&=& Z_0^\mu + \xi_I A_I
\eeqa
and
\be
\xi_I= {g \over {2}}\e_I +{\cal O}\left(M_I^{-4}\right)
\ee

\beq
{\cal L}_{{\it mass\,\, gauge}} = m_Z^2 Z_\mu Z^\mu
+ m_W^2 W^{+ \mu} {W^-}_\mu + m_W^2 W^{- \mu} {W_\mu}^+ +\sum_I m_{Z_I'}^2 {Z_I'}_\mu {Z_I'}^\mu
\eeq

We define
\beqa
\alpha_\mu^{(u)} &=& {1\over 2}(g_2 W_{3 \mu} + {g_Y}A_\mu^Y +
\sum_{I}q^I_u\, g_I \, A^I_{\m})\nonumber \\
\alpha_\mu^{(d)} &=& {1\over 2}(g_2 W_{3 \mu} + {g_Y}A_\mu^Y +
\sum_{I}q^I_d\, g_I \, A^I_{\m}) \nonumber \\
\beta_\mu^{(u)} &=& {1\over 2}(- g_2 W_{3 \mu} + {g_Y}A_\mu^Y +
\sum_{I}q^I_u\, g_I \, A^I_{\m}) \nonumber \\
\beta_\mu^{(d)} &=& {1\over 2}(- g_2 W_{3 \mu} + {g_Y}A_\mu^Y +
\sum_{I}q^I_d\, g_I \, A^I_{\m})
\eeqa
and set $g_2'=g_2/2$.
We have
\beqa
&&\alpha_\mu^{(u,d)}= \nonumber \\
&& \frac{1}{2} ({g_Y} O^A_{A_Y \gamma}+{g_2} O^A_{W_3 \gamma}
+{g_I} O^A_{A_I\gamma} {q_{u,d}^I}) {A_\gamma}_\mu(x) \nonumber \\
&& +\frac{1}{2}({g_Y} O^A_{A_Y Z}+{g_2} O^A_{W_3 Z}+{g_I} O^A_{A_I Z} {q_{u,d}^I}) Z_\mu(x) \nonumber \\
&&+\frac{1}{2} ({g_Y} O^A_{A_Y  Z_I'}+{g_2} O^A_{W_3 Z_I'}+{g_I} O^A_{A_I Z_I'} {q_{u,d}^I}) {Z_I'}_\mu(x)
\eeqa

\beqa
&&\beta_\mu^{(u,d)}= \nonumber \\
&& \frac{1}{2} ({g_Y} O^A_{A_Y \gamma} -{g_2} O^A_{W_3 \gamma}
+{g_I} O^A_{A_I\gamma} {q_{u,d}^I}) {A_\gamma}_\mu(x) \nonumber \\
&& +\frac{1}{2}({g_Y} O^A_{A_Y Z} -{g_2} O^A_{W_3 Z}+{g_I} O^A_{A_I Z} {q_{u,d}^I}) Z_\mu(x) \nonumber \\
&&+\frac{1}{2} ({g_Y} O^A_{A_Y  Z_I'} -{g_2} O^A_{W_3 Z_I'}+{g_I} O^A_{A_I Z_I'} {q_{u,d}^I}) {Z_I'}_\mu(x)
\eeqa

The contribution from the Higgs sector in the quadratic potential is then summarized in the expressions

\beqa
&&{\cal L}_{\it Higgs \,1}= \frac{1}{2} \partial_\mu H^+ \partial^\mu H^- +
\frac{1}{2} \partial_\mu H_0 \partial^\mu H_0 + \frac{1}{2}\partial_\mu h_0 \partial^\mu h_0
\nonumber \\
&&\qquad \qquad + \frac{1}{2} \partial_\mu G^+ \partial^\mu G^-
+\frac{1}{2} \partial_\mu \chi \partial^\mu \chi +\frac{1}{2}\sum_{i=1}^{N_s + 1} \partial_\mu G^0_i \partial^\mu
G^0_i \nonumber \\
&&\qquad \qquad + \frac{1}{2} m_{h_0}^2 + \frac{1}{2} m_{H_0}^2 + \frac{1}{2} m_{\chi}^2 + m_{H^\pm}^2 H^+ H^-
\eeqa

\beqa
{\cal L}_{\it Higgs \,2} &=&\frac{1}{2}\sum_{i=u,d}\left( \alpha_{\mu}^{(i)} \Sigma^{(i)}_{1} +
\beta_{\mu}^{(i)} \Sigma^{(i)}_{2} + {\alpha^{(i)}}^2 \Sigma^{(i)}_{3}+ {\beta^{(i)}}^2 \Sigma^{(i)}_{4} + \Sigma^{(i)}_5\right)
\eeqa

\beqa
{\cal L}_{\it mix}= M_I  O^\chi_{I+2\, i}\partial_\mu G^0_i
\left( O^A_{A_I Z} Z^\mu + O^A_{A_I,Z'_I}{Z'_I}^\mu\right)
\nonumber \\
\eeqa

\beqa
&&\Sigma^{(u)}_{1} = \nonumber \\
&& i H^+ \partial^\mu H^- {\cos^2\beta}-i H^- \partial^\mu H^+ {\cos^2\beta}
+{i {g_2'} O^\chi_{11}
   \chi H^+ W^{- \mu} {\cos\beta}}{} \nonumber \\
&& +{i {g_2'} O^\chi_{1i} G^0_i
   H^+ W^{- \mu} {\cos\beta}}-{{g_2'} {\sin\alpha} h_0 H^+ W^{- \mu}
   {\cos\beta}}{}\nonumber \\
&& +{{\cos\alpha} {g_2'} H_0 H^+ W^{- \mu}
   {\cos\beta}}{} -{i {g_2'} O^\chi_{11} \chi H^- W^{+\mu}
   {\cos\beta}}\nonumber \\
&&-{i {g_2'} O^\chi_{1i} G^0_i H^- W^{+\mu}
   {\cos\beta}}{}-{{g_2'} {\sin\alpha} h_0 H^- W^{+\mu}
   {\cos\beta}}{} \nonumber \\
&& +{{\cos\alpha} {g_2'} H_0 H^- W^{+\mu} {\cos\beta}}{}-i
   {\sin\beta} H^+ \partial^\mu G^- {\cos\beta}
\nonumber \\
&&+i {\sin\beta} H^- \partial^\mu G^+ {\cos\beta}
 -i {\sin\beta} G^+ \partial^\mu H^- {\cos\beta} \nonumber \\
&& +i {\sin\beta} G^- \partial^\mu H^+ {\cos\beta}
 -{i
   {g_2'} O^\chi_{11} {\sin\beta} \chi G^+ W^{- \mu}}{} \nonumber \\
&& -{i {g_2'} O^\chi_{1i}
   {\sin\beta} G^0_i G^+ W^{- \mu}}{}
 +{{g_2'} {\sin\alpha} {\sin\beta} G^+
   h_0 W^{- \mu}}{} \nonumber \\
&& -{{\cos\alpha} {g_2'} {\sin\beta} G^+ H_0
   W^{- \mu}}{}  +{i {g_2'} O^\chi_{11} {\sin\beta} \chi G^-
   W^{+\mu}}{} \nonumber \\
&& +{i {g_2'} O^\chi_{1i} {\sin\beta} G^0_i G^-
   W^{+\mu}}{}
 +{{g_2'} {\sin\alpha} {\sin\beta} G^- h_0
   W^{+\mu}}{} \nonumber \\
&& -{{\cos\alpha} {g_2'} {\sin\beta} G^- H_0 W^{+\mu}}{}
 +i
   {\sin\beta}^2 G^+ \partial^\mu G^- \nonumber \\
&& -i {\sin\beta}^2 G^- \partial^\mu G^+
\eeqa

\beqa
&& \Sigma^{(u)}_{2}= \nonumber \\
&& -{i {g_2'} O^\chi_{11} {\sin\beta} \chi G^+ W^{- \mu}}{}-{i {g_2'}
   O^\chi_{1i} {\sin\beta} G^0_i G^+ W^{- \mu}}{} \nonumber \\
&& +{{g_2'} {\sin\alpha} {\sin\beta}
   G^+ h_0 W^{- \mu}}{} -{{\cos\alpha} {g_2'} {\sin\beta} G^+ H_0
   W^{- \mu}}{} \nonumber \\
&& +{i {\cos\beta} {g_2'} O^\chi_{11} \chi H^+
   W^{- \mu}}{}+{i {\cos\beta} {g_2'} O^\chi_{1i} G^0_i H^+
   W^{- \mu}}{} \nonumber \\
&& -{{\cos\beta} {g_2'} {\sin\alpha} H_0 H^+
   W^{- \mu}}{}+{{\cos\alpha} {\cos\beta} {g_2'} H_0 H^+
   W^{- \mu}}{} \nonumber \\
&& +{i {g_2'} O^\chi_{11} {\sin\beta} \chi G^-W^{+\mu}}{}
 +{i {g_2'} O^\chi_{1i} {\sin\beta} G^0_i G^-
   W^{+\mu}}{} \nonumber \\
&& +{{g_2'} {\sin\alpha} {\sin\beta} G^- h_0
   W^{+\mu}}{} -{{\cos\alpha} {g_2'} {\sin\beta} G^- H_0W^{+\mu}}{}\nonumber \\
&& -{i {\cos\beta} {g_2'} O^\chi_{11} \chi H^-
   W^{+\mu}}{}-{i {\cos\beta} {g_2'} O^\chi_{1i} G^0_i H^-
   W^{+\mu}}{} \nonumber \\
&& -{{\cos\beta} {g_2'} {\sin\alpha} h_0 H^-
   W^{+\mu}}{}+{{\cos\alpha} {\cos\beta} {g_2'} H_0 H^- W^{+\mu}}{}\nonumber \\
&&+2
   O^\chi_{11} {\sin\alpha} h_0 \partial^\mu \chi-2 {\cos\alpha} O^\chi_{11} H_0 \partial^\mu \chi \nonumber \\
&& +2 O^\chi_{1i} {\sin\alpha} h_0 \partial^\mu G^0_i-2 {\cos\alpha} O^\chi_{1i} H_0 \partial^\mu G^0_i \nonumber \\
&& -2
   O^\chi_{11} {\sin\alpha} \chi \partial^\mu h^0
-2 O^\chi_{1i} {\sin\alpha} G^0_i \partial^\mu h^0 \nonumber \\
&& +2{\cos\alpha} O^\chi_{11} \chi \partial^\mu H^0+2 {\cos\alpha} O^\chi_{1i} G^0_i \partial^\mu H^0
\eeqa

\beqa
&& \Sigma^{(u)}_{3}= \nonumber \\
&& H^- H^+ {\cos\beta}^2-{\sin\beta} G^+ H^- {\cos\beta}-{\sin\beta} G^-
   H^+ {\cos\beta} \nonumber \\
&& +{\sin\beta}^2 G^- G^+
\eeqa

\beqa
&& \Sigma^{(u)}_{4}= \nonumber \\
&& {O^\chi_{11}}^2 \chi^2+2 O^\chi_{11} O^\chi_{1i} G^0_i \chi+{O^\chi_{1i}}^2
   {G^0_i}^2+{\sin\alpha}^2 h_0^2 \nonumber \\
&& +{\cos\alpha}^2 H_0^2-2 {\cos\alpha} {\sin\alpha} h_0 H_0
\eeqa

\beqa
&& \Sigma^{(u)}_{5}=  \nonumber  \nonumber \\
&& {g_2'}^2 {H^-} {H^+} {W^{-\mu}} {W^+_\mu} \cos^2\beta \nonumber \\
&& -{g_2'}^2 {\sin\beta} G^+ {H^-} {W^{-\mu}} {W^+_\mu} {\cos\beta}  -{g_2'}^2
   {\sin\beta} G^- {H^+} {W^{-\mu}} {W^+_\mu} {\cos\beta}   \nonumber \\
&& -{g_2'} {O^\chi_{11}} {H^+}
   {W^-} \partial_\mu \chi {\cos\beta} -{g_2'} {O^\chi_{11}} {H^-} {W^{+\mu}} \partial_\mu\chi
   {\cos\beta}  \nonumber \\
&& -{g_2'} {O^\chi_{1i}} {H^+} {W^{-\mu}} \partial_\mu G^0_i {\cos\beta}-{g_2'} {O^\chi_{1i}}
   {H^-} {W^{+\mu}} \partial_\mu G^0_i {\cos\beta}   \nonumber \\
&& -i {g_2'} {\sin\alpha} {H^+} {W^-} \partial_\mu h_0
   {\cos\beta}
+i {g_2'} {\sin\alpha} {H^-} {W^{+\mu}} \partial_\mu h_0 {\cos\beta}  \nonumber \\
&& +i {\cos\alpha} {g_2'}
   {H^+} {W^{-\mu}} \partial_\mu h_0 {\cos\beta}
 -i {\cos\alpha} {g_2'} {H^-} {W^{+\mu}} \partial_\mu h_0
   {\cos\beta}  \nonumber \\
&& +{g_2'} {O^\chi_{11}} \chi {W^{+\mu}} \partial_\mu H^- {\cos\beta}+{g_2'} {O^\chi_{1i}}
   G^0_i {W^{+\mu}} \partial_\mu H^- {\cos\beta}   \nonumber \\
&& -i {g_2'} {\sin\alpha} H_0 {W^{+\mu}} \partial_\mu H^- {\cos\beta}
+i {\cos\alpha} {g_2'} H_0 {W^{+\mu}} \partial_\mu H^- {\cos\beta}  \nonumber \\
&&-{\sin\beta} \partial_\mu G^+
   \partial_\mu H^- {\cos\beta}
 +{g_2'} {O^\chi_{11}} \chi {W^{-\mu}} \partial_\mu H^+ {\cos\beta}  \nonumber \\
&& +{g_2'}{O^\chi_{1i}} G^0_i {W^{-\mu}} \partial_\mu H^+ {\cos\beta}
+i {g_2'} {\sin\alpha} H_0 {W^-}
   \partial_\mu H^+ {\cos\beta}  \nonumber \\
&&-i {\cos\alpha} {g_2'} H_0 {W^{-\mu}} \partial_\mu H^+ {\cos\beta}-{\sin\beta}
   \partial_\mu G^- \partial_\mu H^+ {\cos\beta}   \nonumber \\
&&+{g_2'}^2 {O^\chi_{11}}^2 \chi^2 {W^-}
   {W^+}  \nonumber \\
&&+{g_2'}^2 {O^\chi_{1i}}^2 {G^0_i}^2 {W^{-\mu}} {W^+_\mu}+{g_2'}^2 \sin^2\alpha
   H_0^2 {W^{-\mu}} {W^+_\mu}   \nonumber \\
&&+\cos^2\alpha {g_2'}^2 H_0^2 {W^{-\mu}} {W^+_\mu}+2
   {g_2'}^2 {O^\chi_{11}} {O^\chi_{1i}} \chi G^0_i {W^{-\mu}} {W^+_\mu}  \nonumber \\
&& +{g_2'}^2 \sin^2\beta
   G^- G^+ {W^{-\mu}} {W^+_\mu}-2 {\cos\alpha} {g_2'}^2 {\sin\alpha} H_0 H_0
   {W^{-\mu}} {W^+_\mu}   \nonumber \\
&& +{g_2'} {O^\chi_{11}} {\sin\beta} G^+ {W^-} \partial_\mu \chi+{g_2'}
   {O^\chi_{11}} {\sin\beta} G^- {W^{+\mu}} \partial_\mu\chi  \nonumber \\
&& +{g_2'} {O^\chi_{1i}} {\sin\beta} G^+
   {W^{-\mu}} \partial_\mu G^0_i+{g_2'} {O^\chi_{1i}} {\sin\beta} G^- {W^{+\mu}} \partial_\mu G^0_i  \nonumber \\
&& -{g_2'} {O^\chi_{11}} {\sin\beta} \chi {W^+}
   \partial_\mu G^--{g_2'} {O^\chi_{1i}} {\sin\beta} G^0_i {W^+} \partial_\mu G^-(x )  \nonumber \\
&& +i {g_2'} {\sin\alpha}
   {\sin\beta} H_0 {W^+} \partial_\mu G^-
-i {\cos\alpha} {g_2'} {\sin\beta} H_0 {W^+}
   \partial_\mu G^-  \nonumber \\
&& -{g_2'} {O^\chi_{11}} {\sin\beta} \chi {W^-} \partial_\mu G^+-{g_2'} {O^\chi_{1i}}
   {\sin\beta} G^0_i {W^-} \partial_\mu G^+   \nonumber \\
&& -i {g_2'} {\sin\alpha} {\sin\beta} H_0 {W^-}
   \partial_\mu G^++i {\cos\alpha} {g_2'} {\sin\beta} H_0 {W^-} \partial_\mu G^+  \nonumber \\
&& +i {g_2'} {\sin\alpha} {\sin\beta} G^+ {W^{-\mu}} \partial_\mu h_0  \nonumber \\
&&-i {g_2'}
   {\sin\alpha} {\sin\beta} G^- {W^{+\mu}} \partial_\mu h_0-i {\cos\alpha} {g_2'} {\sin\beta} G^+
   {W^{-\mu}} \partial_\mu h_0   \nonumber \\
&& +i {\cos\alpha} {g_2'} {\sin\beta} G^- {W^{+\mu}} \partial_\mu h_0
\eeqa

\beqa
&& \Sigma^{(d)}_1= \nonumber \\
&& i G^+ \partial_\mu G^- {\cos\beta}^2-i G^- \partial_\mu G^+ {\cos\beta}^2-i {g_2} {O^\chi_{21}} {\chi}
   G^+ {W^-_\mu} {\cos\beta}\nonumber \\
&& -i {g_2} {O^\chi_{2i}} {G^0_i} G^+ {W^-_\mu} {\cos\beta}+{\cos\alpha}{g_2} G^+ {h_0} {W^-_\mu} {\cos\beta}
+{g_2} {\sin\alpha} G^+ {h_0} {W^-_\mu}{\cos\beta} \nonumber \\
&&+i {g_2} {O^\chi_{21}} {\chi} G^- {W^+_\mu} {\cos\beta}+i {g_2} {O^\chi_{2i}} {G^0_i}
   G^- {W^+_\mu} {\cos\beta}+{\cos\alpha} {g_2} G^- {h_0} {W^+_\mu} {\cos\beta}\nonumber \\
&&+{g_2} {\sin\alpha}G^- {h_0} {W^+_\mu} {\cos\beta}+i {\sin\beta} {H^+} \partial_\mu G^- {\cos\beta}-i {\sin\beta} { H^-}\partial_\mu G^+ {\cos\beta} \nonumber \\
&&+i {\sin\beta} G^+ \partial_\mu H^- {\cos\beta}-i {\sin\beta} G^- \partial_\mu H^+
   {\cos\beta}-i {g_2} {O^\chi_{21}} {\sin\beta} {\chi} {H^+} {W^-_\mu}\nonumber \\
&&-i {g_2} {O^\chi_{2i}} {\sin\beta}{G^0_i} {H^+} {W^-_\mu}+{\cos\alpha} {g_2} {\sin\beta} {h_0} {H^+} {W^-_\mu}+{g_2}{\sin\alpha} {\sin\beta} {h_0} {H^+} {W^-_\mu}
+i {g_2} {O^\chi_{21}} {\sin\beta} {\chi} { H^-}{W^+_\mu} \nonumber \\
&& +i {g_2} {O^\chi_{2i}} {\sin\beta} {G^0_i} { H^-} {W^+_\mu}+{\cos\alpha} {g_2} {\sin\beta}{h_0} { H^-} {W^+_\mu} +{g_2} {\sin\alpha} {\sin\beta} {h_0} { H^-} {W^+_\mu} \nonumber \\
&& +i {\sin\beta}^2{H^+} \partial_\mu H^--i {\sin\beta}^2 { H^-} \partial_\mu H^+
\eeqa

\beqa
&& \Sigma^{(d)}_{2}= \nonumber \\
&& i {\cos\beta} {g_2} {O^\chi_{21}} {\chi} G^+ {W^-_\mu}-i {\cos\beta} {g_2} {O^\chi_{2i}} {G^0_i}
   G^+ {W^-_\mu}+{\cos\alpha} {\cos\beta} {g_2} G^+ {h_0} {W^-_\mu} \nonumber \\
&& +{\cos\beta} {g_2} {\sin\alpha}
   G^+ {h_0} {W^-_\mu}-i {g_2} {O^\chi_{21}} {\sin\beta} {\chi} {H^+} {W^-_\mu}-i {g_2}
   {O^\chi_{2i}} {\sin\beta} {G^0_i} {H^+} {W^-_\mu} \nonumber \\
&& +{\cos\alpha} {g_2} {\sin\beta} {h_0} {H^+}
   {W^-_\mu}+{g_2} {\sin\alpha} {\sin\beta} {h_0} {H^+} {W^-_\mu}+i {\cos\beta} {g_2} {O^\chi_{21}}
   {\chi} G^- {W^+_\mu}+i {\cos\beta} {g_2} {O^\chi_{2i}} {G^0_i} G^- {W^+_\mu}\nonumber \\
&& +{\cos\alpha}
   {\cos\beta} {g_2} G^- {h_0} {W^+_\mu}+{\cos\beta} {g_2} {\sin\alpha} G^- {h_0}
   {W^+_\mu}+i {g_2} {O^\chi_{21}} {\sin\beta} {\chi} { H^-} {W^+_\mu} \nonumber \\
&& +i {g_2} {O^\chi_{2i}} {\sin\beta}
   {G^0_i} { H^-} {W^+_\mu}+{\cos\alpha} {g_2} {\sin\beta} {h_0} { H^-} {W^+_\mu}+{g_2}
   {\sin\alpha} {\sin\beta} {h_0} { H^-} {W^+_\mu} \nonumber \\
&& +2 {\cos\alpha} {O^\chi_{21}} {h_0} \partial_\mu{\chi} +2 {O^\chi_{21}}
   {\sin\alpha} {h_0}\partial_\mu{\chi}+2 {\cos\alpha} {O^\chi_{2i}} {h_0} \partial_\mu G^0_i \nonumber \\
&& +2 {O^\chi_{2i}} {\sin\alpha} {h_0}
   \partial_\mu G^0_i-2 {\cos\alpha} {O^\chi_{21}} {\chi} \partial_\mu h_0-2 {\cos\alpha} {O^\chi_{2i}} {G^0_i} \partial_\mu h_0-2{O^\chi_{21}} {\sin\alpha} {\chi} \partial_\mu h_0 \nonumber \\
&& -2 {O^\chi_{2i}} {\sin\alpha} {G^0_i} \partial_\mu h_0
\eeqa

\beqa
&& \Sigma^{(d)}_3= \nonumber \\
&& G^- G^+ {\cos\beta}^2+{\sin\beta} G^+ { H^-} {\cos\beta}+{\sin\beta} G^- {H^+}
   {\cos\beta}+{\sin\beta}^2 { H^-} {H^+}
\eeqa

\beqa
&& \Sigma^{(d)}_4=\nonumber \\
&& {O^\chi_{21}}^2 {\chi}^2+2 {O^\chi_{21}} {O^\chi_{2i}} {G^0_i} {\chi}+{O^\chi_{2i}}^2 {G^0_i}^2
+{\cos\alpha}^2 {h_0}^2+{\sin\alpha}^2 {h_0}^2 \nonumber \\
&& +2 {\cos\alpha} {\sin\alpha} {H_0} {h_0}
\eeqa

\beqa
&& \Sigma^{(d)}_5= \nonumber \\
&& {g_2}^2 G^- G^+ {W^-_\mu} {W^+_\mu} {\cos\beta}^2+{g_2}^2
   {\sin\beta} G^+ { H^-} {W^-_\mu} {W^+_\mu} {\cos\beta}\nonumber \\
&&+{g_2}^2 {\sin\beta} G^- {H^+}
   {W^-_\mu} {W^+_\mu} {\cos\beta}+{g_2} {O^\chi_{21}} G^+ {W^-_\mu}\partial_\mu{\chi} {\cos\beta}+{g_2}
   {O^\chi_{21}} G^- {W^+_\mu} {\chi}' {\cos\beta}\nonumber \\
&& +{g_2} {O^\chi_{2i}} G^+ {W^-_\mu} \partial_\mu G^0_i
   {\cos\beta}+{g_2} {O^\chi_{2i}} G^- {W^+_\mu} \partial_\mu G^0_i {\cos\beta}-{g_2} {O^\chi_{21}} {\chi}
   {W^+_\mu} \partial_\mu G^- {\cos\beta} \nonumber \\
&& -{g_2} {O^\chi_{2i}} {G^0_i} {W^+_\mu} \partial_\mu G^- {\cos\beta}+i {\cos\alpha}
   {g_2} {h_0} {W^+_\mu} \partial_\mu G^- {\cos\beta} \nonumber \\
&&+i {g_2} {\sin\alpha} {h_0} {W^+_\mu} \partial_\mu G^-
   {\cos\beta}-{g_2} {O^\chi_{21}} {\chi} {W^-_\mu} \partial_\mu G^+ {\cos\beta}\nonumber \\
&& -{g_2} {O^\chi_{2i}} {G^0_i}
   {W^-_\mu} \partial_\mu G^+ {\cos\beta}-i {\cos\alpha} {g_2} {h_0} {W^-_\mu} \partial_\mu G^+ {\cos\beta}
-i {g_2}{\sin\alpha} {h_0} {W^-_\mu} \partial_\mu G^+ {\cos\beta} \nonumber \\
&& +i {\cos\alpha} {g_2} G^+ {W^-_\mu} \partial_\mu h_0{\cos\beta}-i {\cos\alpha} {g_2} G^- {W^+_\mu} \partial_\mu h_0 {\cos\beta}+i {g_2} {\sin\alpha} G^+{W^-_\mu} \partial_\mu h_0 {\cos\beta} \nonumber \\
&&-i {g_2} {\sin\alpha} G^- {W^+_\mu} \partial_\mu h_0 {\cos\beta}+{\sin\beta}
   \partial_\mu G^+ \partial_\mu H^- {\cos\beta}+{\sin\beta} \partial_\mu G^- \partial_\mu H^+ {\cos\beta}\nonumber \\
&& +{g_2}^2 {O^\chi_{21}}^2
   {\chi}^2 {W^-_\mu} {W^+_\mu}+{g_2}^2 {O^\chi_{2i}}^2 {G^0_i}^2 {W^-_\mu} {W^+_\mu}\nonumber \\
&&+{\cos\alpha}^2
   {g_2}^2 {h_0}^2 {W^-_\mu} {W^+_\mu}+{g_2}^2 {\sin\alpha}^2 {h_0}^2 {W^-_\mu} {W^+_\mu}+2
   {g_2}^2 {O^\chi_{21}} {O^\chi_{2i}} {\chi} {G^0_i} {W^-_\mu} {W^+_\mu} \nonumber \\
&& +2 {\cos\alpha} {g_2}^2 {\sin\alpha}
   {h_0} {h_0} {W^-_\mu} {W^+_\mu}+{g_2}^2 {\sin\beta}^2 { H^-} {H^+} {W^-_\mu}
   {W^+_\mu}+{g_2} {O^\chi_{21}} {\sin\beta} {H^+} {W^-_\mu} \partial^\mu{\chi}\nonumber \\
&& +{g_2} {O^\chi_{21}} {\sin\beta}
   { H^-} {W^+_\mu} \partial^\mu{\chi}+{g_2} {O^\chi_{2i}} {\sin\beta} {H^+} {W^-_\mu} \partial_\mu G^0_i
+{g_2}{O^\chi_{2i}} {\sin\beta} { H^-} {W^+_\mu} \partial_\mu G^0_i \nonumber \\
&&+i {\cos\alpha}
   {g_2} {\sin\beta} {H^+} {W^-_\mu} \partial_\mu h_0-i {\cos\alpha} {g_2} {\sin\beta} { H^-} {W^+_\mu}
   \partial_\mu h_0+i {g_2} {\sin\alpha} {\sin\beta} {H^+} {W^-_\mu} \partial_\mu h_0 \nonumber \\
&& -i {g_2} {\sin\alpha} {\sin\beta}
   { H^-} {W^+_\mu} \partial_\mu h_0 -{g_2} {O^\chi_{21}} {\sin\beta}
   {\chi} {W^+_\mu} \partial_\mu H^--{g_2} {O^\chi_{2i}} {\sin\beta} {G^0_i} {W^+_\mu} \partial_\mu H^- \nonumber \\
&& +i {\cos\alpha}
   {g_2} {\sin\beta} {h_0} {W^+_\mu} \partial_\mu H^-+i {g_2} {\sin\alpha} {\sin\beta} {h_0} {W^+_\mu}
   \partial_\mu H^--{g_2} {O^\chi_{21}} {\sin\beta} {\chi} {W^-_\mu} \partial_\mu H^+ \nonumber \\
&& -{g_2} {O^\chi_{2i}} {\sin\beta}
   {G^0_i} {W^-_\mu} \partial_\mu H^+-i {\cos\alpha} {g_2} {\sin\beta} {h_0} {W^-_\mu} \partial_\mu H^+
-i {g_2}{\sin\alpha} {\sin\beta} {h_0} {W^-_\mu} \partial_\mu H^+ \nonumber \\
\eeqa

Other terms

\beqa
{\cal L_\theta}=\sum_{i=u,d} v^i\left( \theta^{(i)}_1{\alpha^{(i)}}^2 + \theta^{(i)}_2 {\beta^{(i)}}^2
+\theta^{(i)}_3\alpha^{(i)} + \theta^{(i)}_4 \beta^{(i)} + \theta^{(i)}_5 \right)
\eeqa

\beqa
&&\theta^{(u)}_1=0 \nonumber \\
&&\theta^{(u)}_2=
2 {\sin\alpha} {H_0}-2 {\cos\alpha} {H_0}\nonumber \\
&&\theta^{(u)}_{3\mu}=
{g_2} {\sin\beta} {G^+} {W^-_\mu}-{\cos\beta} {g_2} {H^+} {W^-_\mu}+{g_2} {\sin\beta} {G^-}
   {W^+_\mu}-{\cos\beta} {g_2} {H^-} {W^+_\mu}\nonumber \\
&&\theta^{(u)}_{4\mu}=
{g_2} {\sin\beta} {G^+} {W^-_\mu}-{\cos\beta} {g_2} {H^+} {W^-_\mu}+{g_2} {\sin\beta} {G^-}
   {W^+_\mu}-{\cos\beta} {g_2} {H^-} {W^+_\mu}
+2 {O^\chi_{11}} \partial_\mu{\chi} \nonumber \\
&&\qquad\qquad +2 {O^\chi_{1i}}\partial_\mu{G^0_i}\nonumber \\
&&\theta^{(u)}_5=2 {\sin\alpha} {H_0} {W^-_\mu} {W^+_\mu} {g_2}^2-2
{\cos\alpha} {H_0} {W^-_\mu} {W^+_\mu} {g_2}^2+i{\sin\beta} {W^+_\mu}\partial^\mu {G^-} {g_2}\nonumber \\
&&\qquad \qquad -i {\sin\beta} {W^-_\mu} \partial^\mu{G^+} {g_2}
\eeqa

\beqa
&&\theta^{(d)}_1=0 \nonumber \\
&&\theta^{(d)}_2=2 {\cos\alpha} {H_0}+2 {\sin\alpha} {H_0}\nonumber \\
&&\theta^{(d)}_{3\mu}={\cos\beta} {g_2} {G^+} {W^-_\mu}+{g_2}
{\sin\beta} {H^+} {W^-_\mu}+{\cos\beta} {g_2} {G^-}{W^+_\mu}+{g_2} {\sin\beta} {H^-} {W^+_\mu}\nonumber \\
&&\theta^{(d)}_{4\mu}={\cos\beta} {g_2} {G^+} {W^-_\mu}+{g_2} {\sin\beta} {H^+} {W^-_\mu}
+{\cos\beta} {g_2} {G^-}
   {W^+_\mu}+{g_2} {\sin\beta} {H^-} {W^+_\mu} \nonumber \\
&&\qquad \qquad +2 {O^\chi_{21}}\partial_\mu {\chi}+2 {O^\chi_{2i}} \partial_\mu{G_i}
\nonumber \\
&&\theta^{(d)}_5=2 {\cos\alpha} {H_0} {W^-_\mu} {W^+_\mu} {g_2}^2+2 {\sin\alpha} {H_0} {W^-_\mu} {W^+_\mu} {g_2}^2+i
   {\cos\beta} {W^+_\mu} \partial^\mu{G^-} {g_2} \nonumber \\
&& \qquad \qquad -i {\cos\beta} {W^-_\mu} \partial^\mu{G^+} {g_2}
\eeqa

\newpage



\end{document}